\documentclass[aps,prd,amsmath,amssymb,10pt,letterpaper,balancelastpage,superscriptaddress,nofootinbib,floatfix,notitlepage,twocolumn]{revtex4-2}

\usepackage{graphicx}	
\usepackage{ragged2e}	
\usepackage{bm}		    
\usepackage[sort&compress]{natbib}	 	
	\setcitestyle{square,numbers,comma}	
\usepackage[colorlinks=true,urlcolor=blue,linkcolor=black,citecolor=red]{hyperref}	

\usepackage{xcolor}
\usepackage{fontawesome5}
\definecolor{orcidlogocol}{rgb}{0.65, 0.807, 0.223}
\newcommand{\orcid}[1]{\,\href{https://orcid.org/#1}{\textcolor{orcidlogocol}{\footnotesize\faOrcid}}\,}
\newcommand{\tabref}[2][]{Tab{#1}.~\ref{#2}}		
\newcommand{\figref}[2][]{Fig{#1}.~\ref{#2}}		
\newcommand{\secref}[2][]{Sec{#1}.~\ref{#2}}		
\newcommand{\appref}[2][x]{Appendi{#1}~\ref{#2}}	
\renewcommand{\eqref}[2][]{Eq{#1}.~(\ref{#2})}		
\newcommand{\citeR}[2][]{Ref{#1}.~\cite{#2}}			
\newcommand{\lb}{\ensuremath{\left}}					
\newcommand{\rb}{\ensuremath{\right}}					
\newcommand{\nl}{\nonumber \\* & \quad }				
\newcommand{\dm}{\textsc{dm}}
\newcommand{\pbh}{\textsc{pbh}}
\newcommand{\grb}{\textsc{grb}}

\begin{document}

\title{Picolensing as a Probe of Primordial Black Hole Dark Matter}

\author{Michael A.~Fedderke\orcid{0000-0002-1319-1622}}
\email{mfedderke@perimeterinstitute.ca}
\affiliation{Perimeter Institute for Theoretical Physics, 31 Caroline Street North, Waterloo, Ontario, N2L 2Y5, Canada}
\author{Sergey Sibiryakov\orcid{0000-0001-9972-8875}}
\email{ssibiryakov@perimeterinstitute.ca, sibiryas@mcmaster.ca}
\affiliation{Perimeter Institute for Theoretical Physics, 31 Caroline Street North, Waterloo, Ontario, N2L 2Y5, Canada}
\affiliation{Department of Physics \& Astronomy, McMaster University, 1280 Main Street West, Hamilton, Ontario, L8S 4M1, Canada}

\begin{abstract}
The gravitational-lensing parallax of gamma-ray bursts (GRBs) is an intriguing probe of primordial black hole (PBH) dark matter in the asteroid-mass window, $2\times 10^{-16}M_{\odot} \lesssim M_{\pbh} \lesssim 5 \times 10^{-12}M_{\odot}$.
Recent work in the literature has shown exciting potential reach for this ``picolensing'' signal if a future space mission were to fly two x-/$\gamma$-ray detectors in the \emph{Swift}/BAT class, with inter-spacecraft separation baselines on the order of the Earth--Moon distance. 
We revisit these projections with a view to understanding their robustness to various uncertainties related to GRBs.
Most importantly, we investigate the impact of uncertainties in observed GRB angular sizes on reach projections for a future mission.
Overall, we confirm that picolensing shows great promise to explore the asteroid-mass window; however, we find that previous studies may have been too optimistic with regard to the baselines required.
Detector baselines on the order of at least the Earth--L2 distance would make such a mission more robust to GRB size uncertainties; baselines on the order of an astronomical unit (AU) would additionally enable reach that equals or exceeds existing microlensing constraints up to $M_{\pbh}\sim 2\times 10^{-8}M_{\odot}$.
\end{abstract}
\maketitle

\section{Introduction}
\label{sec:introduction}

The nature and identity of dark matter (DM) remains a mystery.
Despite decades of experimental progress to elucidate whether and how it may couple to the Standard Model (SM), evidence of any non-gravitational interactions the DM may have is elusive.
DM searches are complicated by the fact that there are viable DM candidates that span an enormous range of possible masses, from wavelike bosonic candidates as light as $\sim 10^{-19}$\,eV~\cite{Dalal:2022rmp} (or thereabouts~\cite{Rogers:2020ltq,DES:2020fxi,Irsic:2017yje}) up to supermassive, non-elementary dark objects that can be a sizeable fraction of a solar mass, or even heavier.
No single non-gravitational search technique can target this entire mass range: different techniques must be tailored to different candidates of interest.

One particularly interesting DM candidate is sub-solar-mass primordial black holes (PBHs)~\cite{1971MNRAS.152...75H}; see \citeR[s]{Carr:2020vsd,Green:2020jor,Escriva:2022duf,GREEN2024116494,Carr:2024nlv,Khlopov:2008qy,Belotsky:2014kca} for recent reviews.
At present, there is a window of masses $2\times 10^{-16}M_{\odot} \lesssim M_{\pbh} \lesssim 5 \times 10^{-12}M_{\odot}$ for which such small black holes may still constitute all of the dark matter; see \citeR{Gorton:2024cdm} for a recent review.
This is sometimes called the `asteroid-mass window', because the masses of such objects ($4\times 10^{17}\,\text{g} \lesssim M_{\pbh} \lesssim 1\times 10^{22}\,\text{g}$) are coincidentally broadly comparable to those of Solar-system asteroids.
Lighter PBHs are ruled out because they would Hawking evaporate~\cite{1974Natur.248...30H} too rapidly, contributing an unacceptably large extra component to a number of observables (e.g., the extragalactic $\gamma$-ray flux, the galactic 511\,keV line, Voyager $e^\pm$ measurements, etc.)~\cite{Carr:2009jm,DeRocco:2019fjq,Laha:2019ssq,Clark:2016nst,Mittal:2021egv,Boudaud:2018hqb,Laha:2020ivk,Kim:2020ngi,Dasgupta:2019cae,Coogan:2020tuf,Su:2024hrp,Tan:2024nbx,DelaTorreLuque:2024qms,Korwar:2023kpy} (see also, e.g., \citeR[s]{Ray:2021mxu,Profumo:2024fxq}).
On the other hand, heavier PBHs, at least up to $\sim 30\, M_{\odot}$, are ruled out as the dominant DM component by the absence of microlensing observations~\cite{Macho:2000nvd,EROS-2:2006ryy,Griest:2013aaa,Niikura:2017zjd,Oguri:2017ock,Zumalacarregui:2017qqd,Smyth:2019whb,Leung:2022vcx,Esteban-Gutierrez:2023qcz,Mroz:2024mse,Mroz:2024wag,Mroz:2024wia} (see also \citeR[s]{Bai:2018bej,DeRocco:2023gde,Tamta:2024pow}), with other limits beyond.

Previously, it was claimed~\cite{PhysRevD.86.043001} that some fraction of the asteroid-mass window is constrained by femtolensing~\cite{1992ApJ...386L...5G}; it was however subsequently shown that this observable is not robust to finite GRB source sizes~\cite{Katz:2018zrn}.
There are additionally possible constraints in the window from the destruction of old white dwarfs by the transit of PBH DM~\cite{Graham:2015apa} (see \citeR{Bramante:2015cua} for related ideas), but the robustness of those bounds has also been questioned~\cite{Montero-Camacho:2019jte}.

Gravitational-wave signatures of PBH mergers (see, e.g., \citeR{Profumo:2024okx}), signatures of PBHs captured in extrasolar systems (see, e.g., \citeR{Lehmann:2022vdt}), Lyman-$\alpha$ signatures of PBH evaporation (see, e.g., \citeR{Saha:2024ies}), and other more exotic possibilities (see, e.g., \citeR[s]{Dai:2024guo,Lehmann:2019zgt}) have all recently been studied.

There is no known SM mechanism to produce sub-solar-mass black holes.
PBH production mechanisms therefore usually rely on some non-standard feature or spike in the primordial density power spectrum that seeds structure in the Universe~\cite{Carr:2020vsd,Green:2020jor,Escriva:2022duf,GREEN2024116494,Carr:2024nlv,Khlopov:2008qy}.
When a sufficiently large primordial overdensity re-enters the horizon, it can undergo collapse to a black hole; typically, the smaller the horizon size at re-entry, the smaller the PBH mass.
A PBH in the asteroid-mass window would correspond to horizon re-entry around a time~\cite{Carr:2020vsd,Escriva:2022duf,GREEN2024116494} $t \sim 10^{-18}\,\text{s}\,\times M_{\pbh} / (10^{-12} M_{\odot})$.
Numerous other production scenarios have also been discussed; see \citeR[s]{Carr:2020vsd,Green:2020jor,Escriva:2022duf,GREEN2024116494,Khlopov:2008qy} and references therein.

One promising technique to access this range of PBH masses is gravitational-lensing parallax~\cite{Nemiroff:1995ak,Kolb:1995bu,Marani:1998sh,Nemiroff:1998zi}, also known in this mass range as `picolensing'~\cite{Kolb:1995bu} (after the picoarcsecond Einstein angles that are characteristic of cosmologically distant gravitational lenses in the window).
If two spatially separated detectors observe the same distant source at the same time, the line of sight from one detector to the source may have an impact parameter with respect to an intervening PBH lens that is on the order of the latter's Einstein radius, while the other line of sight does not.%
\footnote{\label{ftnt:bothWithin}%
    As will be made clear in the remainder of the paper, this is actually too na\"ive a criterion.
    For small observer separations, both lines of sight can pass nearby the same lens, but with different impact parameters and still give rise to a differential magnification large enough to be observable.
} %
This results in a large magnification of the source brightness as seen by the first observer, as compared to that observed by the second.
Initial estimates made in the 1990s~\cite{Nemiroff:1995ak,Kolb:1995bu,Marani:1998sh,Nemiroff:1998zi} indicated promising potential reach for this technique as applied to gamma-ray burst (GRB) sources, and two recent detailed studies have bolstered this conclusion~\cite{Jung:2019fcs,Gawade:2023gmt}

GRBs are some of the most energetic events in the Universe, with bright long-GRB events characterized by an isotropic equivalent energy%
\footnote{\label{ftnt:Eiso}%
    The isotropic equivalent energy (as measured at the source) is defined as $E_{\text{iso}} = 4\pi (D^{\text{lum}}_{S})^2 S_{\text{bol}}/(1+z_S)$ where $D^{\text{lum}}_{S}$ is the luminosity distance to the GRB at redshift $z_S$, and $S_{\text{bol}}$ is the total bolometric fluence (energy per area) of the GRB at the observer~\cite{Bloom_2001,Atteia:2017dcj}.  
    See also Appendix B of \citeR{Lien:2016zny}.
} %
on the order of a solar rest mass,%
\footnote{\label{ftnt:brightestGRB}%
    The brightest known long GRB has $E_{\text{iso}} \sim 10^{55}\,\text{erg}$~\cite{Burns_2023} and possesses an optical supernova counterpart~\cite{Blanchard:2024ank}.
} %
$E_{\text{iso}} \sim (\text{few})\times 10^{54}$ erg~\cite{Atteia:2017dcj}; see \citeR[s]{Meszaros_2006,DAVANZO201573,2012Sci...337..932G} for reviews.
GRB emission is however extremely beamed (Lorentz boosts $\Gamma \gtrsim 10^2$), so the actual energy release is only on the order of $E_{\text{emit}} \sim E_{\text{iso}}/\Gamma^2 \sim (\text{few}) \times 10^{50}\,\text{erg}$~\cite{Frail:2001qp}, which is nevertheless comparable to that of a Type Ia supernova.
There is strong evidence (see, e.g., \citeR{Golkhou:2015lsa}) for a bimodal distribution of GRB events: the so-called `short' ($T_{90}<2\,\text{s}$) and `long' ($T_{90}>2\,\text{s}$) GRBs, which are distinguished by the duration $T_{90}$ over which 90\% of the observed GRB event fluence is recorded.
At least some short GRBs are thought to originate from neutron-star mergers; they are typically dimmer and more local.
On other hand, long GRBs are less well understood, but are thought to be associated with supernovae/hypernovae of extremely massive stars (cf.~\citeR{Blanchard:2024ank}); they are brighter and cosmologically distant.

The utility of these transient GRB sources as regards picolensing is twofold: (1) the extreme brightness of long GRBs makes them observable at cosmological distances (i.e., redshifts of a few), so lines of sight to the source traverse a sizeable fraction of the observable universe, increasing the lensing optical depth; and (2) they are bright in the hard x-ray / soft $\gamma$-ray regime for which geometrical optics remains a good approximation to compute PBH lensing effects over most of the asteroid-mass window.
Furthermore, the observed part of the emission regions of GRBs are thought to have transverse angular sizes that also make them among the most compact objects on the sky that are observable at cosmological distances; nevertheless, they are still comparable in angular size to the Einstein angles for lensing in the PBH asteroid-mass window, which can adversely impact lensing magnifications.
The sizes of these emission regions are also not directly measured and are therefore highly uncertain.

Using a moderate, fiducial assumption about GRB emission-region sizes~\cite{Barnacka:2014yja} and assuming a disk intensity model for the GRB emission, \citeR{Gawade:2023gmt} made a comprehensive investigation of the prospects for picolensing as a PBH probe.
Their main conclusion was that, with a sample of $N_{\grb} \sim 10^4$ GRBs and using realistic detector assumptions based on the proposed ISRO \emph{Daksha}~\cite{Bhalerao:2022edb,Bhalerao:2022pon} mission (which would deploy two hard x-ray / soft $\gamma$-ray detectors in the \emph{Swift}/BAT class~\cite{2004ApJ...611.1005G,2005SSRv..120..143B,Sakamoto_2008,2011ApJS..195....2S,Lien:2016zny}), one could probe the majority of the PBH asteroid-mass window if the two detectors were separated by around the Earth--Moon distance, but not if they were placed in low Earth orbit (as assumed for the actual \emph{Daksha} proposal).

Following the methodology of \citeR{Gawade:2023gmt}, we revisit this topic in order to further explore two points: (1) how robust picolensing is to various uncertainties, such as the significant uncertainties in GRB source sizes (here we mean both systematic and statistical modifications to a fiducial source-size assumption, as well as placing cuts on the smallest source size), the choice of the GRB intensity profile on the sky (Gaussian vs.~disk), and detector background levels (to simulate baselines that may require spacecraft in higher-background orbits); and (2) to re-examine the results of \citeR{Jung:2019fcs}, which showed that using a much longer baseline (e.g., Earth--L2, or a few AU) between the detectors could expand the PBH parameter space one can probe and bolster the robustness of the probe to the uncertainties noted above.
We also develop some approximate analytical understanding for the scalings of our results with physical parameters such as the PBH mass $M_{\pbh}$, the observer separation $R_O$, and the source angular size $\theta_S$.

\begin{figure*}[t]
    \centering
    \includegraphics[width=0.75\textwidth]{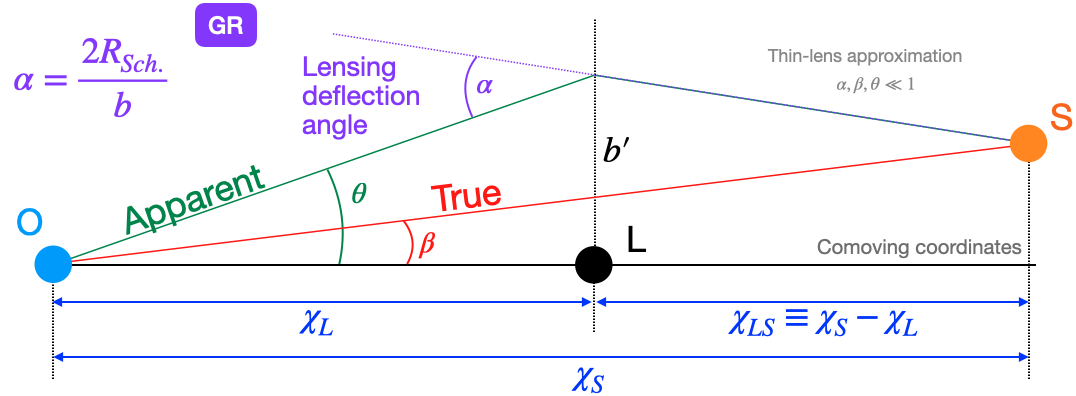}
    \caption{\label{fig:lensing}%
    Lensing geometry (not to scale). 
    This diagram is drawn in comoving coordinates: away from compact mass distributions, light rays thus travel on straight lines even in Friedmann--Lema{\^i}tre--Robertson--Walker (FLRW) spacetime.
    Absent lensing, an observer (O, blue) would see a source (S, orange) at a true angular location $\beta$ with respect to some arbitrary reference direction.
    The source is at comoving distance $\chi_S$.
    Suppose that we place a PBH lens (L, black) in the reference direction, so that the source has a true angular offset $\beta$ from the lens (red line). 
    The lens is at comoving distance $\chi_L$.
    Because of gravitational lensing (shown here in the thin-lens, small-angle approximation), a light ray going past the lens suffers a sharp deflection by an angle $\alpha$.
    The source will therefore appear to the observer to be instead at an apparent angular offset $\theta$ with respect to the lens (green line).
    The lensing deflection angle given by general relativity is $\alpha = 2R_{\text{Sch.}}/b$, where $R_{\text{Sch.}} = 2G_{\text{N}}M$ is the Schwarzschild radius of the PBH lens and $b$ is the physical impact parameter of the light ray past the lens, where $b' = (1+z_L)b$ relates the physical and comoving impact parameters ($z_L$ is the lens redshift).
    This diagram is drawn such that the observer, lens, and source are all coplanar.
    }
\end{figure*}

One of our main conclusions is that GRB size uncertainties are highly significant, and must be accounted for in making picolensing reach projections.
For instance, we agree with the finding of \citeR{Gawade:2023gmt} that, with a moderate fiducial GRB size estimate, separations between two detectors on the order of the Earth--Moon separation may be able to give access to PBH DM in the asteroid-mass window. 
However, when conservative assumptions about GRB sizes are made (consistent with their uncertainties), separation baselines between the detectors up to and including the Earth--Moon distance become insufficient to robustly probe the asteroid-mass window, at least given a sample of $N_{\grb} \sim 3000$ GRBs (roughly twice the size of current \emph{Swift}/BAT catalogue~\cite{Sakamoto_2008,2011ApJS..195....2S,Lien:2016zny,SWIFTBATcurrent,SWIFTBATwhatWeUsed}, and 7--8 times the size of the fraction of the catalogue that has good source-redshift determinations).
On the other hand, with separations of the Earth--L2 distance or larger, most of the PBH parameter space in the asteroid-mass window above $M_{\pbh} \sim 10^{-14}M_{\odot}$ can be \emph{robustly} probed, even under conservative GRB size assumptions.
With a much larger baseline on the order of an AU, we find similarly robust and deeper sensitivity in the window even under conservative assumptions, and also that sensitivity would equal or exceed that of some existing microlensing constraints, permitting new parameter space to be probed all the way up to $M_{\pbh} \sim 10^{-5}M_{\odot}$.
We find little difference in our results from the binary choice between a disk or Gaussian GRB intensity profile (provided that the characteristic size of the Gaussian profile is appropriately chosen in comparison to the disk radius).
Higher detector backgrounds or more demanding signal-to-noise requirements would mildly degrade some of these conclusions.

The remainder of this paper is structured as follows.
In \secref{sec:Lensing} we review the general theory of gravitational lensing (\secref{sec:gravLensing}), its application to picolensing (\secref{sec:picolensing}), the detection criteria for the latter based on \citeR{Gawade:2023gmt}~(\secref{sec:detectionCriteria}), the definition of the cross section and volume for picolensing (\secref{sec:picoSigma}), and finally the lensing probability, as well as how to set exclusion limits (\secref{sec:probAndExclusion}).
In \secref{sec:GRBs}, we discuss the GRB sources envisaged to be observed by a picolensing mission, reviewing existing observations with \emph{Swift}/BAT and other facilities (\secref{sec:SWIFT-BAT}), as well as GRB angular sizes and their uncertainties (\secref{sec:GRBsizes}).
We then provide some more technical details of our computational implementation in \secref{sec:ourComputation}, the results of which we show and discuss in \secref{sec:results}.
We offer some thoughts about the implications of a future detection of picolensing in \secref{sec:beyondConstraints}.
We conclude in \secref{sec:Conclusion}.
In \appref{app:lensingPot} we give some extra details about the lensing potential.
In \appref{app:Gaussian} we discuss the magnification for disk and Gaussian source profiles.
In \appref{app:sigmaGaussianAnalyt} we derive the parametric dependence of the picolensing cross section and volume on relevant physical parameters.
Finally, in \appref{app:differentApproach} we apply a slightly modified statistical procedure and show that our results are robust.

\section{Gravitational (Pico)Lensing}
\label{sec:Lensing}
In this section, we review the basics of gravitational lensing and its application to the picolensing signal of interest in this paper.

\subsection{Gravitational Lensing}
\label{sec:gravLensing}
We review the theory of gravitational lensing for point sources and then for extended sources, in both cases for point lenses.
For an in-depth review of gravitational lensing, see \citeR[s]{Wambsganss:1998gg,Bartelmann:2010fz}.

\subsubsection{Point Sources}
\label{sec:pointSource}
Imagine an observer who views a distant point source through an intervening mass distribution; see \figref{fig:lensing}.
Absent the mass distribution, suppose that the observer would see the source to be at a true angular location%
\footnote{\label{ftnt:vectors}%
    Our convention is to use bold notation for a spatial 3-vector $\bm{\chi}$, bold notation with a subscript $\perp$ (i.e., $\bm{\chi}_\perp$) for a spatial 2-vector in a plane transverse to specific given direction, and the notation $\vec{\theta}$ for a vector of angles locating an object on the celestial 2-sphere. 
    When we write equalities such as $\bm{\chi}_\perp = \chi \vec{\theta}$, we mean these expressions are equal component-by-component.
    Derivatives such as $\vec{\nabla}$ are covariant derivatives with respect to the angular coordinates on the 2-sphere, although for small deflections a flat-sky approximation is sufficient.
} %
$\vec{\beta}$ on the celestial 2-sphere.
Owing to the bending of light rays in the presence of the intervening mass--energy, the observer will instead see the source to be at apparent angular location $\vec{\theta}$.
In flat FLRW spacetime and in the limit of small angular deflections, the true and apparent locations are related by the so-called lensing potential $\psi(\vec{\theta})$ as follows~\cite{Bartelmann:2010fz}:%
\footnote{\label{ftnt:hbarc}%
    We work in units where $\hbar=c=1$.
} %
\begin{align}
     \vec{\beta} &= \vec{\theta} - \vec{\nabla} \psi( \vec{\theta} ) \label{eq:lensEqn1} \\
   \psi &= 2 \int_0^{\chi_S} \frac{\chi_S-\chi'}{\chi_S\cdot\chi'} \phi( \bm{\chi}_\perp = \chi' \vec{\theta} , \chi^3 = \chi' )d\chi' \:,\label{eq:lensingPotential}
\end{align}
where, in the small-angle approximation, $\bm{\chi}_\perp \approx \chi \vec{\theta}$ are the comoving coordinates in the transverse plane at comoving distance $\chi$, the angular gradient is $\vec{\nabla} \approx \chi \bm{\nabla}_{\perp} \equiv \chi \partial/\partial \bm{\chi}_{\perp}$, $\chi_S$ is the comoving distance of the source, and $\phi$ is the Newtonian potential of the intervening mass distribution.%
\footnote{\label{ftnt:care_with_redshifts}%
    Specifically, working in Newtonian gauge, we write \[ ds^2 \approx [a(\eta)]^2 \lb[ (1+2\phi) d\eta^2 - (1-2\phi)d\bm{\chi}^2 \rb]. \]
    Throughout the paper, we keep terms only to linear order in $\phi$.
} %
We have also assumed without loss of generality that the origin of the coordinate system is centered on the observer, and that the 3-axis is aligned to point from the observer to the centroid of the lensing mass distribution (assumed here to be localized).

For a point lens of mass $M$ located at comoving distance $\chi=\chi_L$, we have $\phi = -G_{\text{N}} M / \delta r_{\text{phys}}$ where $\delta r_{\text{phys}}$ is the physical distance away from the lens where the potential is being evaluated.
Because $\phi$ is only significantly different from 0 over a range $\delta r_{\text{phys}} \sim\mathcal{O}( G_{\text{N}}M) \ll a_L \chi_L$ and we assume $\theta \ll 1$ (small source and small deflections), we can approximate $\delta r_{\text{phys}} \approx a_L \sqrt{ (\chi-\chi_L)^2 + \chi_L^2 \theta^2 }$, where $a_L=(1+z_L)^{-1}$ is the scale factor at the location of the lens, which is at redshift $z_L$.
Then, to leading order in $\theta$, we have (see \appref{app:lensingPot} for a derivation)
\begin{align}
    \vec{\nabla} \psi \approx \frac{\theta_E^2}{\theta} \hat{\theta}\: ,
\end{align}
where $\hat{\theta}$ is the unit vector in the direction of $\vec{\theta}$, and the Einstein angle $\theta_E$ is defined as~\cite{1980ApJ...241..507Y}
\begin{align}
    \theta_E^2 &\equiv 4G_{\text{N}}M(1+z_L) \frac{ \chi_{LS} }{\chi_L\cdot \chi_S} \quad [\chi_{LS} \equiv \chi_S -\chi_L] \label{eq:thetaE1}\\
    &= \frac{4G_{\text{N}}M(1+z_L)}{\chi_S} \frac{ 1-\xi }{\xi } \qquad [\xi\equiv \chi_L/\chi_S] \:. \label{eq:thetaE2}
\end{align}
Defining the (physical) angular diameter distance of an object at comoving distance%
\footnote{\label{ftnt:FLRW}%
    In a flat FLRW cosmology, the comoving distance $\chi_i$ is related to the angular diameter distance $\mathcal{D}_i$ by $\mathcal{D}_i = \chi_i / (1 + z_i)$, where $\chi_i \equiv \int_0^{z_i} [H(z)]^{-1} dz$ with $[H(z)]^2 \equiv 8\pi\rho(z)/(3M_{\text{Pl.}}^2)$, where $\rho(z)$ is the (physical) energy density at redshift $z$ and $M_{\text{Pl.}} \equiv 1/\sqrt{G_{\text{N}}}$ is the Planck mass [$G_{\text{N}}$ is Newton's constant]. 
    If $H_0=67.66\,\text{km/s/Mpc}$~\cite{Planck:2018vyg} is the Hubble parameter and $\Omega_j \equiv \rho_j/\rho_c$ is the fractional energy density in component $j$ (with equation-of-state $P_j = w_j \rho_j$) as compared to the critical density $\rho_c = (3H_0^2M_{\text{Pl.}}^2)/(8\pi)$ for a flat universe, then $\rho(z) = \rho_c \sum_j \Omega_j ( 1 + z )^{3(1+w_j)}$. 
    We also define $H_0 \equiv (100 h)\,\text{km/s/Mpc} \equiv h \tilde{H}_0$ and $\omega_j \equiv \Omega_jh^2$, with $h=0.6766$.
} %
$\chi_i$ to be $\mathcal{D}_i = a_i \chi_i$, and defining $\mathcal{D}_{LS} = a_S \chi_{LS} = \mathcal{D}_S - (a_S/a_L) \mathcal{D}_L$, we have the more familiar form of the Einstein angle~\cite{Bartelmann:2010fz,Katz:2018zrn,1980ApJ...241..507Y}:
\begin{align}
    \theta_E = \sqrt{ 4G_{\text{N}}M \frac{ \mathcal{D}_{LS} }{ \mathcal{D}_L \mathcal{D}_S} }\:. \label{eq:compToAppendixExpr}
\end{align}
We can also refine the (physical) Einstein radius (in the lens plane), $R_E \equiv \mathcal{D}_L \theta_E \equiv \chi_E / ( 1 + z_L) = \chi_L \theta_E / ( 1 + z_L ) \Rightarrow \chi_E \equiv \chi_L \theta_E$, leading to
\begin{align}
    R_E &= \sqrt{ 4G_{\text{N}}M \frac{ \mathcal{D}_{LS} \mathcal{D}_L  }{ \mathcal{D}_S}} \:; \\
    \chi_E &= \sqrt{ 4G_{\text{N}}M(1+z_L) \frac{ \chi_{LS}  \chi_L  }{ \chi_S } } \\
    &= \sqrt{ 4G_{\text{N}}M\chi_S(1+z_L)\cdot \xi(1-\xi) }\:.
\end{align}
From \eqref{eq:lensEqn1}, the lens equation is then
\begin{align}
    \beta \hat{\beta} = ( \theta - \theta_E^2/\theta ) \hat{\theta} \: ,
\end{align}
which can only be satisfied if $\hat{\theta} = \hat{\beta}$ and
\begin{align}
    y = x - 1/x \: ,
\end{align}
where we have defined $y \equiv \beta / \theta_E$ and $x\equiv \theta / \theta_E$.
In general there are two solutions of this equation:
\begin{align}
   x = \frac{1}{2} \lb[ y \pm \sqrt{ 4 + y^2 } \rb] \: ,
\end{align}
corresponding to two images of the source, one appearing on either side of the line joining the observer to the lens.
For a source directly behind the lens ($y=0 \Rightarrow x = \pm 1$), the image will actually be a ring at $\theta = \theta_E$.

If we define the area magnification factor to be the ratio of the apparent and true angular sizes of the source,%
\footnote{\label{ftnt:axialSym}%
    Here, we used axial symmetry of the point lens to write $d\phi_{\text{apparent}} = d\phi_{\text{true}}$.
} %
$\mu \equiv d\Omega_{\text{apparent}}/d\Omega_{\text{true}} = \sin \theta d\theta / \sin\beta d\beta \approx (\theta/\beta)( d\theta /d\beta) = (x/y)(dx/dy)$, we have
\begin{align}
    \mu_\pm &= \frac{1}{2}\lb|  1 \pm \frac{y^2+2}{y\sqrt{y^2+4}} \rb|\:.
\end{align}
Because gravitational lensing preserves surface brightness~\cite{Wambsganss:1998gg}, the area magnification is also the apparent brightness magnification of the source owing to the lens.

In the event that the two images are not resolved by the observer (e.g., when $\theta_E \ll 1$) and assuming that geometrical optics holds, so that interference phenomena between the two images are irrelevant (see also \citeR{Katz:2018zrn}), these magnifications will add to give a single combined magnification of the source as viewed by the observer:
\begin{align}
    \mu &= \frac{y^2+2}{\sqrt{y^2(y^2+4)}} \rightarrow \begin{cases} 1/|y| &; |y|\ll 1\:; \\ 1 + 2/y^4 &; |y| \gg 1\:. \end{cases} \label{eq:pointLPointSmag}
\end{align}

\subsubsection{Extended Sources}
\label{sec:extendedSource}
An extended source can be considered to be a collection of point sources with a (true, unlensed) smooth intensity distribution $I(\bm{\chi}_{S,\perp})$ as viewed from the observer plane.
Consider a coordinate system whose $3$-axis points from the observer to the source centroid, and let a lens be located at comoving coordinates $\bm{\chi}_L = \bm{\chi}_{L,\perp} + \chi_L \bm{\hat{e}_3}$. 
The total magnification of the source can be obtained by appropriately convolving the point-source magnification $\mu(y)$ from \eqref{eq:pointLPointSmag} with the source intensity distribution~\cite{Katz:2018zrn}:
\begin{align}
    \bar{\mu}(\bm{\chi}_L) &= \frac{ \int I(\bm{\chi}_{S,\perp} ) \cdot \mu[y(\bm{\chi}_{S},\bm{\chi}_L)]\, d^2\bm{\chi}_{S,\perp}}{\int I(\bm{\chi}_{S,\perp} )\, d^2\bm{\chi}_{S,\perp}}\: , \label{eq:convolveIntensity}
\end{align}
where $y(\bm{\chi}_{S},\bm{\chi}_L) \equiv \beta(\bm{\chi}_{S},\bm{\chi}_L)/\theta_E$ is the true angular offset (in units of the Einstein angle), as viewed from the observer, of the location $\bm{\chi}_{S} = \bm{\chi}_{S,\perp} + \chi_S \bm{\hat{e}_3}$ on the source with respect to the lens:
\begin{align}
    y(\bm{\chi}_{S},\bm{\chi}_L) & \approx \lb| \frac{\bm{\chi}_{L,\perp}}{\chi_E^L} - \frac{\bm{\chi}_{S,\perp}}{\chi_E^S} \rb| \:;
\end{align}
this result is correct in the small-angle approximation.

\begin{figure}[t]
    \centering
    \includegraphics[width=0.75\columnwidth]{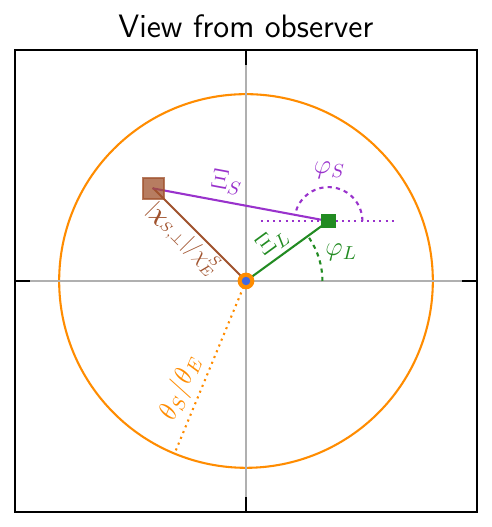}
    \caption{\label{fig:changeOfVariables}%
    Definition of the variables used in \eqref{eq:muBar}.
    An infinitesimal element of an extended source (orange circle) is shown by the brown square; the lens is shown by the green square.
    This diagram is intended to convey angular positional information (normalized to the Einstein angle) of various objects as they would appear to a single observer (blue dot) directing their gaze toward the source centroid (orange dot).
    That is, the lens is shown at the location $\vec{\chi}_L \equiv \bm{\chi}_{L,\perp}/\chi_E^L$ and the infinitesimal element of the source is shown to be at $\vec{\chi}_S \equiv \bm{\chi}_{S,\perp}/\chi_E^S$.
    This figure is accurate in the small-angle approximation.
    Although the source is shown here as being axially symmetric, these definitions and \eqref{eq:muBar} apply for arbitrary (localized) source intensity profiles.
    }
\end{figure}

Computation of the integral at \eqref{eq:convolveIntensity} is simplified if we change variables: 
\begin{align}
    \bm{\chi}_{L,\perp} &\equiv \chi_E^L \Xi_L \begin{pmatrix} \cos\varphi_L \\ \sin\varphi_L \end{pmatrix} \:; \\
    \bm{\chi}_{S,\perp} &\equiv \frac{\chi_S}{\chi_L} \bm{\chi}_{L,\perp} + \chi_E^S \Xi_S \begin{pmatrix} \cos\varphi_S \\ \sin\varphi_S  \end{pmatrix}\:, 
\end{align}
such that $d^2 \bm{\chi}_{S,\perp} = (\chi_E^S)^2 \Xi_S d\Xi_S d\varphi_S$; see \figref{fig:changeOfVariables}.
Then we have
\begin{align}
    y(\bm{\chi}_{S},\bm{\chi}_L) &= \Xi_S
\end{align}
and
\begin{align}
    \bar{\mu}(\bm{\chi}_L) &=  \int_0^\infty d\Xi_S \int_0^{2\pi} d\varphi_S  \times  I_0^{-1} 
    \label{eq:muBar}\\ &\qquad \times  I(\Xi_S,\varphi_S;\Xi_L,\varphi_L ) \cdot \Xi_S \cdot \mu[y=\Xi_S] \nonumber \:,
\end{align}
where $I_0$ is a normalization constant:
\begin{align}
    I_0 \equiv \int_0^\infty d\Xi_S \int_0^{2\pi} d\varphi_S \cdot \Xi_S \cdot I(\Xi_S,\varphi_S;\Xi_L,\varphi_L ) \:.
\end{align}

We also define $ | \bm{\chi}_{L,\perp}| = \chi_E^L \Xi_L \equiv \chi_L \beta_L \Rightarrow \Xi_L = \beta_L/\theta_E \equiv y_L$, where $\beta_L$ is the true angular offset of the lens from the source \emph{centroid}.\\

There are two cases for the source intensity distribution that we will investigate: the Gaussian and the flat disk.

\begin{figure*}[t]
    \centering
    \includegraphics[width=\textwidth]{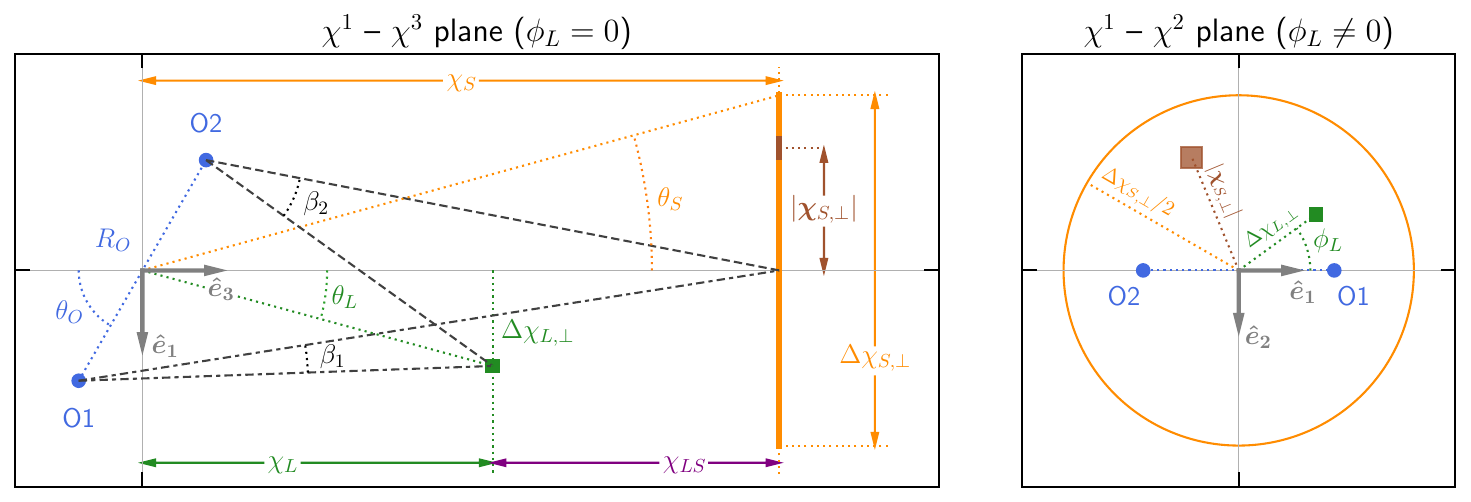}
    \caption{\label{fig:coords}%
    Picolensing geometry (not to scale; $\chi_L \gg \Delta \chi_{L,\perp}$ and $\chi_S \gg \Delta \chi_{S,\perp}$).
    These are comoving coordinates; flat Euclidean geometry may be used, and straight lines represent would-be photon geodesics in the absence of the lens.
    The solid blue dots marked O1 and O2 represent the observers.
    The green square marks a sample lens location; it is shown in the left panel at $\phi_L=0$, and in the right panel at $\phi_L \neq 0$ ($-\pi/2<\phi_L<0$).
    The thick orange bar (left panel) or orange circle (right panel) represents a source that is spatially extended in the transverse direction; an infinitesimal element of the source at transverse location $\bm{\chi}_{S,\perp}$ is marked in brown. 
    Various comoving distances and angles defined in the text are marked.
    Note that $\beta_{1,2}$ are the angular offsets of the lens location, as seen by each observer, compared to the source \emph{centroid} location; for $\phi_L \neq 0$, $\beta_i$ correctly accounts for the offset of the lens location out of the $\chi^1$ -- $\chi^3$ plane that contains O1, O2, and the source centroid [cf.~\eqref{eq:yi}].
    Owing to parallax, the angle $\phi_L$ as defined in this figure differs from the angle $\varphi_L$ defined in \figref{fig:changeOfVariables}; see footnote~\ref{ftnt:phiL}.
    }
\end{figure*}

\paragraph{Gaussian Case}
\label{eq:GaussianCase}
For the Gaussian case, we write 
\begin{align}
    I( \bm{\chi}_{S,\perp} ) &= \frac{1}{2\pi \delta^2}\exp\lb[ - \frac{\lb( |\bm{\chi}_{S,\perp}| / \chi_E^S \rb)^2 }{2\delta^2} \rb] \: ,
\end{align}
where we define $\delta \equiv \Delta \chi_{S,\perp} / ( 2 k_G \chi_E^S )$, where $\Delta \chi_{S,\perp}$ is a characteristic comoving transverse \emph{diameter} of the source and $k_G = \sqrt{ - 2 \ln(1-f_{\text{cover}}) }$ is a constant defined such that a fraction $0<f_{\text{cover}}<1$ of the total source intensity is contained within a disk of \emph{radius} $|\bm{\chi}_{S,\perp}| = \Delta \chi_{S,\perp}/2$:
\begin{align}
    \frac{\int  d^2\bm{\chi}_{S,\perp} I( \bm{\chi}_{S,\perp} ) \Theta(\Delta \chi_{S,\perp}-2|\bm{\chi}_{S,\perp}|)}{\int d^2 \bm{\chi}_{S,\perp} I( \bm{\chi}_{S,\perp} )} = f_{\text{cover}} \: ,
\end{align}
where $\Theta(x)$ is the Heaviside function, defined as
\begin{align}
    \Theta(x) &= \begin{cases}
                    1 &; x\geq 0\:; \\
                    0 &; x<0\:.
                 \end{cases}
\end{align}     
Throughout this paper, we choose $f_{\text{cover}}=0.9$, leading to $k_G\approx 2.1 \Rightarrow \delta \sim \Delta \chi_{S,\perp} / ( 4.3 \chi_E^S )$.
We also define $\Delta \chi_{S,\perp} \equiv 2 \theta_S \chi_S \Rightarrow \delta = \theta_S/(k_G\theta_E)$.

In terms of the redefined variables, we have
\begin{align}
   & I( \Xi_S, \varphi_S ; \Xi_L , \varphi_L ) \\
   &= \frac{1}{2\pi \delta^2} \exp\lb[ - \frac{ \Xi_L^2 + \Xi_S^2 + 2\Xi_S\Xi_L \cos(\varphi_S-\varphi_L ) }{2\delta^2} \rb]\nonumber \: ,
\end{align}
yielding $I_0 = 1$ and 
\begin{align}
    &\bar{\mu}(\bm{\chi}_L) \nonumber \\
    &= \frac{e^{-\Xi_L^2/(2\delta^2)}}{\delta^2} \nl \: \times \int_0^\infty d\Xi_S \frac{e^{-\Xi_S^2/(2\delta^2)} (2+\Xi_S^2) I_0( \Xi_L \Xi_S / \delta^2 ) }{\sqrt{4+\Xi_S^2}}\: \\ 
    &= \frac{e^{-\tilde{z}^2/2}}{\delta} \int_0^\infty d\tilde{x} \frac{e^{-\tilde{x}^2/2} (2+(\delta \cdot\tilde{x})^2) I_0( \tilde{x}\tilde{z} ) }{\sqrt{4+(\delta\cdot \tilde{x})^2}}\:, \label{eq:muG}
\end{align}
where $I_\nu$ is the modified Bessel function of the first kind of order $\nu$, and at the last step we defined $\tilde{x} \equiv \Xi_S/\delta$ and $\tilde{z}=\Xi_L/\delta=k_G \beta_L/\theta_S$.
This integral cannot in general be evaluated in closed form in terms of elementary functions, so we must resort to numerical methods.

There is however a special case that is tractable in closed form: $\Xi_L = 0$, the perfectly aligned lens. 
In that case, we have $\bar{\mu}(\bm{\chi}_L = \chi_L \bm{\hat{e}_3}) = \exp[1/\delta^2] K_1(1/\delta^2)/\delta^2$, where $K_\nu$ is the modified Bessel function of the second kind of order $\nu$.
This has limiting cases $\bar{\mu} \approx 1+1/\delta^2 \sim 1 + 18.4 ( \chi_E^S / \Delta \chi_{S,\perp} )^2$ for $\delta \gg 1$ and $\bar{\mu} \approx \sqrt{\pi/2}/\delta \sim 5.4 ( \chi_E^S / \Delta \chi_{S,\perp} )$ for $\delta \ll 1$.
In contrast to the point-source result at \eqref{eq:pointLPointSmag}, the finite source size regulates the divergence of the magnification at $y=0$.

Furthermore, one can find a set of good approximations to the integral at \eqref{eq:muG} that work in the small-source ($\delta \ll 1$) and large-source ($\delta \gg 1$) limits; see \appref{app:muGApprox}.

\paragraph{Disk Case}
\label{para:DiskCase}
In the case of a disk source, we set
\begin{align}
    I( \bm{\chi}_{S,\perp} ) &= \frac{1}{\pi \delta^2}\Theta( \delta^2 - | \bm{\chi}_{S,\perp} |^2 / (\chi_E^S)^2 )\: ,
\end{align}
where, in this case, we have $\delta \equiv \Delta \chi_{S,\perp} / ( 2 \chi_E^S ) = \theta_S/\theta_E$, where $\Delta \chi_{S,\perp}$ is the comoving transverse \emph{diameter} of the source.
It follows that $I_0 = 1$ and 
\begin{align}
    & I( \Xi_S, \varphi_S ; \Xi_L ,\varphi_L ) \label{eq:IDisk}\\
    &= \frac{1}{\pi \delta^2}\Theta\Big[ \delta^2 - \lb( \Xi_L^2 + \Xi_S^2 + 2\Xi_S\Xi_L \cos(\varphi_S-\varphi_L ) \rb)  \Big]\:. \nonumber
\end{align}

Substituting into \eqref{eq:muBar}, it is possible to perform the $\Xi_S$ integral analytically (see \appref{app:muIntegralDisk}), but the $\phi_S$ integral cannot in general be written in closed form in terms of elementary functions.
\citeR[s]{Jung:2019fcs,Gawade:2023gmt} provide an approximate fitting function for $\bar{\mu}$ based on the results of \citeR{1994ApJ...430..505W}.

Let us examine again the special case of the perfectly aligned lens; in that case, we have $\bar{\mu} \approx \sqrt{ 1 + 4/\delta^2}$.
For $\delta \gg 1$, this gives $\bar{\mu} \approx 1 + 2/\delta^2= 1 + 8(\chi_E^S/\Delta\chi_{S,\perp})^2$; for $\delta \ll 1$, we have $\bar{\mu} \approx 2/\delta = 4 ( \chi_E^S/\Delta \chi_{S,\perp} )$.
These disk expressions differ compared to the Gaussian results by a rescaling of $\Delta\chi_{S,\perp}$ by only an $\mathcal{O}(30$--$50\%)$ factor; we should thus expect similar lensing properties for Gaussian and disk sources given our choice of $f_{\text{cover}}$ for the former.

\subsection{Picolensing}
\label{sec:picolensing}

\begin{figure}[t]
    \centering
    \includegraphics[width=0.495\textwidth]{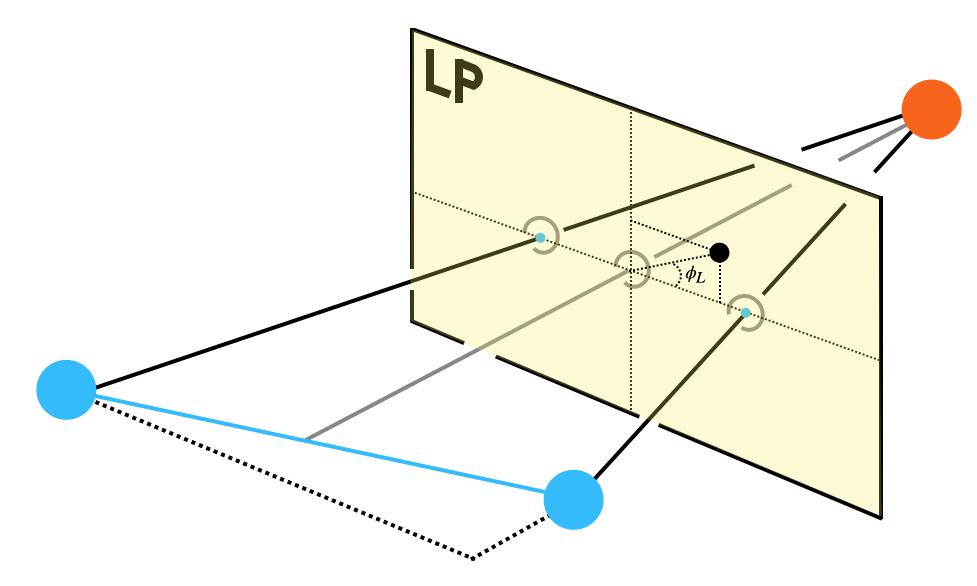}
    \caption{\label{fig:LPgeom}%
    Lens-plane (LP) geometry for the case of picolensing (not to scale).
    The orange circle represents the source; the large blue circles represent the observers.
    The yellow plane marked ``LP'' is (a segment of) the lens plane, which is pierced at the $(x,y)$ coordinate origin by its normal, which is the (gray) line joining the centroid of the source to the midpoint of the separation baseline between the observers (blue line).
    The $x$ [respectively, $y$] direction corresponds to the long [short] axis of the rectangular LP area as drawn; see also \secref{sec:MCsigma}.
    The observer lines of sight to the source centroid (black lines) pierce the lens plane symmetrically (in the small-angle approximation) at the locations marked by the small blue dots.
    A sample lens location in the lens plane is shown by the black dot.
    This diagram is shown in comoving coordinates; the opening angle subtended by the observer separation as viewed from the source should be understood to be many orders of magnitude smaller than shown.
    }
\end{figure}

Now suppose that there are two late-time observers at redshift $z_O = 0$ that are spatially separated by a physical distance $R_{O}$.
With reference to \figref[s]{fig:coords} and \ref{fig:LPgeom}, we define a coordinate system whose origin is centered at the midpoint between the two observers, with a $3$-axis pointing to the centroid of a distant source at comoving distance $\chi_S \gg R_{O}/a_S$, such that the two observers lie in the $1$--$3$ plane; the source centroid therefore has comoving coordinates $\bm{\chi}_S = \chi_S \bm{\hat{e}_3}$.
Specifically, let the observers lie at $\bm{\chi}_{1,\text{phys}} = \frac{1}{2} R_{O} \lb( \sin{\theta_O} \bm{\hat{e}_1} - \cos{\theta_O} \bm{\hat{e}_3} \rb)$ and $\bm{\chi}_{2,\text{phys}} = - \bm{\chi}_{1,\text{phys}}$ (here we can, without loss of generality, restrict to $0\leq \theta_O\leq \pi/2$).
The physical observer separation projected onto the plane perpendicular to the line of sight from the coordinate origin to the source is $R_O' \equiv |(\bm{\chi}_{1,\text{phys}}-\bm{\chi}_{2,\text{phys}})\cdot \bm{\hat{e}_1}| = R_O |\sin\theta_O|$, which is maximized for $\theta_O = \pi/2$ and vanishes for $\theta_O = 0$.
Note that, because $z_O \approx 0$ for both observers, the physical coordinates of the observers are also their comoving coordinates.

Suppose that a point lens is located at comoving coordinates%
\footnote{\label{ftnt:phiL}%
    Owing to geometrical parallax arising from differing coordinate-system definitions, the angle $\phi_L$ defined in this section and in \figref[s]{fig:coords} and~\ref{fig:LPgeom} is \emph{not} the same as the angle $\varphi_L$ defined in \secref{sec:extendedSource} and on \figref{fig:changeOfVariables}.
    The origin of the coordinate system in this section and \figref[s]{fig:coords} and \ref{fig:LPgeom} is located at the midpoint between the observers, whereas the origin of the coordinate system in \secref{sec:extendedSource} and \figref{fig:changeOfVariables} was centered on a single observer.
} %
$\bm{\chi}_L = \chi_L \bm{\hat{e}_3} + \Delta \chi_{L,\perp} \lb( \cos\phi_L \bm{\hat{e}_1} + \sin\phi_L \bm{\hat{e}_2} \rb)$, where $\chi_L < \chi_S$ and $\Delta\chi_{L,\perp} \equiv \chi_L \theta_L \ll\chi_L$.
Provided that $R'_O\neq 0$, the two observers see different true positions of the lens with respect to the source (i.e., there is a geometrical parallax).
With $\chi_E \equiv \chi^L_E \equiv \chi_L \theta_E$ the comoving Einstein radius in the lens plane, $\chi_E^O \equiv \chi^L_E \cdot \chi_S / \chi_{LS}$ the comoving Einstein radius projected into the observer plane,%
\footnote{\label{ftnt:projections}%
    The comoving Einstein radius projected into a given plane $P$ can be obtained using Euclidean geometry in conformal--comoving coordinates as $\chi_E^P = \mathcal{G}_P \chi_L \theta_E$.
    The corresponding projected physical Einstein radius is $R_E^{\text{proj.}} \equiv \chi_E^P/(1+z)$ where $z$ is the redshift of the plane projected to.
    For projections into the lens or source plane, $\chi_E^P$ is simply the comoving transverse size that subtends an angle equal to the Einstein angle, as viewed by the observer, yielding $\mathcal{G}_{L} = 1$ and $\mathcal{G}_{S} = \chi_S / \chi_L$.
    The projection into the observer plane is however the comoving (or physical, since these are equivalent at $z_O=0$) transverse size that subtends the same angle as $\chi_E^L$ \emph{when viewed from the source}, yielding $\mathcal{G}_{O} =\chi_S / \chi_{LS}$.
} %
and working in the limits $\Delta \chi_{L,\perp},R'_O \ll \chi_L,\chi_S$, the true angular offsets $y_i = \beta_i/\theta_E$ (for $i=1,2$) of the source centroid relative to the lens location, as seen by each observer, are given by
\begin{align}
    y_i^2 &= \lb( \frac{\Delta \chi_{L,\perp}}{\chi_E^L } \rb)^2 + \lb( \frac{  R'_O/2 }{\chi_E^O} \rb)^2 + \sigma_i \frac{  R'_O }{ \chi_E^O }  \frac{\Delta \chi_{L,\perp}}{\chi_E^L}  \cos\phi_L\:, \label{eq:yi}
\end{align}
where $\sigma_1 = -1$ and $\sigma_2 = +1$.
For a fixed $\Delta \chi_{L,\perp}$ and $R'_O$, the true lens offsets from the source centroid for each observer thus range over 
\begin{align}
     \lb| \frac{\Delta \chi_{L,\perp}}{\chi_E^L } - \frac{ R'_O/2 }{\chi_E^O} \rb| \leq y_i \leq \lb| \frac{\Delta \chi_{L,\perp}}{\chi_E^L } + \frac{ R'_O/2 }{\chi_E^O} \rb|
\end{align}
as $\phi_L$ is varied, with the maximum for one observer coinciding with the minimum for the other.
For lenses near one of the two lines of sight, the corresponding observer will see $y \sim 0$, while the other observer will see $y\sim R'_O / \chi_E^O$. 

As a result, for observers that are widely separated compared to the observer-plane Einstein radius, $R'_O \gtrsim \chi_E^O$, there are locations in the lens plane where a lens would give a large magnification to a distant source for one observer, while it would be near unity for the other observer.
That is, there is also a lensing parallax: two spatially separated observers can see the same source at significantly different intensity owing to the differential magnification.

\tabref{tab:REproj} provides some benchmark parameter points to build intuition for how far apart observers should be to access this effect: observer separations on the order of the solar radius (roughly twice the Earth--Moon distance) are \emph{coincidentally} on the order of the observer-plane Einstein radius for lens masses $M \sim 10^{-12}M_{\odot}$ (provided those lenses are cosmologically distant).
Because these length scales are reasonably human-accessible, this in principle makes this technique a good one to access the PBH asteroid-mass window, $2\times 10^{-16}M_{\odot} \lesssim M_{\pbh} \lesssim 5 \times 10^{-12}M_{\odot}$, provided that suitable sources can be found.
This effect is known as `picolensing' in this mass range because the corresponding Einstein angles for a cosmologically distant lens are roughly on the order of picoarcseconds (i.e., $\sim 10^{-12}\,\text{arcsec} = 4.85\times 10^{-18}\,\text{rad}$), give or take an order of magnitude or two.

In this work, we will consider a few different observer-separation scenarios; see \tabref{tab:baselines}.

\begin{table}[t]
\begin{ruledtabular}
    \begin{tabular}{lllll}
    $\bm{\chi_L/\chi_S}$ & $\bm{z_L}$ &  $\bm{\theta_E}$ \textbf{[arcsec]} & \textbf{Plane} &  $\bm{R^{\text{proj.}}_E/R_{\odot}}$ \\ \hline \hline
       &&&   Source  & 1.1 \\
    0.25 &  0.20  & $2.9\times 10^{-12}$    &   Lens     & 0.45 \\
        &&&   Observer & 0.72 \\ \hline 
        &&&   Source   &  0.68 \\
        0.50 &  0.43 & $1.9\times 10^{-12}$ &   Lens     & 0.47 \\
        &&&   Observer  & 1.4 \\ \hline
        &&&   Source  & 0.42 \\
        0.75 &  0.69 &  $1.1\times 10^{-12}$ &   Lens      & 0.38 \\
        &&&   Observer & 2.5 \\
    \end{tabular}
\end{ruledtabular}
\caption{\label{tab:REproj}%
    For a PBH lens of mass $M_{\pbh} = 10^{-12}M_{\odot}$ and for various fractional lens distances to the source $\chi_L/\chi_S$  and corresponding lens redshifts $z_L$, we show the Einstein angle $\theta_E$ for the lens/source configuration, and the (physical) Einstein radius $R^{\text{proj.}}_E$ projected into several relevant planes (see footnote \ref{ftnt:projections}).
    The source redshift is fixed to be $z_S = 1$.
    We normalize the projected Einstein radii to the solar radius $R_{\odot} \approx 6.96\times 10^{5}\,\text{km}$~\cite{ParticleDataGroup:2024cfk} because it is coincidentally a convenient scale.
    }
\end{table}

\begin{table}[t]
\begin{ruledtabular}
    \begin{tabular}{lcll}
    \textbf{Scenario}   & \textbf{Abbrev.}  &   \textbf{Baseline} $\bm{R_{O}}$ &  $\bm{R_O/R_{\odot}}$  \\ \hline
    Low Earth Orbit     & LEO   &   $1.40\times 10^{4} \,\text{km}$ & 0.020 \\
    Earth--Moon         & EM    &   $3.84 \times 10^{5} \,\text{km}$ & 0.55 \\
    Lagrange Point 2    & L2    &   $1.50 \times 10^{6} \,\text{km}$ & 2.15 \\
    Astronomical Unit   & AU    &   $1.50 \times 10^{8}\, \text{km}$ & 215
    \end{tabular}
\end{ruledtabular}
\caption{\label{tab:baselines}%
    Observer separations (``baselines'') used for various scenarios.
    The LEO baseline is assumed to be $R_{O} = 2 \times( R_{\oplus} + 650\,\text{km} )$, where $R_{\oplus} \approx 6371 \,\text{km}$ is the mean radius of Earth~\cite{SSphys}, reflecting two spacecraft $180^{\circ}$ apart in a $\sim 650\,\text{km}$-high orbit.
    The EM case is taken to have one spacecraft at the distance of the semimajor axis of the Earth--Moon system~\cite{SSelem} and the other in low Earth orbit.
    The L2 case is assumed to have one spacecraft near the L2 Lagrange point and the other near Earth; the L2 point is a distance of $\sim \text{AU} \times ( M_{\oplus}/ (3M_{\odot} ))^{1/3} \approx 10^{-2}\text{AU}$ from Earth~\cite{MoultonCelestial}, where the solar mass $M_{\odot} \approx 2\times 10^{30}\,\text{kg}$~\cite{ParticleDataGroup:2024cfk}.
    The AU case would be achievable by, e.g., having spacecraft offset by $60^\circ$ around a Sun-centered orbit at a radius equal to the semimajor axis of Earth's orbit.
    The solar radius $R_{\odot}$ is again used as a convenient normalizing length scale.
    }
\end{table}

\subsubsection*{The Validity of Geometrical Optics}
\label{sec:geomOptics}
The discussion we have given here relies on the validity of geometrical optics to derive the summed image magnification at \eqref{eq:pointLPointSmag}.
This is equivalent to the condition~\cite{Katz:2018zrn} $\omega \Delta t \gtrsim 1$, where $\Delta t$ is the differential time delay (both geometric and gravitational) for the two images that we assumed summed incoherently at \eqref{eq:pointLPointSmag}, and $\omega$ is the angular frequency of the probe light.
Since $\Delta t \sim 2 R_{\text{Sch.}} (1+z_L)$ [with $R_{\text{Sch.}}=2G_{\text{N}}M$ the Schwarzschild radius of the lens] is a typical time delay for angular deviations on the order of the Einstein angle~\cite{Katz:2018zrn},%
\footnote{\label{ftnt:SchwarzschildNotEinstein}%
    Note that the time delay is on the order of the \emph{Schwarzschild} radius, not the Einstein radius, despite the fact that the impact parameter of such a light ray with respect to the lens is of order the lens-plane Einstein radius.
} %
this leads to a rough lower bound on $M_{\pbh}$ for which geometrical optics is valid for a given probe energy $E=\omega$~\cite{Katz:2018zrn,Gawade:2023gmt}:
\begin{align}
    M_{\pbh} \gtrsim \frac{M_{\text{Pl.}}^2 }{4E} = 2 \times 10^{-15}M_{\odot} \times \lb( \frac{ 15 \,\text{keV} }{ E } \rb)\:, \label{eq:GOL}
\end{align}
where we took $z_L=0$ in this estimate.
Note that this estimate should only be understood at the order-of-magnitude level, and also that the technique does not necessarily abruptly stop working below this bound.
Rather, \eqref{eq:GOL} delimits the regime in which one needs to be concerned with wave-optics effects that are likely to reduce the signal strength; cf.~\citeR{Katz:2018zrn} (see also \citeR{Takahashi:2003ix}).

In the next section we discuss the detectability of picolensing in the regime where geometrical optics is a good approximation.

\subsection{Detection Criteria}
\label{sec:detectionCriteria}
The mere existence of a lensing parallax is not however sufficient: we need to also be able to detect this difference.
Various approaches have recently been used to do this: \citeR{Jung:2019fcs} used a simple magnification contrast ratio to diagnose detectability, while \citeR{Gawade:2023gmt} defined an SNR based on Poisson statistics of photon counts in two detectors subject to background noise.
We follow the approach of \citeR{Gawade:2023gmt}, which we review here.

Consider two detectors ($i=1,2$) which record photons from a source at rate $\Gamma_S^i$ on average, with average background rates $\Gamma_B^i$.
In a time $T$ (which for simplicity we assume to be common to both detectors), these detectors will record $S_i$ source photons, $B_i$ background photons, and $N_i = S_i+B_i$ total photons.
On average, we will have $\langle S_i \rangle = \Gamma_S^i \cdot T$ and $\langle B_i \rangle = \Gamma_B^i \cdot T$.

Define the expected signal to be the difference of the expected source photons in each detector: $\Delta S \equiv | \langle S_1 \rangle - \langle S_2 \rangle |$, and the observable to be the difference in the number of photons in the two detectors: $\Delta N \equiv | N_1 - N_2|$.
The signal-to-noise ratio $\varrho$ for this signal is then given by $\varrho \equiv  \Delta S / \sigma_{\Delta N}$ where $\sigma_{\Delta N}$ is the uncertainty on $\Delta N$.
Assuming that the counting statistics are Poisson, we have 
\begin{align}
    \sigma_{\Delta N}
    &= \sqrt{ [\sigma(N_1)]^2 + [\sigma(N_2)]^2 } \\
    &= \sqrt{ \langle N_1 \rangle + \langle N_2 \rangle }\\
    &= \sqrt{ \langle S_1 \rangle + \langle S_2 \rangle + \langle B_1 \rangle + \langle B_2 \rangle }\:.
\end{align}
It follows that the expected signal-to-noise ratio is
\begin{align}
    \varrho &= \frac{ | \langle S_1\rangle - \langle S_2 \rangle | }{\sqrt{ \langle S_1 \rangle + \langle S_2 \rangle + \langle B_1 \rangle + \langle B_2 \rangle }}\:. \label{eq:SNR1}
\end{align}

Now suppose that the source-photon count rate results from a source that yields an \emph{unmagnified} photon flux density $F_S$ (i.e., photons per second per unit area) at the detectors, and that each detector has an effective detection area $A_{\text{eff}}^i$.
The effective area is a combination of the physical area of the detector $A_{\text{phys}}^i$, geometrical projection effects that account for the orientation of the detector plane with respect to the line of sight to the source, and any additional suppression effects that can occur because of, e.g., detector shielding by surrounding material: $A_{\text{eff}}^i = \epsilon^i(\vartheta) A_{\text{phys}}^i \cos\vartheta$, where $\vartheta$ is the angle between the detector-plane normal and the line of sight to the source, and $\epsilon^i(\vartheta)$ captures any additional shielding effects, which are in general orientation dependent. 
We then have $\Gamma_S^i(\vartheta)|_{\text{unmag.}} = F_S A_{\text{eff}}^i(\vartheta)$; in principle $\vartheta$ can also vary during an observation, so $\vartheta = \vartheta(t)$.
Although in an actual data analysis it would of course be important to keep track of (and correct for) the differing and time-dependent effective areas of both detectors when observing a specific source, in order to simplify our analysis we will assume that we can replace the effective area by its temporally and angularly averaged value $A_{\text{eff}}^i(\vartheta(t)) \rightarrow \bar{A}_{\text{eff}}^i$, giving an average unmagnified source count rate $\bar{\Gamma}_S^i|_{\text{unmag.}} = F_S \bar{A}_{\text{eff}}^i$.

If the two observers each see the source with respective magnifications $\mu_i$, then the (average) observed source count rates will be $\bar{\Gamma}_S^i= \mu_i F_S \bar{A}_{\text{eff}}^i$.
Although handling the case with detectors of differing averaged effective area is in principle straightforward, we will for the sake of simplicity of the analysis assume that the two detectors have the same averaged effective area: $\bar{A}_{\text{eff}}^1 = \bar{A}_{\text{eff}}^2 \equiv A_{S}$.

The case of the background counts is handled similarly to the unmagnified source case, although in general the effective area for the background $A_B^i$ differs from that for the source (even in the average sense); see \citeR{Bhalerao:2022edb}. 
We will simply write $\Gamma_B^i = F_B A_B^i$ and further assume that the two detectors have the same effective area for backgrounds $A_B^1=A_B^2\equiv A_B$.

Under these assumptions, it follows that~\cite{Gawade:2023gmt}
\begin{align}
    \varrho &= \frac{ | \mu_1 - \mu_2 | \bar{S} }{ \sqrt{ 2\bar{B} + (\mu_1 + \mu_2 ) \bar{S} }}\:, \label{eq:SNR}
\end{align}
where
\begin{align}
    \bar{S} &\equiv F_S A_{S} T\:; & 
    \bar{B} &\equiv F_B A_B T\:.
\end{align}
For a transient source, $T$ should be taken to be the characteristic timescale over which the source flux density is of order $F_S$.
Notice that $\varrho \propto |\mu_1 - \mu_2|$: the SNR is indeed sensitive to the differential magnification.
Note also that $\varrho\propto\sqrt{T}$.

Unless otherwise specified, in this paper we fix the criterion for a single positive picolensing detection to be $\varrho \geq \varrho_*$ where $\varrho_* = 5$ (i.e., we set a $5\sigma$ detection threshold for individual picolensing events).

The SNR developed here has assumed that the noise is entirety dominated by photon shot noise, without considering detector systematics.
Of course, the multiple spacecraft involved in making this measurement would need to have well-calibrated relative response functions.
These could be calibrated on the ground prior to flight for an instrument dedicated for this measurement; in-flight calibration by simultaneous observation of non-transient x-/$\gamma$-ray sources would also allow the relative response (and changes thereof) to be monitored over time. 
The systematics would probably need to be known as well as (or better than) the level of the differential magnification that is being searched for.

\subsection{Picolensing Cross Section, Volume, and Optical Depth}
\label{sec:picoSigma}
Ultimately, we wish to convert a picolensing SNR into a probability for picolensing in order to draw conclusions about whether a given lens density can be probed.

The SNR $\varrho$ is a complicated (and in general nonlinear) function of many parameters that characterize the lens, observers, and source (see \tabref{tab:params} summarizing their definitions): 
\begin{align}
    \varrho = \varrho( M , \chi_L , \bm{\chi}_{L,\perp} , R_O , \theta_O , F_B , A_B , A_S , T , F_S , \theta_S , \chi_S )\:.
\end{align}
In order to make progress, let us first hold fixed all of the source properties ($F_S$, $\theta_S$, $\chi_S$), the size and orientation of the observer-separation baseline relative to the source location ($R_O$, $\theta_O$), all of the detector properties ($F_B$, $A_B$, $A_S$), and the integration time $T$.
With these parameters held fixed, the SNR $\varrho = \varrho(\bm{\chi}_{L,\perp}; M, \chi_L)$ becomes a function only of the transverse location of the lens in the lens plane $\bm{\chi}_{L,\perp}$, the lens mass $M$, and the comoving distance to the lens plane $\chi_L$.
We assume a $\delta$-function lens-mass distribution.

If we specify a certain SNR detection threshold $\varrho_*$, then the condition $\varrho(\bm{\chi}_{L,\perp};M,\chi_L) \gtrsim \varrho_*$ implicitly defines the region(s) of the lens plane at $\chi = \chi_L$ where a lens of mass $M$ would give rise to the detection of a picolensing signal with an SNR at or above that threshold.
In general, symmetries dictate there are either two such disjoint regions in the lens plane, or no such region; we call the union of all such regions the ``picolensing region''.
The total comoving area of the picolensing region, $\sigma(\chi_L,M;\varrho_*)$, is defined to be the (comoving) ``picolensing cross section''~\cite{Jung:2019fcs,Gawade:2023gmt} for that lens distance, lens mass, and detection threshold $\varrho_*$ (with all other parameters held fixed).
If there is no such region that leads to significant picolensing detection, we set $\sigma(\chi_L,M;\varrho_*)=0$.
We detail the computation of $\sigma$ via Monte Carlo methods in \secref{sec:MCsigma}.

Now, suppose the lenses have a comoving number density $n_{\text{lens}}(\chi_L,M)$.
The optical depth of lenses capable of causing a detectable  picolensing event (at or above SNR threshold $\varrho_*$) for this source and detector configuration is then%
\footnote{\label{ftnt:multipleLensing}%
    While this specific definition is always valid for \emph{some} definition of~$\sigma$, we can only use our single-lensing expressions for $\varrho$ to evaluate~$\sigma$ if $\tau \lesssim 1$; otherwise, there is more than one lens expected to lie within the picolensing volume and one must consider the effects of those multiple lenses on the differential magnification signal, which would lead to a different SNR.
} %
\begin{align}
    \tau(M;\varrho_*) &= \int_0^{\chi_S} n_{\text{lens}}(\chi_L,M) \cdot\sigma(\chi_L,M;\varrho_*) d\chi_L\:.
\end{align}
Defining $n_{\text{lens}}^0(M) \equiv n_{\text{lens}}(\chi_L=0,M)$, we can write~\cite{Jung:2019fcs,Gawade:2023gmt}
\begin{align}
    \tau(M;\varrho_*) &\equiv n_{\text{lens}}^0(M) \cdot  \mathcal{V}(M;\varrho_*)\:, \label{eq:opticalDepth}
\end{align}
where we have defined the (comoving) ``picolensing volume'' $\mathcal{V}$ as the line-of-sight integral over the picolensing cross section, weighted by the fractional lens number-density distribution:
\begin{align}
    \mathcal{V}(M;\varrho_*) &\equiv \int_0^{\chi_S} \hat{n}(\chi_L,M)\cdot \sigma(\chi_L,M;\varrho_*) d\chi_L\:; \label{eq:Vlens} \\
    \hat{n}(\chi_L,M) &\equiv n_{\text{lens}}(\chi_L,M)\Big/n_{\text{lens}}^0(M)\:.
\end{align}

In the case where the lenses exhibit a constant comoving number density, we would have $\hat{n}=1$ and~\cite{Jung:2019fcs,Gawade:2023gmt}
\begin{align}
    \mathcal{V}(M;\varrho_*) = \int_0^{\chi_S} \sigma(\chi_L,M;\varrho_*) d\chi_L\:. \label{eq:VlensConst} \\  [\text{constant comoving lens density}]\;\; \nonumber
\end{align}
This approximation is obviously valid for cold, pressureless matter (i.e., `dust') that does not clump significantly along the line of sight.
Of course, real PBH DM does clump and cluster; however, given the very long lines of sight to typical sources that we will consider, the integral in \eqref{eq:Vlens} averages over the resulting under- and over-densities of clumped PBH DM.
The result of such averaging gives approximately the same optical depth to a given source as if the PBH DM instead had a constant and uniform comoving density equal to its average value.
However, clumping modifies the statistics of the lens distribution~\cite{Jung:2019fcs}; i.e., the lens locations become correlated.
Nevertheless, in the limit of small overall optical depth $\tau \ll 1$, as well as small optical depth within any collapsed structure, \citeR{Jung:2019fcs} showed that one can to a very good approximation still treat the clumped lens locations as uncorrelated with only a small error.
Therefore, \eqref{eq:VlensConst} can actually be employed more generally in \eqref{eq:opticalDepth} to compute the optical depth for real PBH DM lenses (which can in turn be treated as having uncorrelated positions), so long as all optical depths are small and the distances to the sources are cosmologically large.

\subsection{Lensing Probability and Exclusion Limits}
\label{sec:probAndExclusion}
The optical depth $\tau(M,\varrho_*)$ is defined as the expected number of lenses of mass $M$ that would give rise to a $\varrho \gtrsim \varrho_*$ picolensing event for a given set of detector and source parameters, assuming that there is only one lens in the picolensing volume.
We will make projections for exclusion limits based on the absence of any such lenses in the picolensing volume. 
Strictly speaking, the possibility of having multiple lenses in the picolensing volume complicates matters, but such events are rare in the relevant limit of small optical depth.
As discussed in \appref{app:differentApproach}, multiple-lens events make a negligibly small difference to our results.

Barring the above caveat and assuming Poisson statistics, the probability of having $\mathfrak{n}$ lenses in the picolensing volume is 
\begin{align}
    \text{Pr}(\mathfrak{n}) &= \frac{ \tau^{\mathfrak{n}} e^{-\tau} }{  \mathfrak{n}! } \: .
\end{align}
Then the probability of \emph{not} having any lenses within the picolensing volume for a given single source is $P^{(1)}_{\text{no lens}} = \text{Pr}(\mathfrak{n}=0) = \exp[-\tau(M,\varrho_*)]$.
On the other hand, the probability of there being \emph{at least one} lens in the picolensing volume for a single source is $P^{(1)}_{\text{lensing}} = 1 - P^{(1)}_{\text{no lens}} = 1-\exp[-\tau(M,\varrho_*)]$.
The probability of there being \emph{exactly one} lens in the picolensing volume is $P^{(1)}_{\text{single lens}} = \text{Pr}(\mathfrak{n}=1) = \tau(M,\varrho_*) \exp[ -\tau(M,\varrho_*) ]$.

Suppose we independently observe $\aleph$ sources in total, each of which has its own optical depth for picolensing $\tau_j(M,\varrho_*) = n_{\text{lens}}^0(M) \cdot \mathcal{V}_j(M,\varrho_*)$ for $j=1,\ldots,\aleph$. 
Then the probability that no lenses will appear in any of the picolensing volumes is
\begin{align}
    P^{(\aleph)}_{\text{no lens}} &= \Pi_{j=1}^{\aleph} P^{(1),j}_{\text{no lens}} \label{eq:statisticalLimitCriterion} \\
    &=\Pi_{j=1}^{\aleph} \exp[-\tau_j(M,\varrho_*)] \\
    &= \exp[ -  n_{\text{lens}}^0(M)\sum_{j=1}^{\aleph} \mathcal{V}_j(M,\varrho_*)] \\
    & \equiv \exp[ - \aleph\cdot n_{\text{lens}}^0(M) \cdot\overline{\mathcal{V}}(M,\varrho_*)]\:,
\end{align}
where we have at the last step defined the population-averaged comoving picolensing volume: 
\begin{align}
    \overline{\mathcal{V}}(M,\varrho_*) \equiv \frac{1}{\aleph} \sum_{j=1}^{\aleph} \mathcal{V}_j(M,\varrho_*)\:. \label{eq:VlensAvg}
\end{align}

To set an exclusion-limit projection, we wish to find the largest $n_{\text{lens}}^0(M)$ such that having no lenses within the picolensing volume is a low-probability event, occurring with probability $1 - \alpha$, where $0\leq \alpha\leq 1$ is the confidence level (i.e., a 95\% confidence level is $\alpha_{95} = 0.95$).
We set $P^{(\aleph)}_{\text{no lens}} = 1-\alpha$ and invert to find
\begin{align}
    n^{0,\text{limit}}_{\text{lens}}(M;\alpha,\varrho_*) &= - \frac{ \ln(1-\alpha) }{ \aleph \cdot \overline{\mathcal{V}}(M,\varrho_*) }\;. \label{eq:n0LensLimit}
\end{align}
Note that $-\ln(1-\alpha)>0$ and that $-\ln(1-\alpha_{95})\approx 3$.

This limit has the following interpretation: whenever $n_{\text{lens}}^0(M) \gtrsim n^{0,\text{limit}}_{\text{lens}}(M;\alpha,\varrho_*)$, the probability of having no lenses (capable of causing detectable $\varrho\geq \varrho_*$ single-lens picolensing) in any of the $\aleph$ picolensing volumes is less than or equal to $(1-\alpha)$.
Stated differently, when $n_{\text{lens}}^0(M) \gtrsim n^{0,\text{limit}}_{\text{lens}}(M;\alpha,\varrho_*)$ there is a probability greater than or equal to $\alpha$ that there will be \emph{at least one} such lens in  \emph{at least one} picolensing volume.
Having no such lenses would thus be a significant outlier observation; on this basis, we would exclude $n_{\text{lens}}^0(M)\gtrsim n^{0,\text{limit}}_{\text{lens}}(M;\alpha,\varrho_*)$ at confidence level $\alpha$.

Strictly speaking, owing to the aforementioned possibility of there being multiple lenses in the picolensing volume, the above criterion is subtly different from not having any events with a detectable differential magnification.
An alternative criterion would require instead the absence of any picolensing volumes containing \emph{exactly one} lens~\cite{Jung:2019fcs,Gawade:2023gmt}.
We apply that alternative criterion in \appref{app:differentApproach} and show that it makes a negligible difference to our results.

It is customary to express the exclusion-limit  projection \eqref{eq:n0LensLimit} in terms of the fraction that PBH lenses constitute of the average DM density of the universe.
The present-day dark-matter energy density, which is equal to its comoving energy density, can be written (see footnote \ref{ftnt:FLRW}) as
\begin{align}
    \rho^0_\dm = \Omega_\dm \rho_c &= 3 \Omega_\dm H_0^2M_{\text{Pl.}}^2/(8\pi)\\
    &= 3 \omega_\dm (H_0/h)^2M_{\text{Pl.}}^2/(8\pi)\:.
\end{align}
Let us define the PBH fraction $f_{\pbh}$ as
$\rho_{\text{lens}}  \equiv M n_{\text{lens}}^0(M)= f_{\pbh} \rho^0_\dm$; then,
\begin{align}
    n^0_{\text{lens}}(M) &= f_{\pbh} \omega_\dm (M_{\odot}/M) \lambda^{-3}\:;\\[1ex]
    \lambda^{-3} &\equiv 3 (H_0/h)^2M_{\text{Pl.}}^2/(8\pi M_\odot)\:.\label{eq:lambdaDefn}
\end{align}
Therefore, the exclusion limit would be projected to be
\begin{align}
    f_{\pbh}^{\text{limit}}(M;\alpha,\varrho_*,\aleph) &= - \frac{  \lambda^3 \ln(1-\alpha) }{ \omega_\dm \cdot \aleph \cdot \overline{\mathcal{V}}(M,\varrho_*) }  \frac{M}{M_{\odot}}\:.
\end{align}

Suppose now that these $\aleph$ sources are merely a \emph{representative sample} of the true source population and that we wish to make a projection for where the limit would lie if we were instead to observe $N_{\grb}$ sources and see no picolensing.
We can then rescale the projected exclusion limit as 
\begin{align}
    f_{\pbh}^{\text{limit}}(M;\alpha,\varrho_*,N_{\grb}) 
    &= - \frac{  \lambda^3 \ln(1-\alpha) }{ \omega_\dm N_{\grb} \overline{\mathcal{V}}(M,\varrho_*) } \frac{M}{M_{\odot}}\:. \label{eq:fDMlimit}
\end{align}
In order to employ this result, we need only compute the average comoving picolensing volume $\overline{\mathcal{V}}(M,\varrho_*)$ for a representative sample of sources using \eqref{eq:VlensConst}.

One final thing to note here is that the average picolensing optical depth for a generic value of $f_{\pbh}$ can be written as 
\begin{align}
    \bar{\tau}(M,\varrho_*;f_{\pbh}) &= n_{\text{lens}}^0(M) \cdot \overline{\mathcal{V}}(M,\varrho_*)\nonumber \\
        &= \frac{ - \ln(1-\alpha) }{  N_{\grb}  } \frac{f_{\pbh}}{f_{\pbh}^{\text{limit}}(M;\alpha,\varrho_*,N_{\grb})} \: , \label{eq:opticalDepth2}
\end{align}
which is of course independent of $N_{\grb}$ at some fixed value of $f_{\pbh}$ because $f_{\pbh}^{\text{limit}} \propto N_{\grb}^{-1}$.
There are two points of general applicability that this expression allows us to make.
First, we note that for $f_{\pbh} = f_{\pbh}^{\text{limit}}$, we have $\bar{\tau}(M,\varrho_*, f_{\pbh} = f_{\pbh}^{\text{limit}} ) = - \ln(1-\alpha) / N_{\grb}   \sim 3  / N_{\grb} $ (the latter case for $\alpha = 0.95$). 
Provided that $N_{\grb} \gg 3$, we therefore have $\tau \ll 1$: at the limits, we are by construction always in the optically thin regime where our single-lensing results are valid (at least for the average picolensing volume).
Second, for $f_{\pbh} = 1$, we have $\bar{\tau}(M,\varrho_*, f_{\pbh} = 1 ) = - \ln(1-\alpha) / ( N_{\grb} f_{\pbh}^{\text{limit}} )   \sim 3  / ( N_{\grb} f_{\pbh}^{\text{limit}} )$.
For this to remain in the single-lensing regime for the average picolensing volume, we need $\bar{\tau}(M,\varrho_*, f_{\pbh} = 1 )\lesssim 1\Rightarrow f_{\pbh}^{\text{limit}} \gtrsim 3/N_{\grb}$. 
We will see that essentially all of our projections satisfy this bound. 
However, if the limit we expect to set is significantly stronger than this, then there is a possibility that regions of parameter space near $f_{\pbh} \sim 1$ will be in the multiple-lensing regime (for the average picolensing volume), which has a qualitatively different signal to single picolensing.

Of course, individual sources may have picolensing volumes significantly larger than the average, so individual sources can still violate the $\tau \lesssim 1$ requirement even if the average picolensing volume over the full population does not. 
However, as long as no such source makes an $\mathcal{O}(1)$ contribution to the average picolensing volume (i.e., $\mathcal{V}_{\text{max}} \lesssim \aleph \overline{\mathcal{V}}$), such violations are not expected to have a large impact on our results.
Viewed in a different light, if we were to be in the $\bar{\tau}\gtrsim 1$ regime (or if $\tau>1$ for a non-negligible collection of individual sources), we would expect a large number of positive picolensing detections if the DM were PBHs; see \secref{sec:beyondConstraints} for further discussion.

\section{Gamma-ray Bursts (GRBs) as a Picolensing Target}
\label{sec:GRBs}
As mentioned in \secref{sec:introduction}, GRBs are good targets for the observation of picolensing signals.
In order to utilize them, we need to know some of their basic properties: (a) their redshift; (b) their duration; (c) their brightness; (d) their physical size, as this tells us whether they act as point or extended sources for the purposes of the picolensing computation at a given PBH mass; and (e) their observed spatial intensity distribution in the plane transverse to the line of sight.
In this section, we review these properties.

\subsection{GRB Observations with \emph{Swift}/BAT}
\label{sec:SWIFT-BAT}
Following \citeR{Gawade:2023gmt}, we base our study of GRB properties on the population of real GRBs that have been observed by the Burst Alert Telescope (BAT) on the Niel Gehrels \emph{Swift} spacecraft~\cite{2004ApJ...611.1005G,2005SSRv..120..143B}.
The BAT consists of a $5240\,\text{cm}^2$ detector plane of CdZnTe scintillators with a coded mask that permits source localization.
The instrument has an on-axis peak effective area (including various efficiency factors) for the coded-mask response that is on the order of $1400\,\text{cm}^2$ in the 15--150\,keV band~\cite{SWIFTtechnical,BATdigest}.%
\footnote{\label{ftnt:rawAeff}%
    The raw, on-axis effective area, accounting only for the coded-mask transmission to the detector plane, is on the order of $2400\,\text{cm}^2$ in the 30--100\,keV band~\cite{2011ApJS..195....2S,BATdigest}.
} %

\emph{Swift}/BAT releases periodic catalogues~\cite{Sakamoto_2008,2011ApJS..195....2S,Lien:2016zny} of observed GRB events as well as an online catalogue that is updated live~\cite{SWIFTBATcurrent}.%
\footnote{\label{ftnt:currentNumbers}%
    As of September 18, 2024, the catalogue consists of 1663 total GRBs, 422 of which have measured redshifts~\cite{SWIFTBATcurrent}.
} %
We base our study on the data available at \citeR{SWIFTBATwhatWeUsed}, which is complete through the end of 2023.
This sample of GRBs contains a total of 1587 GRBs, but only 409 of these have all the necessary data for our study (mostly, the others are missing reliable redshift determinations).

For these 409 GRBs, the \emph{Swift}/BAT catalogue contains the following information: (a) the source redshift%
\footnote{\label{ftnt:avgzS}%
    The average source redshift in the catalogue is $\bar{z}_S = 2.1$.
} %
$z_S$; (b) $T_{90}$, the temporal duration over which 90\% of the event fluence (energy per area) is detected, and which we use as the observation duration parameter $T$ in \eqref{eq:SNR}; and (c) spectral fits for the observed source photon flux density.
However, as GRBs are not resolved, \emph{Swift}/BAT gives no information on (d) their angular sizes, or (e) the GRB spatial intensity profile on the sky, the latter of which we will therefore assume to be either a disk or a Gaussian.

Before discussing the question of GRB sizes, which is the key uncertainty in picolensing, let us first summarize the spectral fit information for the source flux densities, as well as some information about detector backgrounds.
We also mention other missions with available data.

\subsubsection{Swift/BAT GRB Spectra}
\label{sec:SWIFT-BAT_spectra}
For each GRB observed, the \emph{Swift}/BAT catalogue provides two different fits for the observed GRB spectrum: a power-law (PL) or cutoff power-law (CPL), as well as a statement as to which of the two fits to the data is better.

In terms of observer-frame energy $E$, these spectral fits are defined, respectively, as 
\begingroup
\allowdisplaybreaks
\begin{align}
    f_{\text{PL}}(E) &\equiv K_{50}^{\text{PL}} \lb( \frac{E}{ 50\,\text{keV}} \rb)^{\alpha_{\text{PL}}}\:; \\[3ex]
    f_{\text{CPL}}(E) &\equiv K_{50}^{\text{CPL}} \lb( \frac{ E }{ 50\,\text{keV} } \rb)^{\alpha_{\text{CPL}}} \nl\quad \times \exp\lb( - \frac{E(2+\alpha_{\text{CPL}})}{E^{\text{peak}}_{\text{CPL}}} \rb) \:,
\end{align}
\endgroup
where 
\begin{align}
f_{\text{(C)PL}}(E) \equiv \frac{dN_{\gamma}}{dA \, dt\, dE} %
\end{align}
is the differential photon flux density per unit energy.
The parameter $K_{50}^{\text{(C)PL}}$ gives the normalization of the spectrum at $E=50\,\text{keV}$ in units of photons/cm${}^2$/s/keV, $\alpha_{\text{(C)PL}}$ is a dimensionless power-law index, and $E^{\text{peak}}_{\text{CPL}}$ has units of energy.
The \emph{Swift}/BAT catalogues specify $K_{50}^{\text{(C)PL}}$, $\alpha_{\text{(C)PL}}$, and (where relevant) $E^{\text{peak}}_{\text{CPL}}$.
We adopt whichever of the PL or CPL spectra is judged by \emph{Swift}/BAT to be the better fit for each given GRB in the catalogue.

In order to obtain the source flux density $F_S$ that enters in \eqref{eq:SNR}, one integrates the relevant $f_{\text{(C)PL}}(E)$ over the detector bandpass $E_{\text{min}}\leq E \leq E_{\text{max}}$:
\begin{align}
    F_S = \int_{E_{\text{min}}}^{E_{\text{max}}} f_{\text{(C)PL}}(E)\, dE\:, \label{eq:sourceFluxDensity}
\end{align}
which is in units of photons/cm${}^2$/s.
For \emph{Swift}/BAT, $E_{\text{min}} = 15\,\text{keV}$ and $E_{\text{max}} = 150\,\text{keV}$.
We adopt this same detector bandpass $[15,150]\,\text{keV}$ for our projections.

\citeR{Gawade:2023gmt} considered a bandpass of [20,200]\,\text{keV} (see the discussion of the \emph{Daksha} mission in \secref{sec:otherMissions} below). 
We have checked explicitly that this small difference in assumed bandpasses makes only a 10--20\% difference to our results, which is an irrelevantly small difference in light of other uncertainties.
\citeR{Gawade:2023gmt} also studied the source flux densities that would follow from assuming a Band function~\cite{1993ApJ...413..281B} for the emission in the GRB frame, but concluded this gives rise to only modest, $\mathcal{O}(10\%)$ differences in picolensing sensitivity as compared to the (C)PL results.
We therefore do not investigate Band function-derived source flux densities in this paper.

\subsubsection{Detector Backgrounds}
\label{sec:backgrounds}
Space-based x-/$\gamma$-ray detectors are subject to backgrounds from a number of origins: other sources on the sky, diffuse backgrounds, instrumental backgrounds, cosmic rays and cosmic-ray activation, Earth-related backgrounds, the Sun, etc.; see, e.g., \citeR{2020A&A...640A...8B} for a detailed discussion in the context of the \emph{Fermi}/GBM.

In our study, we take a fiducial value for the single-detector background flux density to be $F_B = 10\,\text{cm}^{-2}\,\text{s}^{-1}$, while the effective areas we assume are $A_S=1300\,\text{cm}^2$ and $A_B=2400\,\text{cm}^2$ (see also \tabref{tab:params} and the discussion in \secref{sec:ourComputation}).
We chose these parameters for two reasons: (1) they match the parameters assumed in \citeR{Gawade:2023gmt} and therefore facilitate comparison with that work, and (2) they give a single-detector sensitivity broadly in line with that of the \emph{Swift}/BAT instrument. 
To see the latter point, let us compute the $5\sigma$ flux-density sensitivity.
By an argument similar to that in \secref{sec:detectionCriteria}, the single-detector SNR is 
\begin{align}
    \varrho^{(1)} &= \frac{F_S A_S T}{\sqrt{ F_S A_S T + F_B A_B T }}
    \sim \frac{F_S A_S T}{\sqrt{ F_B A_B T }}\:,
\end{align}
where, in the latter step, we assumed $F_S A_S \ll F_B A_B$, as can be verified \emph{a posteriori}.
Setting $\varrho^{(1)}=5$ and inverting to find the source sensitivity, we obtain 
\begin{align}
    F^{(1),\varrho=5}_S &= 5\sqrt{ \frac{ F_B  A_B }{  A_S^2  T } } \sim 6\times 10^{-2}\,\text{cm}^{-2}\,\text{s}^{-1} \times \sqrt{ \frac{100\,\text{s}}{T} }\:.
\end{align}
Assuming 15\,keV photons, this would correspond to a 5$\sigma$ sensitivity of $60\,\text{milliCrab}\times \sqrt{(100\,\text{s})/T}$, which is within an $\mathcal{O}(1)$ factor of the \emph{Swift}/BAT sensitivity~\cite{2005SSRv..120..143B,SWIFTtechnical}.%
\footnote{\label{ftnt:crabUnit}%
    $1\,\text{Crab} = 2.4\times 10^{-8}\,\text{erg}\,\text{cm}^{-2}\,\text{s}^{-1}$ and $1\,\text{keV} \approx 1.602\times 10^{-9}\,\text{erg}$.
} %

Detector backgrounds in space can however vary significantly.
For instance, even instruments in low Earth orbit have significantly increased background rates during their passage through the South Atlantic Anomaly (and for some time thereafter)~\cite{2020A&A...640A...8B} owing to the concentration of high-energy charged particles found there.
Because we are also making projections for larger spacecraft separations (see \tabref{tab:baselines}) which imply that the detectors cannot (both) be in low-background regions of low Earth orbit, we also study the impacts on picolensing sensitivity for a future mission under different constant background-rate assumptions.
Specifically, we assume a `low background' case where we set $F_B = 3 \,\text{cm}^{-2}\,\text{s}^{-1}$ (roughly $1/3$ of the fiducial value) and a `high background' case where we set $F_B = 100 \,\text{cm}^{-2}\,\text{s}^{-1}$ (10 times the fiducial value).
Viewed alternatively, the impact of these variations on the projections gives an understanding of how effective future-mission background rejection or shielding would need to be in order to achieve target sensitivity.

\subsubsection{Other Missions}
\label{sec:otherMissions}
\emph{Swift}/BAT is not the only satellite in orbit capable of detecting GRBs.

The \emph{Fermi} satellite's Gamma-Ray Burst Monitor (GBM)~\cite{Meegan_2009} covers a wider energy range ($\sim 8\,\text{keV}$ to $\sim 40\,\text{MeV}$ combined over both of the GBM detector types, NaI and BGO) with an average effective area of%
\footnote{\label{eq:perDetectorA}%
    Note that \citeR{Bissaldi:2008df} gives on-the-ground \emph{Fermi} calibration information showing a peak on-axis effective area on the order of $\mathcal{O}(100\,\text{cm}^2)$ for a single ``flight module'' (FM) NaI detector in this energy range.
    There are 12 such FMs in the GBM, so this per-detector area must be multiplied by 12.
    However, averaging over off-axis response then drops the total effective area back down by a factor of $\sim 2$.
} %
$\mathcal{O}(600\,\text{cm}^2)$ for the 12~NaI detectors over an energy range of roughly $10$--$100\,\text{keV}$~\cite{Bissaldi:2008df}, and a field of view that covers about 70\% of the sky at any one time (i.e., almost everything except regions occulted by Earth).
\emph{Fermi} has recorded a sample%
\footnote{\label{ftnt:Fermi}%
    Numbers are correct as of September 19, 2024.
} %
of 2356 GRBs over $\sim 10$\,\text{years} of datataking~\cite{Paciesas_2012,Goldstein_2012,Gruber_2014,vonKienlin_2014,Bhat_2016,vonKienlin_2020}.
Data from other spacecraft (e.g., INTEGRAL~\cite{2005A&A...438.1175R}, BATSE~\cite{Paciesas:1999tp}, Konus-Wind~\cite{KonusWind}, etc.) are also available.
Although these data could all be used to augment the \emph{Swift}/BAT catalogue in our study, we judge that the 409 GRBs from the \emph{Swift}/BAT catalogue that have all the necessary data is a sufficiently large and representative sample of GRBs for the purposes of making projections.

Interestingly, there are many smaller x-/$\gamma$-ray sensitive instruments located on other space-based or even interplanetary spacecraft, some of which are a few AU apart from others.
These are used together in the `Interplanetary Network' (IPN)~\cite{Palshin:2013jgt,Svinkin:2022yhs,IPN} for, e.g., GRB sky localization via timing delay.%
\footnote{\label{ftnt:thanks}%
    The authors thank Aaron Tohuvavohu for bringing the IPN to our attention. 
} %
However, these existing detectors may not be ideal for a picolensing search, as their relative response calibrations are somewhat poorly known, their effective areas are typically much smaller than dedicated x-/$\gamma$-ray detectors in LEO, and their detector bandpasses may not be ideal for our application.
Whether a picolensing search could already be undertaken using IPN data is however a point worthy of some further study.

Additionally, there are a number of proposed future x-/$\gamma$-ray space missions.
One we mention in particular is the \emph{Daksha} proposal~\cite{Bhalerao:2022edb,Bhalerao:2022pon}, as it formed the basis of the study in \citeR{Gawade:2023gmt}.
It is envisaged as a pair of satellites, which would in principle enable its usage for picolensing.
Broadly speaking, each \emph{Daksha} satellite would be a detector in the same class as \emph{Swift}/BAT, with a planar detector area of $\sim 2400\,\text{cm}^2$, an average effective area of $\sim 1300\,\text{cm}^2$, and a sensitive range of $20$--$200\,\text{keV}$ for that effective area.
It would also have a field of view roughly 6 times larger than that of \emph{Swift}/BAT, comparable to that of \emph{Fermi}/GBM, enabling a higher rate of GRB detection than either (because of its larger effective area than \emph{Fermi}/GBM~\cite{Bhalerao:2022pon}).
The drawback for picolensing of the current \emph{Daksha} proposal is, however, a short baseline: both satellites are proposed to be placed in low Earth orbit.

\subsection{GRB Sizes}
\label{sec:GRBsizes}
The transverse size of the GRB emission region visible at the detector is the final parameter of importance in determining the observability of the picolensing signal.
We remind the reader that, as a general rule,%
\footnote{\label{ftnt:violationOfGeneralRule}%
    With some exceptions for sources with angular sizes on the order of $\theta_E$; see \figref{fig:sigmaChiL} and also \citeR{Gawade:2023gmt}.
} %
small GRB source angular sizes are favourable for picolensing detection, whereas extended GRB source angular sizes can severely suppress the signal (here, small vs.~extended is judged on the scale of $\theta_E$).
GRBs are however too small for any direct measurements of their angular sizes to be made, so they must be inferred indirectly; see the discussions in \citeR[s]{Barnacka:2014yja,Golkhou:2015lsa,Katz:2018zrn}.

A proxy for the physical size of the emission region can be obtained from the so-called observed ``minimum variability timescale'' $t_{\text{var}}$ of the GRB~\cite{Barnacka:2014yja}, from which one can estimate that the physical size of the emission region (``fireball'') is very roughly~\cite{Golkhou:2015lsa} $D_{\text{emit}}^{\text{phys}} \sim \Gamma^2 \Delta t_{\text{var}}/(1+z_S)$, where $\Gamma \sim 10^2$ is the typical Lorentz boost of the source in the detector frame.
However, owing to relativistic beaming effects~\cite{Barnacka:2014yja,Katz:2018zrn}, the transverse physical size of the patch of the emission region that is actually visible to a distant observer is instead $D_S^\text{phys} \sim D_{\text{emit}}^{\text{phys}}/\Gamma \sim \Gamma \Delta t_{\text{var}}/(1+z_S)$.

\citeR{Barnacka:2014yja} gives some empirical evidence for a very rough scaling relationship $D_S^\text{phys} \sim T_{90}/(1+z_S)$, implying a rough empirical relationship $T_{90} \sim \Gamma \Delta t_{\text{var}}$.
By contrast, \citeR{Golkhou:2015lsa} shows little evidence for such an overall trend in the values of $D_{\text{emit}}^{\text{phys}}$ with $T_{90}$, although there is a clear bimodal distribution for $D_{\text{emit}}^{\text{phys}}$ for short vs.~long GRBs; whether or not there is such a trend for $D_S^{\text{phys}}$ is not clear.
Moreover, even though these estimates may give a rough idea of GRB sizes, there is also a large scatter in estimated GRB sizes around any deterministic estimate based on $T_{90}$: \citeR[s]{Barnacka:2014yja,Golkhou:2015lsa} both show that uncertainties on $D_S^{\text{phys}}$ or $D_{\text{emit}}^{\text{phys}}$ for individual GRBs can be \emph{some orders of magnitude} coming from different physically motivated estimates.
One thing is clear: all of these GRB size estimates are \emph{highly} uncertain in (a) the overall normalization in any deterministic GRB size estimate based on $T_{90}$, (b) the scatter for individual GRB sizes around any such deterministic estimate, and (c) whether or not a single overall trend in observed sizes with $T_{90}$ even exists.

Based on the empirical trends in \citeR{Barnacka:2014yja}, the study of \citeR{Gawade:2023gmt} proceeded by assigning the physical diameter of the observable (i.e., post-beaming) transverse GRB size to be 
\begin{align}
    D_S^\text{phys} &= \frac{k_S T_{90}}{ 1 + z_S  }\label{eq:DSphys} 
    \equiv \frac{ \Delta \chi_{ S,\perp} }{  1 + z_S } \:; \\[1ex]
    \Rightarrow \Delta \chi_{ S,\perp} &= k_S T_{90}\: \label{eq:chiT90}\:,
\end{align}
where $k_S$ is a constant governing the overall systematic shift of the GRB size distribution.
\citeR{Gawade:2023gmt} took this to be $k_S = 2$; i.e., they took the observable physical source transverse \emph{radius} to be $\sim T_{90} / ( 1 + z_S )$.
The corresponding source angular half-size (i.e., the angle subtended by the source physical radius) is
\begin{align}
    \theta_S &= \frac{k_S}{2} \frac{ T_{90} }{ \chi_S }\:.
\end{align}

\begin{figure}[t]
    \centering
    \includegraphics[width=\columnwidth]{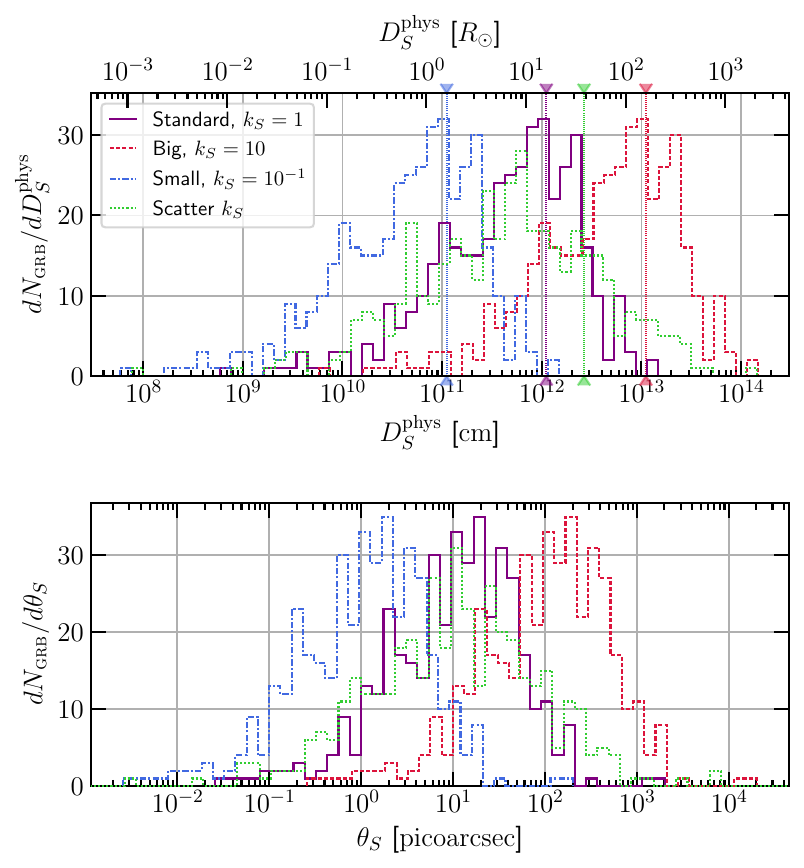}
    \caption{\label{fig:rShist}%
    Assigned GRB sizes for the 409 \emph{Swift}/BAT GRBs on which our study is based, computed using observed values of $T_{90}$ and $z_S$~\cite{SWIFTBATwhatWeUsed}.
    \textsc{Upper panel:} Histogram of physical transverse GRB source diameters $D_S^{\text{phys}} = k_S T_{90}/(1+z_S)$ under the ``standard'' (fiducial), ``big'' (conservative), ``small'' (aggressive), and ``scatter'' (single realization) GRB source-size assumptions; see discussion in text.
    Units are marked in centimetres on the lower axis and (for convenience) in solar radii on the upper axis.
    The mean of each distribution is denoted by the dotted vertical line marked with arrows.
    \textsc{Lower panel:} Corresponding observed source angular half-sizes $\theta_S \equiv D_S^{\text{phys}} / (2\mathcal{D}_S) =  (k_S/2)\times (T_{90} / \chi_S)$ in units of picoarcseconds. 
    Recall: $1\,\text{picoarcsec} = 4.85\times 10^{-18}\,\text{rad}$.
    }
\end{figure}

To quantify the uncertainty associated with the GRB source sizes, in this work we will consider four different scenarios for the source sizes based on \eqref{eq:chiT90} [see \figref{fig:rShist}]: (1) a ``standard'' (fiducial) deterministic size assumption, which we take to have $k_S = 1$; (2) a ``big'' (conservative) deterministic assumption, which we take to have $k_S = 10$, (3) a ``small'' (aggressive) deterministic assumption, which we take to have $k_S = 10^{-1}$; and, finally, (4) we break the deterministic dependence on $T_{90}$ by randomly sampling $k_S$ for each individual GRB from the distribution%
\footnote{\label{ftnt:distroMotivation}%
    We do not have a specific physical motivation for this distribution.
    We choose it merely as a proxy to compare the effect of a random error in determination of GRB sizes, to the impact of the correlated shifts assumed for the ``big'' and ``small'' cases.
} %
$\log_{10} k_S \sim \mathcal{U}[-1,1)$, which we denote as the ``scatter'' case. 
For the scatter case, computational limitations restrict us to considering one single realization of randomly drawn values of $k_S$; the GRB size and angular-size distributions for this single realization are the ``scatter'' results shown in \figref{fig:rShist}.
Moreover, we study the dependence of our results on placing various minimum size cuts on $D_S^{\text{phys}}$ in order to test the sensitivity of the results to the smallest assigned GRB sizes.
In making comparisons to results of \citeR{Gawade:2023gmt}, we at times also set $k_S = 2$ to match their assumptions; we clearly identify when this is done.

Absent a better determination (or modeling) of GRB emission region sizes, we consider this bracketing of the source sizes by plus or minus an order of magnitude around a fiducial value (i.e., cases 1--3 above) that is at least somewhat physically motivated (and for which there is some empirical evidence~\cite{Barnacka:2014yja}) gives at least an idea of the possible range of results one could expect under either aggressive or conservative assumptions about a systematic shift in GRB sizes.
Case 4 allows us to also understand the impact of the scatter in GRB source sizes.
This approach allows us to judge the sensitivity of the picolensing observable to source-size uncertainties, which in turn allows us to understand its robustness when conservative assumptions are made.

\begin{table*}[t]
\begin{ruledtabular}
    \begin{tabular}{lp{5cm}p{5cm}p{5cm}}
    \textbf{Parameter}   & \textbf{Description}  &   \textbf{Value} &  \textbf{Comment}  \\ \hline \hline
    $\omega_\dm$    &   Current average DM density     &  0.11933  & $\omega_\dm = \Omega_\dm h^2$. See \citeR{Planck:2018vyg}. \\ \hline
    $z_S$       &   Source redshift     &   Selected per GRB from \citeR{SWIFTBATwhatWeUsed} &   --- \\ \hline
    $\chi_S$    &   Comoving distance to source    &   Computed from $z_S$ &   See footnote \ref{ftnt:FLRW}. \\ \hline
    $T$         &   Integration duration    &   $T = T_{90}$, selected per GRB from \citeR{SWIFTBATwhatWeUsed}       &    --- \\ \hline
    $[E_{\text{min}},E_{\text{max}}]$   &   Detector-frame bandpass  &   [15,150]\,\text{keV}    &   See also comments in \secref{sec:SWIFT-BAT_spectra}. \\ \hline
    $F_S$       &   Source flux density      &    Constructed per GRB from spectral fits in \citeR{SWIFTBATwhatWeUsed} and the detector bandpass  & Power-law or cut-off power law spectrum; see also \secref{sec:SWIFT-BAT_spectra}.    \\ \hline
    --- & Source transverse intensity profile &   Disk or Gaussian    &   $f_{\text{source}}=0.9$ for the Gaussian.    \\ \hline
    $D_S^{\text{phys}}$     &   Observed GRB transverse diameter (physical)    & Constructed per GRB as  $D_S^{\text{phys}} = k_S T_{90} / (1+z_S)$    & We take $k_S = 0.1,1,10$ or randomly distributed as $\log_{10}k_S \sim \mathcal{U}[-1,1)$ to study uncertainty (also use $k_S = 2$ to compare with \citeR{Gawade:2023gmt}). We also perform analysis with hard lower limits on $D_S^{\text{phys}}$. See \secref{sec:GRBsizes} for discussion. \\ \hline
    $\theta_S$  & Angular half-size of the source   &   $\theta_S = (\Delta \chi_{S,\perp} )/(2\chi_S)$   & $\Delta \chi_{S,\perp} \equiv D_S^{\text{phys}}(1+z_S)$ \\ \hline
    $A_S$   &   Detector effective area for signal  &   $1300\,\text{cm}^2$   &   Matches \citeR{Gawade:2023gmt}. \\ \hline
    $A_B$   &   Detector effective area for background  &   $2400\,\text{cm}^2$   &   Matches \citeR{Gawade:2023gmt}. \\ \hline
    $F_B$   &   Background flux density  &  $10\,\text{cm}^{-2}\,\text{s}^{-1}$ & Unless otherwise stated; matches  \citeR{Gawade:2023gmt}. \\ \hline
    $R_O$   &   Detector baseline separation &   Varied &   See \tabref{tab:baselines}. \\ \hline
    $\theta_O$  & Observer-baseline--source orientation angle    &   Sampled and averaged over   &   See discussion in \secref{sec:numericalGeneral}. \\ \hline
    $M$     &   Lens mass   &   Varied     &    --- \\ \hline
    $\chi_L$ & Comoving distance to lens plane   & Varied    &  Picolensing cross-section results are integrated over the lens plane distance to get the picolensing volume; see \secref{sec:picoSigma}. \\ \hline
    $z_L$   &   Lens redshift    & Computed from $\chi_L$   &   See footnote \ref{ftnt:FLRW}. \\ \hline
    $\bm{\chi}_{L,\perp}$   & Lens location in the transverse lens plane (comoving) &   Varied  & Lens locations in the lens plane are sampled to obtain the picolensing cross section; see \secref{sec:MCsigma}. \\ \hline
    $\varrho_*$    &   Single-event picolensing detection threshold    &   $\varrho_* = 5$    &   We also do a limited study with $\varrho_* = 10$, where noted.  \\ \hline
    $\alpha$    &   PBH $f_{\pbh}$ exclusion confidence level    &   $\alpha = 0.95$     &    95\% confidence exclusion.  \\ \hline
    $N_{\grb}$  & Number of GRBs to which projections are scaled    &   $3\times 10^{3}$    & Unless otherwise stated (e.g., $N_{\grb} = 10^4$ used for comparison to \citeR{Gawade:2023gmt}).
    \end{tabular}
\end{ruledtabular}
\caption{\label{tab:params}%
    A summary of the parameters used in our picolensing analysis.
    }
\end{table*}

\section{Computational Procedure}
\label{sec:ourComputation}
In principle, the computation of picolensing that we have described above is straightforward to implement.
The only real computational challenge is finding the picolensing region and cross section $\sigma$ defined in \secref{sec:picoSigma}, in order to construct the comoving picolensing volume $\mathcal{V}$ defined at \eqref{eq:VlensConst}.
This is mostly because the source-averaged magnifications seen by each observer, $\bar{\mu}_{i}$, which enter the computation via the lensing SNR $\varrho$ at \eqref{eq:SNR}, are nonlinear functions of the location of the lens in the transverse lens plane.
Moreover, for extended sources, these magnifications are not generally analytically expressible in closed form in terms of simple functions; see \secref{sec:extendedSource}.
This means that the picolensing region can have a reasonably complicated shape; see \figref{fig:sigmaExample} later for some examples.
Accordingly, the easiest way to compute the picolensing cross section is numerically, using standard Monte Carlo (MC) methods~\cite{Gawade:2023gmt}. 

In this section, we first discuss how we handle the representative sample of 409 \emph{Swift}/BAT GRBs on which our study is based, we then give a general overview of our computational implementation, and finally we detail our MC cross-section computation.

We collect in \tabref{tab:params} some important parameters that we assume in our study.

\subsection{Sampling the GRB Population}
\label{sec:numericalGeneral}
The procedure we follow to compute the picolensing cross section and picolensing volume is as follows.

We use the $N_{\grb}^{\text{\emph{Swift}/BAT}} = 409$ \emph{Swift}/BAT GRBs~\cite{SWIFTBATwhatWeUsed} in the representative real-GRB dataset discussed in \secref{sec:SWIFT-BAT} to make population-informed projections (see also \citeR{Gawade:2023gmt}).
For each real GRB, we compute the source flux density $F_S$ per \eqref{eq:sourceFluxDensity} and fix the observation time for that GRB to be $T=T_{90}$.

Our computation based on these GRBs is then best phrased as a raster parameter scan over the lens mass $M$, the choice of the observer-separation baseline $R_O$ from among the options in \tabref{tab:baselines}, the binary choice of GRB source-intensity profile (disk vs.~Gaussian) [see \secref{sec:extendedSource}], and the method of choosing the GRB size parameter $k_S$ (see \secref{sec:GRBsizes}).
For each fixed point in this parameter space, we perform the following procedure.

We simulate the isotropic distribution of GRBs on the sky (see, e.g., \citeR{1992Natur.355..143M}) that a future mission would see by placing each of the $N_{\grb}^{\text{\emph{Swift}/BAT}} = 409$ \emph{Swift}/BAT GRBs at four different random locations on the sky, instead of at their true locations.
To do so, for each \emph{Swift}/BAT GRB, we randomly draw $N_\theta = 4$ source--observer orientation angles $\theta_O$ by selecting a uniformly distributed random number $c_\theta \sim \mathcal{U}[0,1)$, and inverting for $|\sin \theta_O| = \sqrt{1-c_\theta^2}$ in order to construct the projected observer-separation baseline $R_O' = R_O |\sin\theta_O|$.
Note that this procedure can also be considered to be an averaging over the time-dependent detector-baseline orientations that would result from their orbital motion~\cite{Gawade:2023gmt}.

We then treat the $\aleph = N_{\theta} \times N_{\grb}^{\text{\emph{Swift}/BAT}} = 4\times 409 = 1636$ resulting combinations of sky locations and real observed GRB parameters as representative ``GRB samples'' that a future mission might see.
For each of these $k=1,\ldots,\aleph$ GRB samples, we compute the picolensing volume $\mathcal{V}_k$.

\subsection{Computing the Picolensing Volume}
\label{sec:numericalLensingVolume}
To compute the picolensing volume for each source $k$, we first compute the picolensing cross section over some grid of lens-plane distances $\chi_L$ that lie between the observer plane and the source, which lies at comoving distance $\chi_S^k$.

\begin{figure*}[t]
    \centering
    \includegraphics[width=\textwidth]{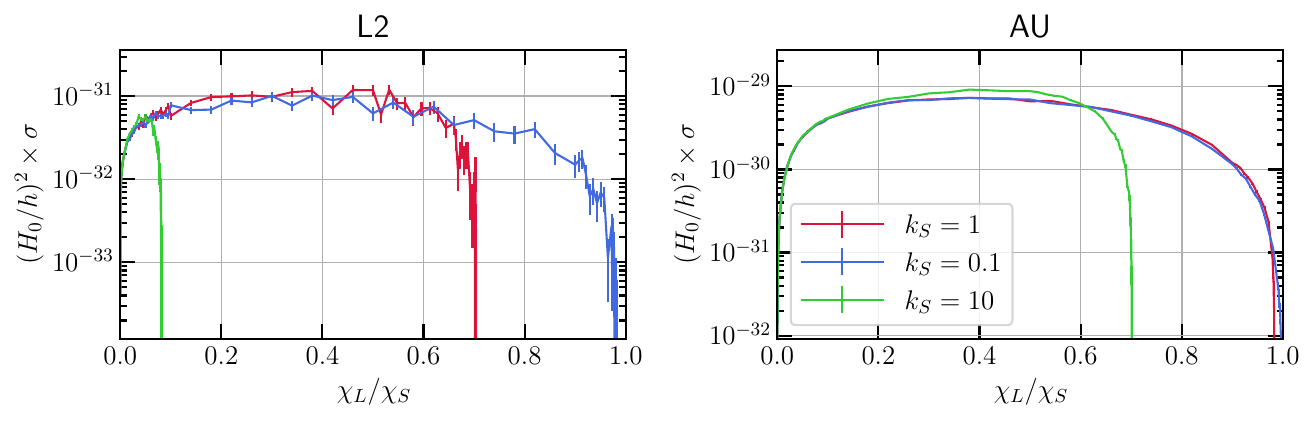}
    \caption{\label{fig:sigmaChiL}%
    Example picolensing cross sections $\sigma$ and their dependence on $\chi_L/\chi_S$ for a GRB with the parameters of GRB051001 ($z_S = 2.46$, $T_{90}=190.3$\,s, $F_S = 0.129\,\text{cm}^{-2}\text{s}^{-1}$) for an L2 (left panel) or AU (right panel) observer-separation baseline $R_O'=R_O|\sin\theta_O|$, and a Gaussian source profile with the values of $k_S$ indicated in the legend in the right panel; cf.~\eqref{eq:chiT90}.
    The lens mass is set to be $M=10^{-7}M_{\odot}$.
    Note, for instance for the AU baseline, that a larger source size can allow $\sigma$ to actually \emph{increase} compared to its value of smaller source sizes at fixed $\chi_L$ (an observation also made in \citeR{Gawade:2023gmt}).
    Both panels show that the cross section for larger source sizes can however go to zero far before the source is reached.
    The results for the L2 baseline show more significant statistical fluctuations than for the AU baseline because the same lens density was used to sample the picolensing region for both cases, but the picolensing cross section is much smaller in the former case (note the differing vertical scales between the two panels).
    The vertical scale in both panels has the picolensing cross section normalized to the Hubble radius squared, $H_0^{-2}$ (up to a factor of $h^2 \approx 0.46$~\cite{Planck:2018vyg}), because this is a convenient scaling for \eqref{eq:fDMlimit} [cf.~\eqref{eq:lambdaDefn}]; of course, the comoving picolensing cross section is \emph{extremely} small on the scale of the squared comoving radius of the observable Universe.
    }
\end{figure*}

Let $\xi = \chi_L/\chi_S^k$ in this section.
We select an initial fixed grid of lens-plane distances $\chi_L$ in the range $\xi_{\text{min}} < \xi < \xi_{\text{max}}$, where $\xi_{\text{min}} = 1 - \xi_{\text{max}} = 10^{-5}$ (we limit the range in this way for computational reasons; the impact on the final results is negligible).
This initial grid is not taken to be uniform over the entire $\chi_L$ range, but instead has a denser regular spacing near the detectors ($\xi<0.1$) and the source ($\xi>0.9$), while it has a coarser regular spacing in between.%
\footnote{\label{ftnt:samplingLogic}%
    The logic behind this is clear from \figref{fig:sigmaChiL}: provided that $\sigma(\chi_L)$ does not go to zero significantly before $\xi \sim 1$, its general behaviour is to increase rapidly at small $\xi \sim 0$, vary quite smoothly over most of the range of $0.1\lesssim \xi \lesssim 0.9$, and then go to zero again rapidly near $\xi \sim 1$.
    Our initial grid is chosen to sample more densely in regions where we expect \emph{a priori} that $\sigma(\chi_L)$ varies more rapidly.
} %
At each $\chi_L$ thus chosen in this grid, we compute the picolensing cross section $\sigma(\chi_L)$; we detail the actual cross-section computation separately in the next subsection.

In order to capture fine features in the $\chi_L$-dependence of $\sigma(\chi_L)$, we then dynamically update the grid of sampled $\chi_L$ points by running the following refinement procedure twice in succession.

First, there can be cases where $\sigma(\chi_L)$ peaks close to the detectors and then falls rapidly to zero, leading to most of the cross-section evaluations performed thus far being zero to within the computational tolerance.
In order to not accidentally omit such cases, we check whether most of the cross-section evaluations very near the source, $\xi_{\text{min}} \leq \xi \leq \xi_{\text{check}}$, are indeed greater than zero (within the computational tolerance).%
\footnote{\label{ftnt:checkPointFirstPass}%
    On the first pass through the refinement procedure, $\xi_{\text{check}} \approx 0.05$.
    The initial grid construction guarantees that there are at least three (but possibly more) $\sigma(\chi_L)$ samples on $\xi_{\text{min}} \leq \xi \lesssim 0.05$.
    On the second pass of the refinement procedure, the value of $\xi_{\text{check}}$ may be dynamically updated to be instead a smaller value.
} %
If this is not the case, we define a large number of additional $\chi_L$ grid points on an even more densely spaced regular sub-grid spanning $\xi_{\text{min}} \leq \xi \leq \xi_{\text{check}}$ and \emph{substitute} this new sub-grid in place of the previous $\chi_L$ grid points in the range $\xi_{\text{min}} < \xi < \xi_{\text{check}}$.
We then compute $\sigma(\chi_L)$ for the interior points in the new sub-grid of $\chi_L$ sample points, and substitute these freshly computed cross-section values in place of any previously computed values over that range.

We subsequently perform a second check within each refinement pass.
We examine in turn each of the $\chi_L$ grid points throughout the interior of the thus-far sampled grid (including any grid points added immediately by the procedure described in the preceding paragraph).
Should the value of $\sigma(\chi_L$) at a given $\chi_L$ grid point vary downward by more than a factor of 4 from the $\sigma(\chi_L)$ values at either the preceding or succeeding $\chi_L$ grid points, or if it should vary by more than a factor of 3 from the value of $\sigma(\chi_L)$ averaged over those neighbouring grid points, we flag that grid point for refinement. 
After all interior grid points are thus examined, we define a set of additional, more finely spaced grid points in the vicinity of each grid point flagged for refinement, and compute $\sigma(\chi_L)$ on these newly added points.
We then add in these newly sampled grid points and cross-section computations to the overall set of sampled points.
Note that on each pass through the refinement procedure, we add additional grid points under this second check only once at the end of the pass; this is not a recursive process within each pass.

At the completion of the refinement passes, this dynamically constructed grid of $\chi_L$ sample points consists of a collection of regularly spaced sub-grids of points that cover different, contiguous ranges of $\chi_L$ over the entire sampled range $\xi_{\text{min}} \leq \xi \leq \xi_{\text{max}}$.
For each point in the grid, we have a computed value of $\sigma(\chi_L)$.
Typically, this procedure ends up yielding on the order of 50 to 100 $\chi_L$ grid points and cross-section evaluations in total for each source $k$; the exact number varies dynamically for each GRB sampled.
Some examples of the $\chi_L$-dependence of $\sigma(\chi_L)$ are shown in \figref{fig:sigmaChiL}.

To then obtain $\mathcal{V}_k$, we employ Simpson's rule to first integrate $\sigma(\chi_L)$ over each of the regular sub-grids that have been defined within the overall sample of $\chi_L$ points.
We then take the simple sum of these sub-grid integral results to obtain $\mathcal{V}_k$ as defined at \eqref{eq:VlensConst}.

The average picolensing volume defined at \eqref{eq:VlensAvg} is then computed over all $\aleph$ GRB samples, which achieves averaging over both the GRB source-population properties and sky locations (i.e., the orientation angles $\theta_O$). 
Thus armed, it is trivial to extract $f_{\pbh}^{\text{limit}}$ per \eqref{eq:fDMlimit} for any desired number of observed GRBs that follow the same source-property distributions as the \emph{Swift}/BAT sample~\cite{SWIFTBATwhatWeUsed}.

\subsection{Monte Carlo Picolensing Cross-Section Computation}
\label{sec:MCsigma}
In order to obtain the picolensing cross section $\sigma(\chi_L)$ at each $\chi_L$ grid point discussed in the preceding subsection, we proceed as follows.

Let $(x,y)$ label comoving coordinates in the transverse lens plane, measured in units of the comoving Einstein radius in the lens plane.
These coordinates for the transverse lens plane are chosen such that the line joining the midpoint of the observer-separation baseline to the centroid of the source (i.e., the normal vector that defines the lens plane) pierces this coordinate system at the origin; see \figref[s]{fig:coords} and \ref{fig:LPgeom}.
Additionally, the plane that contains the source centroid and the two detectors is assumed to intersect the lens plane along the line $y=0$.
As a result, in the (highly justified) small-angle approximation, the lines of sight from the two observers to the centroid of the source will pierce the lens plane at locations $x_{i}^{\text{LP LoS}} \equiv ( \pm x_O , 0 )$, where $x_O = R'_O / ( 2 \chi_E^O ) > 0$, and $i = 1$ (respectively, $i=2$) corresponds to the $+$ ($-$) sign.
Let the radius of the source in the lens plane in these coordinates be $x_S \equiv \Delta \chi_{S,\perp}/ ( 2\chi_S \theta_E ) = \theta_S/\theta_E$.

Note that by assumption of an axially symmetric source intensity profile and in the small-angle approximation, the geometry of the lens plane is such that there are independent reflection symmetries $x \rightarrow -x$ and $y \rightarrow -y$.
That is, if we placed lenses at $(x,y)=(\pm|x|,\,  \pm |y|)$ [independent sign choices], identical results would be obtained for the lensing magnifications $\mu_{1,2}$ with respect to each observer. 
As such, we could in principle restrict our attention to lenses located at $(x,y)$ with $x>0$ and $y>0$, and use symmetry to extend to the rest of the plane; in our computation, we only exploited this freedom to restrict to $x>0$, but we considered lenses with both signs of $y$.

Consider a rectangular region (``sample box'') in the lens plane bounded by $x_{\text{min}} \leq x \leq x_O + \ell$ and $|y|\leq \ell$, where $x_{\text{min}} \equiv \max[0,x_O - \ell]$ and $\ell \equiv 1.25\cdot x_S + 5$.
The physical interpretation of this choice (at least for $x_{\text{min}} > 0$) is that we consider a sample box with a side length that is equal to 125\% of the full angular size subtended by the extended source in the lens plane, plus an additional perimeter around that box that is 5 Einstein radii wide.
Given that the lensing magnification for either a point or extended source falls off dramatically when the lens location is separated from any part of the source by more than a few Einstein radii, this is expected to be a sufficiently large region to capture all locations in the lens plane where a significant (pico)lensing signal is expected to exist.%
\footnote{\label{ftnt:boundaryCheck}%
    We also check explicitly that no sampled lens location that forms part of the picolensing region ever comes within 5\% of $\ell$ from the edges of the sample box.
    If $x_{\text{min}}>0$, we perform this check over all of the box edges.
    On the other hand, if $x_{\text{min}}=0$, we allow lenses arbitrarily close to $x=0$ because, after reflection about the vertical symmetry axis, $x=0$ is not an external boundary of the sample region in this case.
} %
For $x_{\text{min}} = 0$, the sample box is truncated in the direction of the observer-separation vector projected into the lens plane, but this does not change that this truncated sample box is still sufficient to capture all locations where a relevant picolensing signal is expected to exist.
The scaled area of this sample box is $\mathcal{A} \equiv 2\ell( x_O + \ell - x_{\text{min}} ) \leq 4 \ell^2$.

\begin{figure*}[t]
    \centering
    \includegraphics[width=0.495\textwidth]{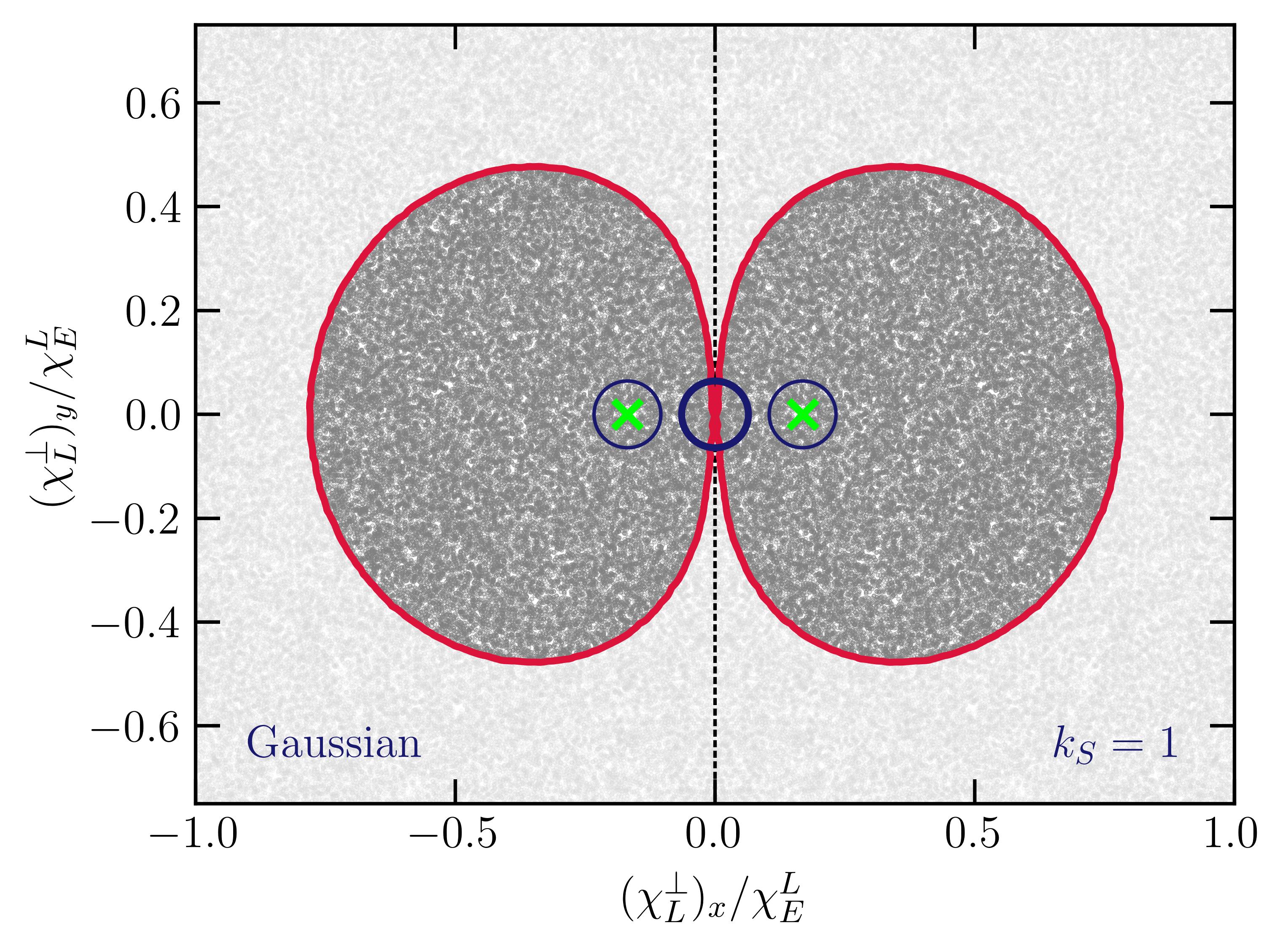} \hfill
    \includegraphics[width=0.495\textwidth]{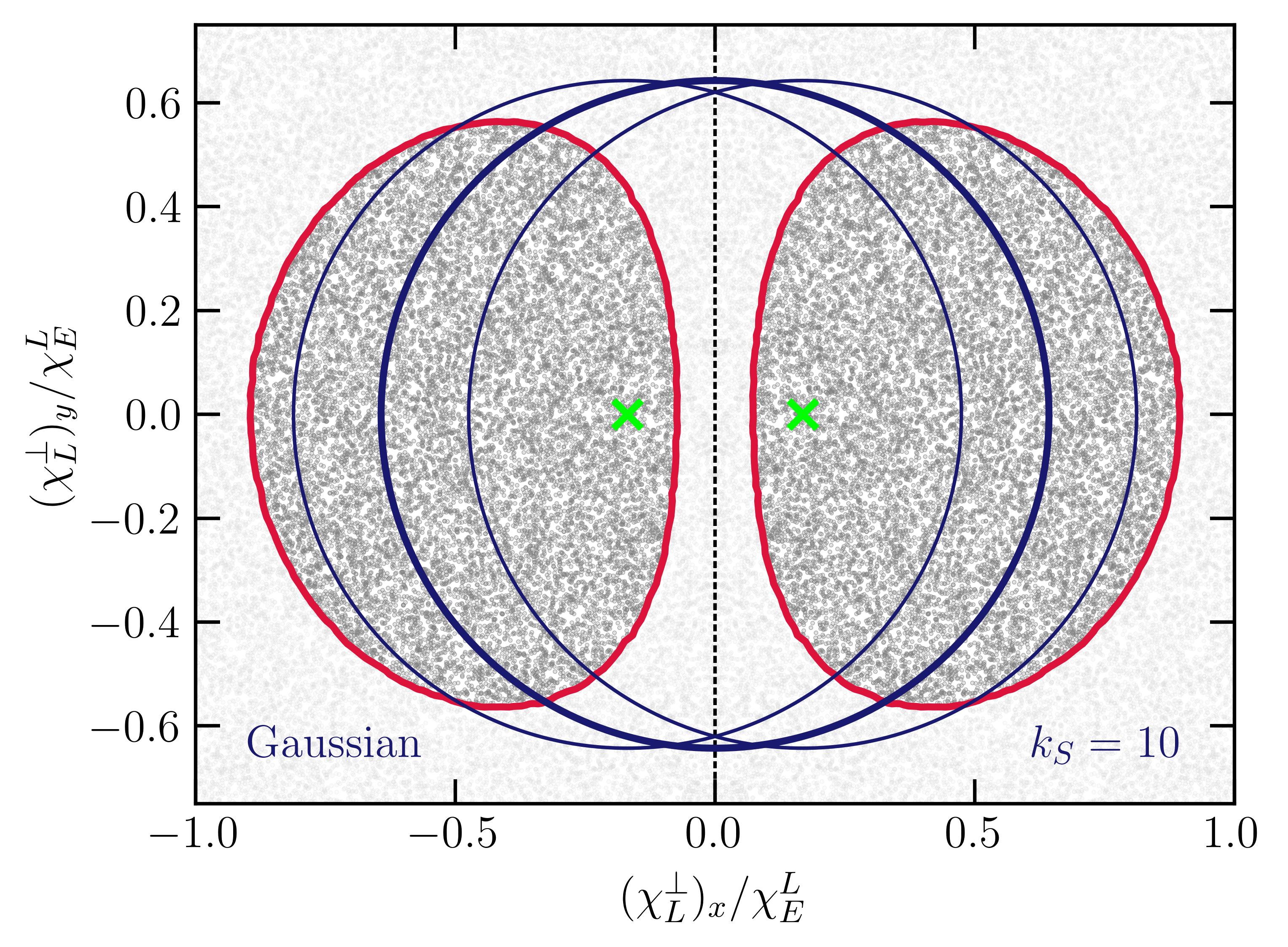}\\ 
    \includegraphics[width=0.495\textwidth]{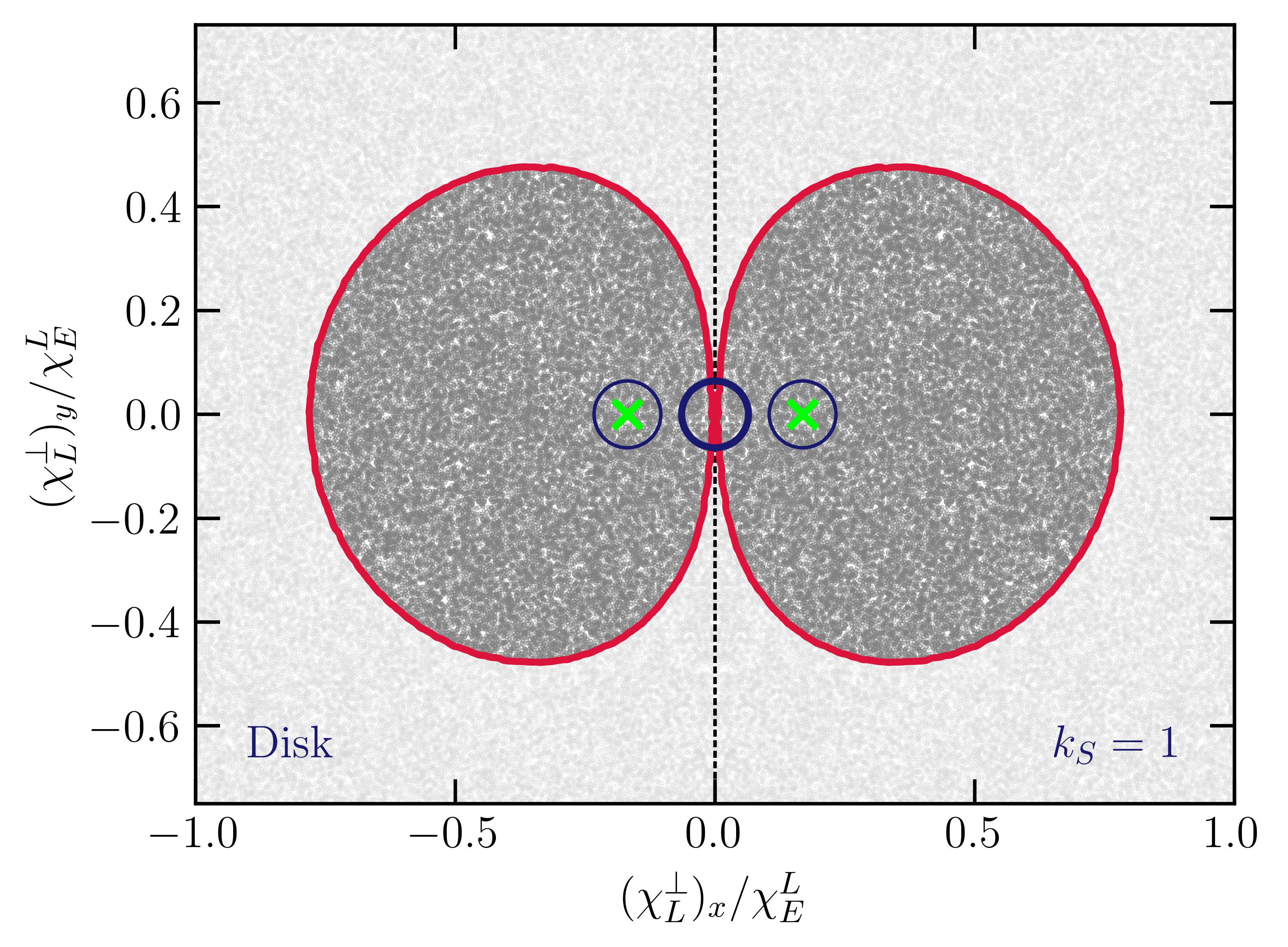} \hfill
    \includegraphics[width=0.495\textwidth]{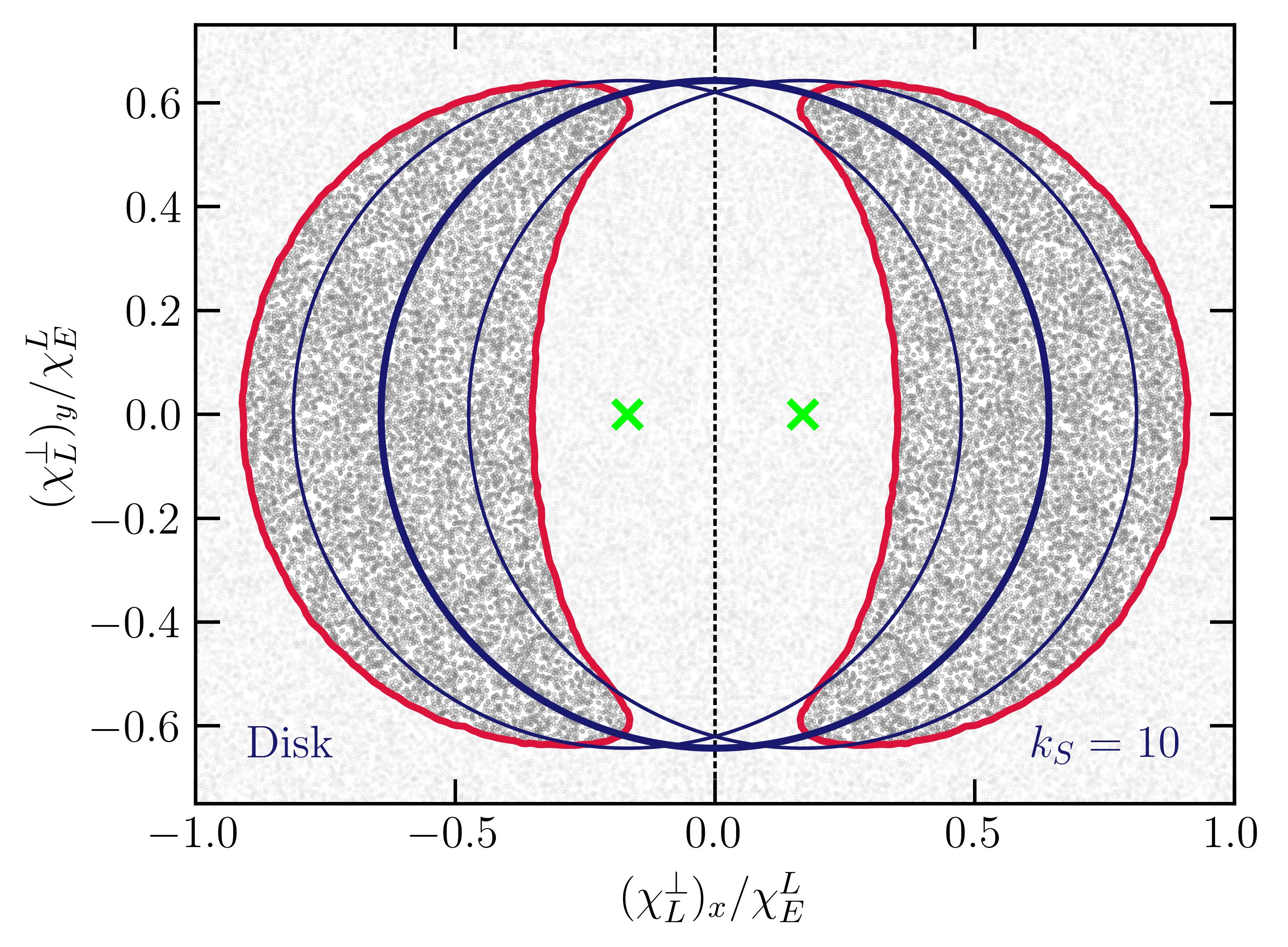}
    \caption{\label{fig:sigmaExample}%
    Example picolensing regions for the lens plane at $\chi_L/\chi_S = 0.5$ for a GRB with the parameters of GRB051001 ($z_S = 2.46$, $T_{90}=190.3$\,s, $F_S = 0.129\,\text{cm}^{-2}\text{s}^{-1}$), for the values of $k_S=1$ (left column) or $k_S = 10$ (right column) [cf.~\eqref{eq:chiT90}], and the Gaussian (top row) or disk (bottom row) profiles.
    The lens mass is set to be $M=10^{-7}M_{\odot}$ throughout.
    The axes are plotted in comoving transverse coordinates in the lens plane, normalized to the comoving Einstein radius: i.e., $(\chi_L^{\perp})_x/\chi_E^L = x$ and $(\chi_L^{\perp})_y/\chi_E^L = y$, where $x,y$ are as defined in the text of \secref{sec:MCsigma}.
    The red line is an approximate, smoothed boundary of the picolensing region, containing sampled lens locations (darker gray dots) that have a picolensing SNR above the detection threshold $\varrho_* = 5$.
    The lens-plane intersections of the lines of sight to the source centroid for two observers separated by $R_O' = R_O |\sin\theta_O| = 1\,\text{AU}$ are shown by the green crosses, while the source size projected into the lens plane as viewed from the midpoint between the observers (respectively, from each observer location) is denoted by the thick (thin) blue circle(s).
    The vertical dotted black line denotes the symmetry axis: the right side of each image is actually computed; the left side is mirrored from the right for the purposes of visualization.
    }
\end{figure*}

Within this sample box, we then pick $\mathcal{N}$ random lens locations whose $x$ and $y$ coordinates are each independently uniformly randomly distributed over the relevant dimensions that define the sample box.
For each sampled lens location, we compute the magnifications of the source with respect to each observer, $\bar{\mu}_{1,2}$, and the picolensing SNR $\varrho$.
We count the number of sampled lens locations $\mathcal{N}_*$ for which $\varrho \geq \varrho_*$.
In these scaled units, this permits us to assign the scaled comoving cross section to be $\tilde{\sigma} = (\mathcal{N}_*/\mathcal{N})\mathcal{A}$.
In real units, the comoving picolensing cross section for the given SNR threshold is then $\sigma(\chi_L) = \chi_E^2 \tilde{\sigma}$.

We find that, for most situations, $\mathcal{N} = 2.5\times 10^{4}$ lenses at each value of $\chi_L$ is a sufficiently large sample to obtain a statistically meaningful evaluation of $\tilde{\sigma}$.
However, at large $M$, our sampling procedure can become inefficient because the observer separations we consider become smaller than the Einstein radius in the observer plane;%
\footnote{\label{ftnt:smallerNlens}%
    As further evidence of convergence of the cross-section estimate with respect to the number of lenses simulated, we also ran light-weight computations with $\mathcal{N} = 2\times 10^{3}$ lenses.
    For low lens masses (small Einstein radii compared to the lens-plane-projected observer separation) the results obtained were statistically in agreement with those for $\mathcal{N} = 2.5\times 10^{4}$ lenses; for large mass lenses, the $\mathcal{N} = 2.5\times 10^{4}$ results indicated larger picolensing cross sections than for the $\mathcal{N} = 2\times 10^{3}$ results. 
    We re-ran some of these results with $\mathcal{N} = 10^5$ lenses and observed no further statistically significant changes. 
} %
for situations where we suspect (or have evidence) that the cross section is undersampled with $\mathcal{N} = 2.5\times 10^{4}$, we re-run the computation with $\mathcal{N} = 10^{5}$ to achieve convergence.%
\footnote{\label{ftnt:unsamplingIsConservative}%
    Note also that in the situation where a picolensing region is statistically undersampled, a \emph{smaller} cross section is usually inferred than would be the case for sufficient MC sampling, which is conservative with regard to picolensing detectability or exclusion. 
} %

Some example picolensing regions, computed for the purposes of visualization with a much higher density of sample lens locations than is typically utilized in our actual computations, are shown in \figref{fig:sigmaExample}.

\section{Results and Discussion}
\label{sec:results}
Having described the methodology of the computation, let us proceed to present and discuss the results.

\begin{figure}[t]
    \centering
    \includegraphics[width=\columnwidth]{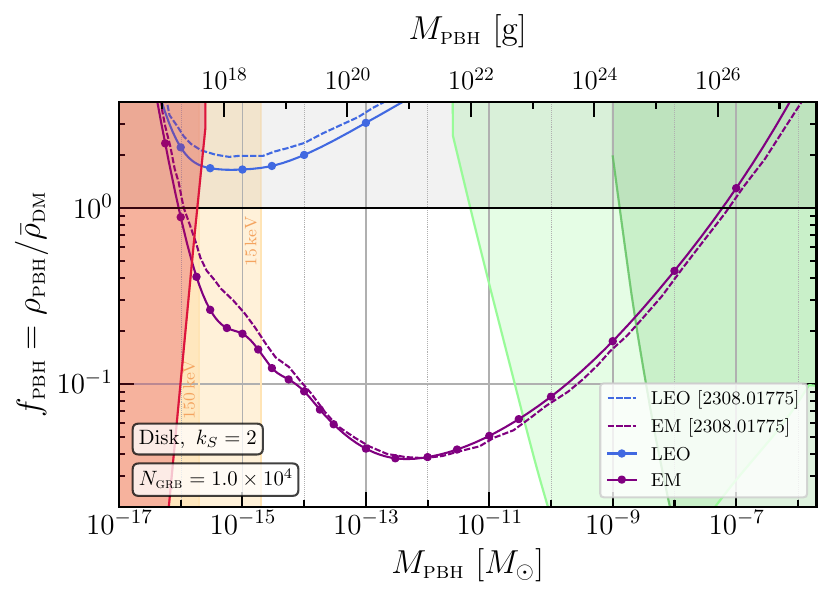}
    \caption{\label{fig:comp}%
    Comparison with results in the existing literature. 
    All plotted lines are projections for 95\%-confidence exclusions on the fraction of the DM mass density in PBHs $f_{\pbh} \equiv f_{\pbh}^{\text{limit}}$ that could be placed based on the non-observation of any $5\sigma$ (i.e., $\varrho \geq \varrho_*=5$) picolensing events when $N_{\grb}=10^4$ GRBs with properties similar to those in the \emph{Swift}/BAT GRB catalogue are observed, for both LEO (blue) and EM (purple) baselines; see \tabref{tab:baselines}. 
    The results of \citeR{Gawade:2023gmt} are shown as dashed lines.
    Our results are shown as the marked points and solid lines: they are computed only at the marked data points; the solid lines are log-log cubic spline interpolations to those computed points. 
    We adopt parameters from \tabref{tab:params}, except that here we assume a disk source profile with $k_S=2$ and $N_{\grb}=10^4$ in order to match the assumptions of \citeR{Gawade:2023gmt}. 
    Note that while \citeR{Gawade:2023gmt} assumed a detector-frame bandpass of $20$--$200\,\text{keV}$, our results are shown for a bandpass of $15$--$150\,\text{keV}$; this makes only a small, $\mathcal{O}(10\%)$ difference to the results at most masses (see discussion in \secref{sec:SWIFT-BAT_spectra}).
    The red shaded region is excluded by observables related to the Hawking evaporation of the PBHs~\cite{Carr:2009jm,DeRocco:2019fjq,Laha:2019ssq,Clark:2016nst,Mittal:2021egv,Boudaud:2018hqb,Laha:2020ivk,Kim:2020ngi,Dasgupta:2019cae,Coogan:2020tuf,Su:2024hrp,Tan:2024nbx,DelaTorreLuque:2024qms,Korwar:2023kpy}, while the lighter green shaded region is excluded by microlensing constraints~\cite{Macho:2000nvd,EROS-2:2006ryy,Griest:2013aaa,Niikura:2017zjd,Oguri:2017ock,Zumalacarregui:2017qqd,Smyth:2019whb,Leung:2022vcx,Esteban-Gutierrez:2023qcz,Mroz:2024mse,Mroz:2024wag}; both sets of constraints were taken from the \textsc{PBHbounds} compilation of limits~\cite{PBHbounds}.
    Separately, the darker green shaded region shows recent OGLE microlensing exclusion limits from high-cadence observations of the Magellanic Clouds~\cite{Mroz:2024wia} (see the \hyperref[noteAdded]{Note Added}).
    The gray shaded region indicates too much DM: $f_{\pbh}>1$.
    The light- and dark-beige shaded regions show where geometrical optics is expected to be a poor approximation [cf.~\eqref{eq:GOL}] for probes at 15\,keV and 150\,keV, respectively; projections in these regions are not accurate as they assume geometrical optics, but they are kept for illustrative purposes.
    We assume a $\delta$-function PBH mass distribution.
    }
\end{figure}

First, we make a direct comparison of our results with \citeR{Gawade:2023gmt}; see \figref{fig:comp}. 
To do so, we adopt assumptions as close as possible to those used in \citeR{Gawade:2023gmt}: a disk source profile with $k_S = 2$, $N_{\grb}=10^4$, and a LEO or EM baseline.
The only differences in the assumptions are: (a) we use a slightly enlarged catalogue of \emph{Swift}/BAT GRBs as compared to \citeR{Gawade:2023gmt}, (b) our assumed detector-frame bandpass is $15$--$150\,\text{keV}$, whereas \citeR{Gawade:2023gmt} used $20$--$200\,\text{keV}$ (see discussion in \secref{sec:SWIFT-BAT_spectra}), and (c) there is a minor difference in handling the statistics for making the exclusion-limit projections (see \secref{sec:probAndExclusion} and \appref{app:differentApproach}).
Although the agreement evidenced in \figref{fig:comp} is not exact, the differences are only at the level of an $\mathcal{O}(10\%)$ factor over most of the mass range probed; this is sufficiently good to validate our independent computational implementation against the results of \citeR{Gawade:2023gmt}.

\begin{figure}[t]
    \centering
    \includegraphics[width=\columnwidth]{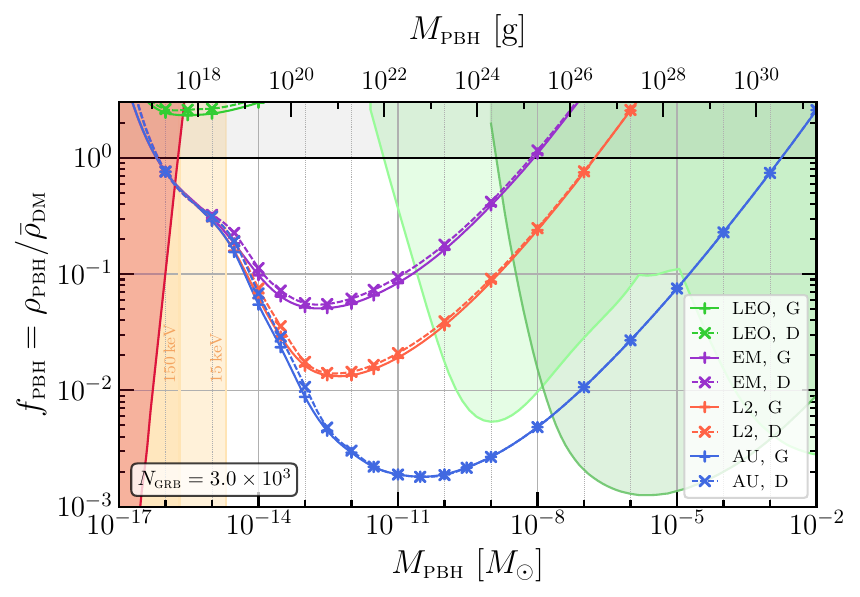}
    \caption{\label{fig:baseline}%
    Our computation for the observer-separation baselines $R_O$ listed in \tabref{tab:baselines} with standard source sizes ($k_S = 1$).
    The shaded regions have the same meaning as in \figref{fig:comp}.
    The lines also show the same nature of 95\%-confidence exclusion projections described for \figref{fig:comp}; however, here, the crosses (respectively, pluses) show our computed results for a disk [D] (respectively, Gaussian [G]) source profile, with the dashed (respectively, solid) lines being log-log cubic spline interpolations of the computed data points. 
    We adopt the parameters in \tabref{tab:params}, and assume here a more conservative observed population of $N_{\grb}=3000$ GRB sources.
    }
\end{figure}

Next, in \figref{fig:baseline}, we show projections for our ``standard'' source-size assumption ($k_S = 1$), for longer observer baselines than considered in \citeR{Gawade:2023gmt} (although see \citeR{Jung:2019fcs}), under both Gaussian or disk source intensity profile assumptions.
For these, and all subsequent, results we also use a more conservative assumption of $N_{\grb}=3\times 10^3$ observed GRBs, which is slightly larger than the \emph{Fermi}/GBM 10-year catalogue~\cite{vonKienlin_2020}.

\begin{figure*}[t]
    \centering
    \includegraphics[width=\textwidth]{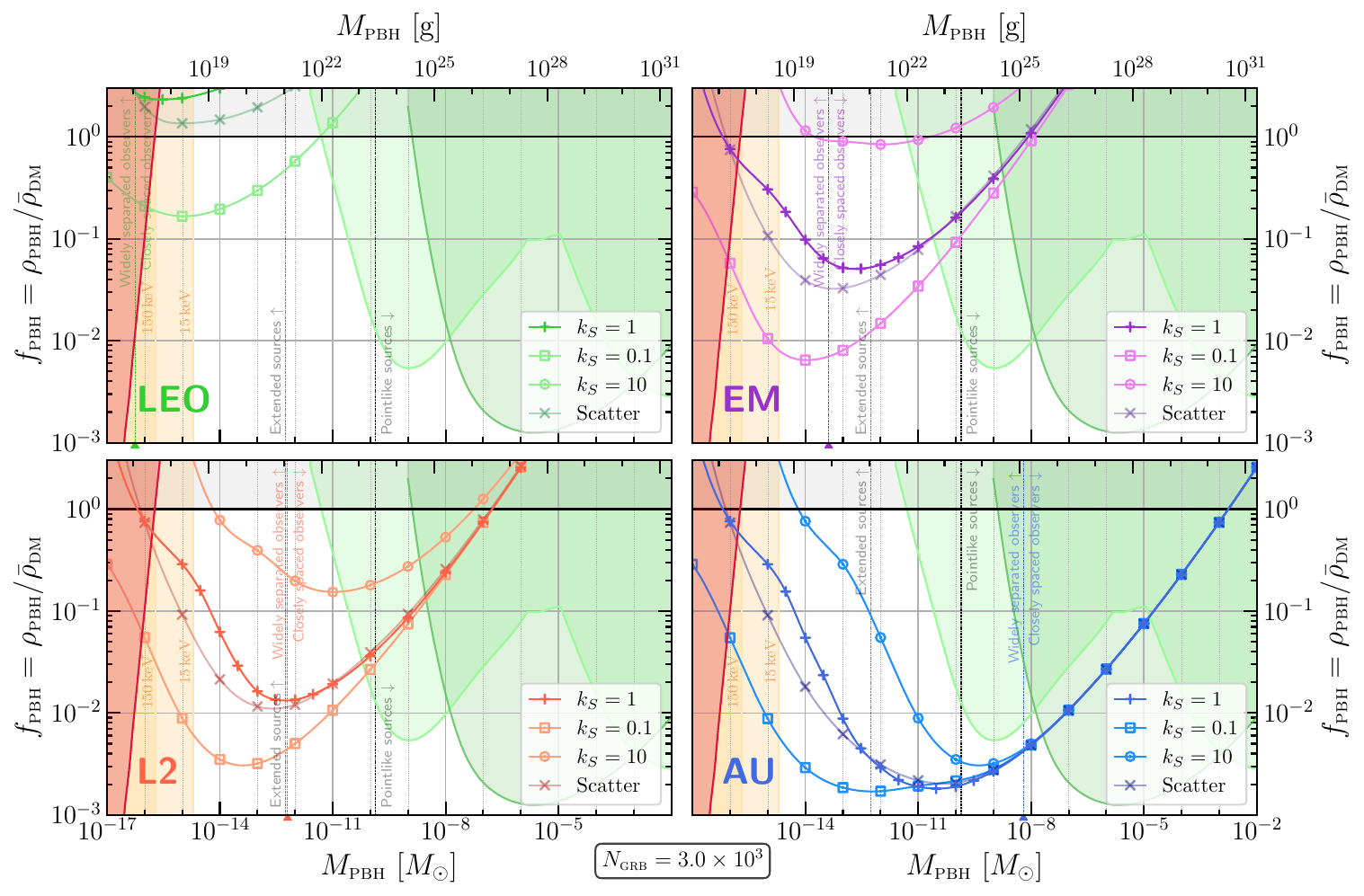}
    \caption{\label{fig:sourceSize}%
    Variation of results with source-size distribution assumptions, for different baselines $R_O$ as marked in each panel (see \tabref{tab:baselines}).
    The shaded regions in each panel are as defined in \figref{fig:comp}.
    Likewise, the marked points and lines are projected 95\%-confidence exclusions of the same nature as those detailed in \figref{fig:comp}, but are drawn here only for a Gaussian source profile for different values of $k_S$, as marked in the legends.
    The lines marked ``Scatter'' are for a single realization of assigning a random value of $k_S$ with $\log_{10} k_S \sim \mathcal{U}[-1,1)$ to each of the 409 GRBs in the portion of the \emph{Swift}/BAT catalogue on which our results are based. 
    We adopt the parameters in \tabref{tab:params}, and assume here an observed population of $N_{\grb}=3000$ GRB sources.
    The ``pointlike sources'' [roughly, $\theta_S \ll \theta_E$ for $k_S=1$], ``extended sources'' [$\theta_S \gg \theta_E$ for $k_S=1$], ``closely spaced observers'' [$R_O \ll 2\chi^O_E$], and ``widely separated observers'' [$R_O \gg 2\chi^O_E$] annotations (and the associated demarcating vertical lines) are explained in more detail \protect\hyperlink{text:annotationExplanationSourceSize}{in the main text}.
    }
\end{figure*}

From \figref{fig:baseline}, we draw five conclusions, which are valid assuming $k_S = 1$: (1) LEO baselines are insensitive as a probe of asteroid-mass PBH DM; (2) EM baselines could be sensitive to the whole of the asteroid-mass window above the lower-mass limitation imposed by the failure of geometrical optics; (3) Earth--L2 baselines, while quantitatively superior to the EM baseline, do not provide qualitatively new reach compared to the EM baselines; (4) an AU baseline would provide qualitatively superior reach: not only would it probe the whole asteroid-mass window deeply, it could also exceed the reach of existing microlensing probes at higher PBH masses, allowing us to probe new parameter space all the way up to $M_{\pbh} \sim 2\times 10^{-8} M_{\odot}$---a single experiment would thus access new parameter space over some $\sim 7$ orders of magnitude in mass; and (5) there is very little difference in the results for a disk profile vs.~those for a Gaussian profile that contains 90\% of its intensity in the same radius as the disk, making the results insensitive to this binary choice.

When the GRB size uncertainties are considered, however, some of these conclusions change.
The results shown in \figref{fig:sourceSize} make abundantly clear that variations in the source-size distribution can have a very significant impact on projected sensitivity for picolensing in the PBH asteroid-mass window.
On the one hand, for the aggressively small source-size assumption, $k_S = 0.1$, all baselines become significantly more sensitive for masses in and around the asteroid-mass window.
Specifically, the biggest qualitative change is that the LEO baseline would gain the ability to probe the window.
On the other hand, for the conservatively large source-size assumption, $k_S = 10$, the EM baseline would just barely have the ability to exclude $f_{\pbh} = 1$ in some small part of the window, which calls into question how robust the picolensing results from a mission with an EM baseline would be.

Even for $k_S = 10$, however, the L2 and AU baselines maintain their ability to robustly probe a significant fraction of the asteroid-mass window at $f_{\pbh} = 1$: nearly the entirety of the window where geometrical optics is a good approximation (for the 15--150\,\text{keV} energy range) is accessible.
An AU baseline would moreover permit such a mission to probe more deeply into sub-component DM parameter space toward the upper edge of the window.

Interestingly, note that the lower value of the PBH mass for which the projected exclusion limit crosses $f_{\pbh}=1$ is approximately the same for either the L2 and AU baselines: $M\sim 6\times 10^{-15}M_{\odot}$.
This is because PBHs of this mass have an Einstein radius projected into the observer plane that is small compared to either of these observer separations.
For observer baselines sufficiently deep into this ``widely separated observers'' regime, the picolensing volume becomes approximately independent of the observer separation baseline because, for most of the comoving distance between the observers and the source, the picolensing region is simply two disjoint and independent near-circles (with one approximately centered around each observer line of sight) with radii that depend on the Einstein radius and/or the source size, but not on the observer separation; see more detailed discussion in \appref[ces]{app:PointS_LargeRsep} and \ref{app:ExtendedS_LargeRsep}.

We also see in \figref{fig:sourceSize} that the ability of the AU baseline to probe below existing microlensing constraints out to higher masses ($M \sim 2\times 10^{-8}M_{\odot}$) is essentially unaffected by the aggressive, standard, or conservative source sizes.
This is because, for these larger masses, the majority of the sources behave as pointlike (i.e., $\theta_S \ll \theta_E$), even for $k_S = 10$.

Also evident from the lines labeled ``Scatter'' in \figref{fig:sourceSize} is that an intrinsic scatter of plus or minus an order of magnitude in GRB source sizes around the deterministic estimate $D_{S}^{\text{phys}} = T_{90}/(1+z_S)$ is not really an issue: the results are much more sensitive to a systematic shift of the whole GRB source-size distribution by a factor of 10 in either direction than they are to this scatter.
At small $M$, we see that these ``Scatter'' results are slightly stronger than those for $k_S = 1$. We attribute this to the somewhat larger number of small GRBs obtained in our single realization of the $k_S$ scatter, as compared to the $k_S=1$ results; cf.~\figref{fig:rShist}.  
As we investigate in more detail below, the smallest GRBs dominate the small-$M$ sensitivity (see also \appref{app:ExtendedS_LargeRsep}).
Note however that because the lines labeled ``Scatter'' are just a single realization of such scatter effects, one must be slightly cautious in making more general conclusions from them.
A study of the distributional nature of the projections arising from an ensemble of random draws of the $k_S$ values (with each draw as described in \secref{sec:GRBsizes}) would be computationally intensive and lies beyond the scope of this paper.

The AU-baseline results of \figref{fig:sourceSize} also demonstrate the analytical scalings derived in detail in \appref{app:sigmaGaussianAnalyt}:\\[1ex]
{\indent}(1) \emph{Pointlike sources, widely separated observers}---An effect that is most obvious in the $k_S = 0.1$ results is that the projected $f^{\text{limit}}_{\pbh}$ becomes independent of the PBH mass $M$ in the intermediate-mass regime.
This is the regime where the majority of sources are pointlike on the scale of the source-plane-projected Einstein radius (i.e., $\theta_S \ll \theta_E$), while at the same time the observers are widely separated on the scale of the observer-plane-projected Einstein radius (i.e., $R_O' \gg 2\chi^O_E$), at least for most values of $\chi_L/\chi_S$.
In this regime, the only angular scale that can be relevant for lensing is the Einstein angle, so the picolensing cross section must therefore scale as $\sigma \propto \chi_E^2 \propto M \Rightarrow \mathcal{V} \propto M$, which leads directly to an $M$-independent $f^{\text{limit}}_{\pbh}$ because $f^{\text{limit}}_{\pbh}\propto M/\overline{\mathcal{V}}$.
See \appref{app:PointS_LargeRsep} for more detail.\\[1ex]
{\indent}(2) \emph{Pointlike sources, closely spaced observers}---At large masses, the observer-plane-projected Einstein radius becomes larger than the observer separation for most values of $\chi_L/\chi_S$ for most sources (i.e., $R_O' \ll 2\chi^O_E$).
In this regime, the cross section becomes parametrically suppressed by a factor of $R_O'/\chi_E$ as compared to the case of widely separated observers: $\sigma \propto \chi_E R_O$; see \appref{app:PointS_SmallRsep} for a detailed demonstration. 
This leads to $\sigma \propto \sqrt{M} R_O \Rightarrow \mathcal{V} \propto \sqrt{M} R_O$, leading to the scaling $f^{\text{limit}}_{\pbh} \propto M / \overline{\mathcal{V}} \propto \sqrt{M} / R_O $ in the bound; see again \appref{app:PointS_SmallRsep} for more detail. 
The $\sqrt{M}$ scaling is evident directly in the L2 and AU panels of \figref{fig:sourceSize} (and to some extent also in the EM panel).
The $R_O^{-1}$ scaling is actually most easily observed by comparing the various projections shown in \figref{fig:baseline} for $M \gtrsim 10^{-9}M_{\odot}$ and using the baseline values from \tabref{tab:baselines}.\\[1ex]
{\indent}(3) \emph{Extended sources, widely separated observers}---At small masses (and long baselines $\sim\text{AU}$), it is apparent that the simultaneous transformation $D_S^{\mathrm{phys}} \rightarrow \kappa D_S^{\mathrm{phys}}$ and $M \rightarrow \kappa^2 M$ leaves $f^{\text{limit}}_{\pbh}$ unchanged; this is most clearly seen comparing the $k_S = 1$ results for the AU baseline at $M /M_{\odot} = 10^{-15}$, $10^{-14}$, and $10^{-13}$ with the cognate results for $k_S = 0.1$ at $M/M_{\odot} = 10^{-17}$, $10^{-16}$, and $10^{-15}$, respectively; and those for $k_S = 10$ at $M/M_{\odot} = 10^{-13}$, $10^{-12}$, and $10^{-11}$, respectively.
The origin of this scaling is more complicated, and is explained in detail in \appref{app:ExtendedS_LargeRsep}.
Heuristically, however, one can understand this result with reference to a few facts that hold in this regime: (a) for a lens located anywhere within the area in the lens plane that is subtended by the source angular size (as seen by each observer), the SNR for picolensing detection is the same, and moreover depends only on the size of the source compared to the Einstein angle [see \eqref{eq:thresholdSNRcase3}] (the SNR is also effectively zero elsewhere in the lens plane); (b) as a result, so long as the SNR is above threshold for the given parameters, the picolensing cross section in this regime scales proportionally with the source angular size squared: $\sigma(\chi_L) \propto ( \theta_S \chi_L )^2$ [cf.~\eqref{eq:sigmaCase3}].
Moreover, (c) the simultaneous rescaling under discussion here sends $\theta_S \rightarrow \kappa \theta_S$ and $(\theta_E \propto \sqrt{M}) \rightarrow \kappa \theta_E$, thus preserving the source angular size as compared to the Einstein radius.
Therefore, the picolensing cross section at any lens distance $\chi_L$ is simply rescaled by an overall factor of $\kappa^2$, but nothing else changes about the picolensing volume.
The upshot is that, if all GRBs are in this regime, the average picolensing volume scales as $\overline{\mathcal{V}} \rightarrow \kappa^2 \overline{\mathcal{V}}$.
However, the simultaneous rescaling of the factor of $M$ in the numerator of \eqref{eq:fDMlimit} compensates this rescaling of the picolensing volume, keeping the projected limit $f^{\text{limit}}_{\pbh}$ fixed.\\[1ex]
{\indent}Note in particular that the last result implies that the lowest mass PBH to which picolensing is a sensitive probe, $M_{\text{pico}}^{\text{min}}$, depends on the systematic-shift parameter $k_S$ governing GRB source sizes [cf.~\eqref{eq:chiT90}] as $M_{\text{pico}}^{\text{min}}\propto k_S^2$ (so long as one stays in the appropriate regime of source size and observer separation).

\begin{figure*}[!t]
    \centering
    \includegraphics[width=\textwidth]{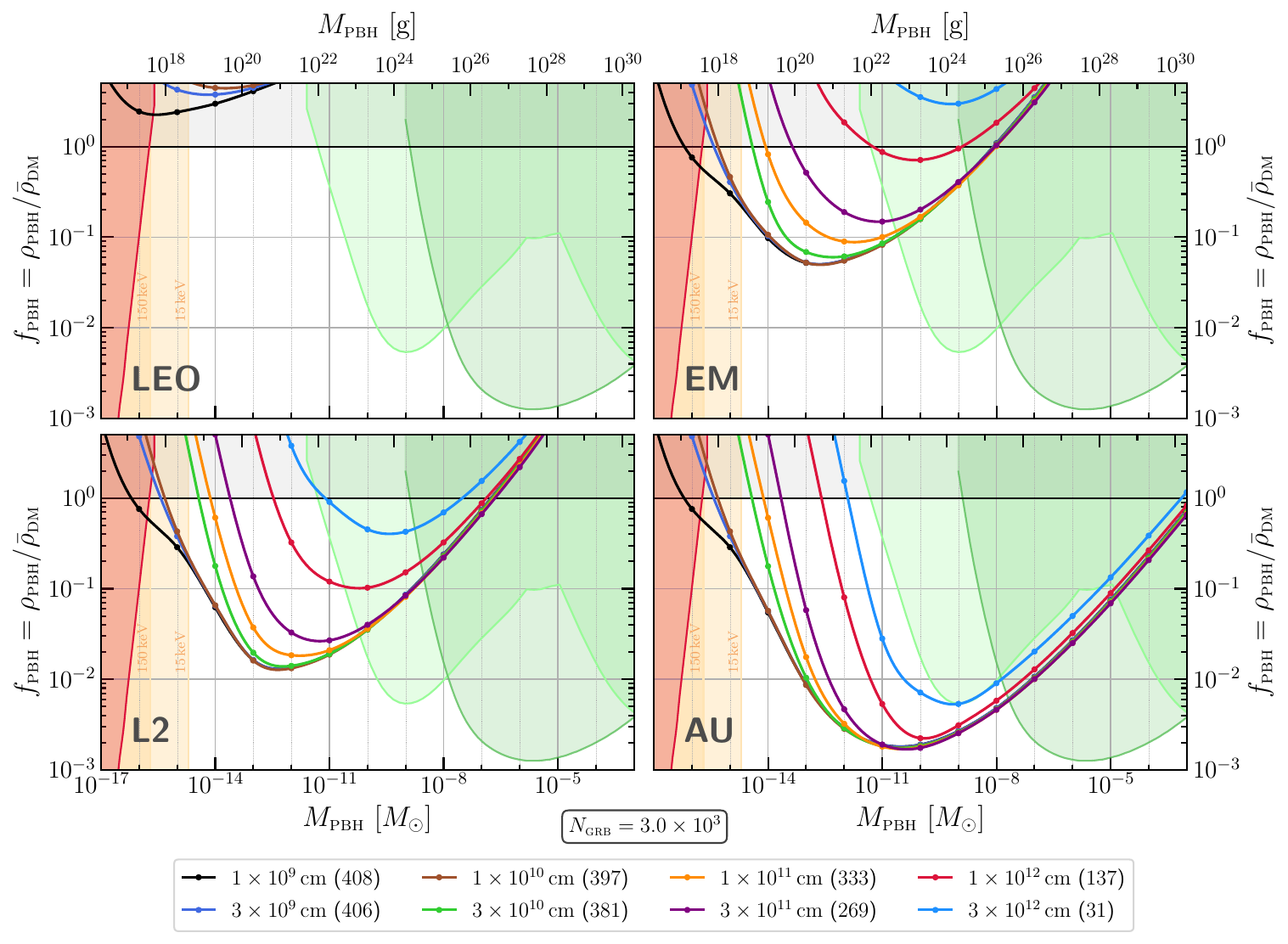}
    \caption{\label{fig:sizeExclusions}%
    Variation of results with a hard lower cutoff placed on the GRB size distribution for the various baselines $R_O$ annotated on each panel; see \tabref{tab:baselines}. 
    The shaded regions are as defined in \figref{fig:comp}.
    The marked points and lines are projected 95\%-confidence exclusions of the same nature as those detailed in \figref{fig:comp}.
    We adopt the parameters in \tabref{tab:params}, and assume here an observed population of $N_{\grb}=3000$ GRB sources.
    Results are shown for a Gaussian source profile with $k_S = 1$.
    However, we exclude from the GRB population obtained from the \emph{Swift}/BAT catalogue any GRBs that have assigned physical sizes $D_S^{\text{phys}}$ that are smaller than the values of $D_{\mathrm{cut}}$ given in the legend.
    For example, if the legend has the line marked as $1\times 10^{11}\,\text{cm}$, we compute the average comoving picolensing volume taking into account only those sources in the \emph{Swift}/BAT catalogue to which we have assigned $D_S^{\text{phys}}> D_{\mathrm{cut}} = 1\times 10^{11}\,\text{cm}$, assume that this truncated distribution is representative of the true GRB population, recompute the average picolensing volume for the truncated distribution, and then scale limits to $N_{\grb}=3000$ such GRBs.
    The number in brackets in the legend shows how many of the 409 \emph{Swift}/BAT GRBs on which our study in based (see \secref{sec:SWIFT-BAT}) survive the respective physical-size cuts.
    }
\end{figure*}

\hypertarget{text:annotationExplanationSourceSize}{%
The relevant mass ranges in which these different regimes apply are also displayed in \figref{fig:sourceSize}.%
} %
In each panel of that figure, there are two vertical dot-dashed black lines which indicate the PBH masses for which%
\footnote{\label{ftnt:recallSourceSize}%
   Recall from \secref{sec:extendedSource} that $\theta_S = D_S^{\text{phys}}/(2\mathcal{D}_S) = \Delta \chi_{S,\perp}/(2\chi_S)$, is the source angular half-size (i.e., the angle subtended by the radius of the source at the observer).
} %
$\theta_E = \theta_S(D_S^{\text{phys}})$ for either (a) $D_S^{\text{phys}}=R_{\odot}$ [vertical line at smaller mass], or (b) $D_S^{\text{phys}} = 16R_{\odot}$ [vertical line at larger mass], assuming a lens at $\chi_L/\chi_S = 0.5$ and a source at $z_S = 1$. 
The source size assumed for (b) is roughly the average value of $D_S^{\text{phys}}$ for $k_S=1$ (see \figref{fig:rShist}).
While this is very rough because lens distances are scanned over and sources are at different redshifts, the region labeled ``Extended sources'' to the left of the first line is where all sources with an observable transverse physical size larger than a solar radius are expected to be extended on the scale of the source-plane-projected Einstein radius, whereas the region labeled ``Pointlike sources'' to the right of second line is where the majority of sources (assuming $k_S = 1$) are pointlike on the scale of the source-plane-projected Einstein radius.

Also shown in \figref{fig:sourceSize} by the vertical dot-dashed coloured line (and marked with the carat on the lower mass axis) on each panel is the PBH mass for which%
\footnote{\label{ftnt:recallOPRE}%
    Recall from \secref[s]{sec:pointSource} and \ref{sec:picolensing} that $\chi^O_E = \theta_E \cdot \chi_L \chi_S / \chi_{LS}$ is the observer-plane-projected Einstein radius.
} %
$R_O = 2\chi^O_E$ for a lens with $\chi_L/\chi_S=0.5$ for $z_S =1$.
While this is again very rough because lens distances are scanned over, effective baselines $R_O' = R_O |\sin\theta_O|$ depend on $\theta_O$ which is averaged over, and sources are at different redshifts, the region to the left of this line marked ``Widely separated observers'' is where the observers are well separated on the scale of the observer-plane-projected Einstein radius, whereas the region to the right marked ``Closely spaced observers'' is the opposite regime.

We additionally study the sensitivity of the results to a hard lower cutoff on the minimum allowed source size; see \figref{fig:sizeExclusions}.
Here, we adopt the standard ($k_S = 1$) source sizes, but exclude all GRBs in the catalogue of 409 \emph{Swift}/BAT GRBs we have considered that have an assigned physical size $D_S^{\mathrm{phys}} \leq D_{\mathrm{cut}}$, for various values of $D_{\mathrm{cut}}$.
We then assume that the \emph{remaining} GRBs with $D_S^{\mathrm{phys}} > D_{\mathrm{cut}}$ are  representative of the true GRB source population, recompute $\overline{\mathcal{V}}$ with this truncated GRB catalogue, and use this recomputed average picolensing volume to make projections for $N_{\grb} = 3\times 10^3$ observed GRBs that are assumed to have the same parameter distributions as the truncated catalogue.

The purpose of this study is to understand how sensitive the results are to the smallest imputed GRBs sizes, in case the estimate $D_S^{\mathrm{phys}} = T_{90}/(1+z_S)$ sets these to be anomalously small compared to real physical GRBs.
As one can see in \figref{fig:sizeExclusions}, excluding the smallest sources has its largest impact at the smallest PBH masses. 
For all three of the EM, L2, and AU baselines, however, the overall qualitative picture of the results in most of the asteroid-mass window where geometrical optics is a reasonably good approximation is largely unaffected until $D_{\mathrm{cut}}$ exceeds $\sim(1$--$3)\times 10^{10}\,\text{cm}$.
Even the results with $D_{\mathrm{cut}} \sim 3\times 10^{11}\,\text{cm}$ differ significantly from those without the cut only for $M \lesssim 10^{-13}M_{\odot}$ for the L2 and AU baselines.
Moreover, the mass range over which $f_{\pbh}=1$ can be probed for these longer baselines is only marginally impacted.
We thus conclude that the results for the longer baselines (i.e., L2 and AU) for $M \gtrsim 10^{-13}M_{\odot}$ are not very sensitive to the exclusion of physical observed GRB source sizes smaller than $\sim R_{\odot} \sim 7\times 10^{10}\,\text{cm}$.
Furthermore, while the depth in $f_{\pbh}$ to which one can probe at smaller $M$ is significantly modified by the exclusion of observed physical GRB source sizes smaller than $R_{\odot}$, PBH DM making up 100\% of the DM (i.e., the line $f_{\pbh}=1$) could still nevertheless be probed using these larger baselines over most of the asteroid-mass window where geometrical optics is a good approximation.

\begin{figure}[t]
    \centering
    \includegraphics[width=\columnwidth]{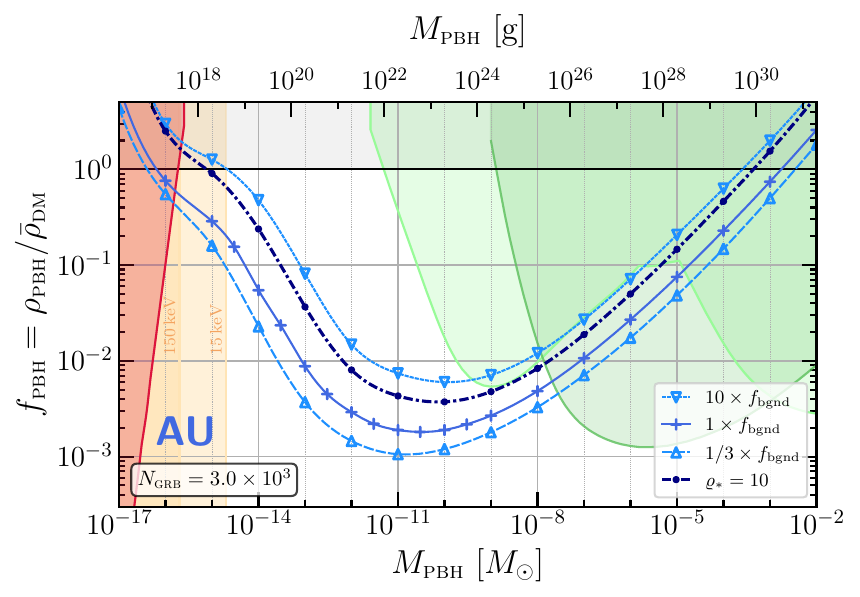}
    \caption{\label{fig:bgnd}%
    Variation of results with changing the uniform background level $F_B$ or the SNR threshold $\varrho_*$, for an AU baseline. 
    The shaded regions are as defined in \figref{fig:comp}.
    The marked points and lines are projected 95\%-confidence exclusions of the same nature as those detailed in \figref{fig:comp}.
    Results are shown for a Gaussian source profile with $k_S = 1$.
    We adopt the parameters in \tabref{tab:params} (except for $F_B$ or $\varrho_*$), and assume here an observed population of $N_{\grb}=3000$ GRB sources.
    The lines marked $``\# \times f_{\text{bgnd}}$'' where $\#$ is a number, show how the exclusion projections change as the background level is varied from the standard result ($F_B = 1\times f_{\text{bgnd}} = 10\,\text{cm}^{-2}\,\text{s}^{-1}$) either up by a factor of 10 to $F_B = 100\,\text{cm}^{-2}\,\text{s}^{-1}$, or down by a factor of $\sim 3$ to $F_B = 3 \,\text{cm}^{-2}\,\text{s}^{-1}$.
    The navy-blue line marked $\varrho_* = 10$ shows how projections change if the standard background level is fixed, but the SNR threshold for individual picolensing event detection is instead increased from $\varrho_* = 5$ to $\varrho_* = 10$.
    }
\end{figure}

Our final study considers the sensitivity of the results to changing either the constant detector background level or the picolensing SNR threshold $\varrho_*$, for the AU baseline; see \figref{fig:bgnd}.
Allowing the background count rate to vary down to $F_B = 3\,\text{cm}^{-2}\text{s}^{-1}$ ($``1/3 \times f_{\text{bgnd}}$'') improves reach marginally ($\propto \sqrt{F_B}$ at larger $M$, but $\propto F_B$ at smaller $M$).
On the other hand, increasing the background by an order of magnitude to $F_B = 100\,\text{cm}^{-2}\text{s}^{-1}$ ($``10 \times f_{\text{bgnd}}$'') degrades the sensitivity by a factor of 3--10 (again, $\propto \sqrt{F_B}$ at larger $M$, but $\propto F_B$ at smaller $M$).
Qualitatively, the main point of concern here is that if \emph{Swift}/BAT-class detectors located in orbits that allow for AU baselines were to necessarily experience much higher background rates than are experienced over most of the time in low Earth orbit (at least outside the South Atlantic Anomaly), this may impede the ability of the AU baseline to probe the region of new parameter space below existing microlensing constraints, roughly $3\times 10^{-10} M_{\odot}\lesssim M \lesssim 2\times 10^{-8}M_{\odot}$.
On the other hand, given that the exact specifications of the instruments needed to make this measurement would be subject to a future design process, these results can also be viewed as specifying a design target for background suppression or rejection.

Finally, if we were to increase the individual picolensing detection threshold SNR from $\varrho_* = 5$ to $\varrho_* = 10$, the projected reach would also be marginally degraded, more significantly at small $M$ as compared to larger $M$, with similar qualitative impacts as for the higher-background scenario.
One way to view this study of setting a higher statistical-only SNR is as a way to mock up a  projection for the case where systematic uncertainties on the picolensing signal are taken to be comparable to the Poisson photon-counting statistical uncertainties.

\section{Beyond Constraints}
\label{sec:beyondConstraints}
Optimistically, beyond simply constraining parameter space, it is also worthwhile to consider what inferences one might be able to draw from a positive picolensing detection.

The question that would arise from a confirmed differential magnification detection is: can one identify the origin of the signal?
There are really two separate questions: (1) are there Standard Model foregrounds that could give rise to a differential magnification that mocks up the DM signal? and (2) if only DM can explain the signal, can one unambiguously identify the kind of DM that has been detected, as well as its properties?

There are two ways that SM matter could give rise to a differential brightness signal: (a) it could act as a lens in the same manner as a PBH would, provided that the matter lies well within its own Einstein radius; or (b) it could fully or partially obscure the source along one line of sight.

Regarding (a): only asteroid-mass SM objects significantly outside our own Solar System could act as relevant lenses: e.g., a $3\,\text{g/cm}^3$ object with mass $10^{-12}M_{\odot}$ would need to be a distance of at least $16\,\text{pc}$ to lie within its own Einstein radius.%
\footnote{\label{ftnt:withRE}%
    This estimate comes from demanding that $4\pi\rho_{\text{obj}} R_E^3/3 \gtrsim M$, where $\rho_{\text{obj}}$ is the object's assumed mass density, leading to \[\mathcal{D}_L \gtrsim \lb(3/(4\pi\rho_{\text{obj}})\rb)^{2/3}\Big/ \lb(4G_{\text{N}}M^{1/3}\rb)\:,\] in the limit where $\mathcal{D}_L \ll \mathcal{D}_S$.
} %
Let us discuss separately two populations of asteroids that could create a lensing foreground: (i) asteroids outside the Milky Way (MW) and (ii) asteroids inside the MW.%
\footnote{\label{ftnt:OGLEplanets}%
    We note that extrasolar \emph{planets} in the MW have been detected via microlensing of background stars (see, e.g., \citeR{2023arXiv231007502M}, and also \citeR{Bond:2004qd} for the first such detection by OGLE), some of which are free-floating/rogue planets (see, e.g., \citeR[s]{2023arXiv231007502M,Mr_z_2018,Mr_z_2020a,Mr_z_2020b,Ryu_2021}).
    However, these detected planets are in the sub-Earth to super-Jupiter mass classes (i.e., far above the asteroid-mass window), and they are lensing extremely numerous background stars (cf.~$N_{\textsc{grb}} \sim 3\times 10^3$ in our projections).
    We also note that these planets were detected in fields imaged toward the MW galactic bulge, and are thus either in the bulge or in the plane of the disk on the line of sight to the bulge~\cite{Mr_z_2018,Ryu_2021}, which are regions of high baryon abundance.} %

We need an estimate of the fraction of baryonic mass in such asteroids.
We grant that our Solar System is an exceptional location, both as compared to a random location in the MW and even more so as compared to a random location in the Universe, but here we will use our Solar System as a guide to estimate this fraction.
In the Solar System, the total mass of asteroids in the Main and Kuiper belts is on the order of $ 10^{-7}M_{\odot}$~\cite{Pitjeva_2018}.%
\footnote{%
    The Oort cloud may be even more massive, by some $\sim 2$ orders of magnitude~\cite{1983A&A...118...90W}, but it is thought to consist mostly of much lighter, comet-mass objects (these properties are also quite uncertain).
    We omit consideration of it here, but our qualitative conclusions would be unaltered were we to include it.
} %
This mass distribution is dominated by large asteroids and/or dwarf planets: e.g., Ceres \cite{NASAasteroids}, which has a mass $\sim 9.4\times 10^{20}\,\text{kg} \sim 5\times 10^{-10}M_{\odot}$~\cite{PARK2019812}, constitutes roughly 40\% of the total $\sim 1.3\times 10^{-9} M_{\odot}$ mass of the Main Belt~\cite{Pitjeva_2018,NASAasteroids}; similarly, $\sim40\%$ of the (more massive) Kuiper Belt mass is in the 31 largest trans-Neptunian belt objects~\cite{Pitjeva_2018}. 
Therefore, less than $10^{-7}$ of the Solar System mass is in asteroids of masses $10^{-11} M_{\odot} \lesssim M \lesssim 10^{-10} M_{\odot}$ for which our picolensing projections for an AU baseline are most sensitive ($f_{\textsc{dm}} \sim 10^{-3}$).
However, to be conservative we will take $10^{-7}$ as our estimate of the fraction of mass in relevant-mass asteroids as compared to the mass of all baryons \emph{in stars} (the mass of the Solar System is dominated by the Sun).%
\footnote{\label{ftnt:extrapolatePowerLaw}%
    An alternative estimate based on the power-law free-floating planet mass distribution \emph{in the MW} given in Eq.~(8) of \citeR{DeRocco:2023gde} can be advanced, albeit with significant uncertainties.
    That mass distribution is based on microlensing events of terrestrial or larger planet-sized objects.
    If we na\"ively extrapolate it (far beyond its intended validity) over several orders of magnitude in mass to ask what fraction of solar mass we might expect per star in free-floating asteroids in the $10^{-11}M_{\odot} \lesssim M \lesssim 10^{-10}M_{\odot}$ mass range, we would obtain estimates that are anywhere from $\sim 10^{-7}$ to $\sim 3\times 10^{-3}$.
    This range of values reflects the large uncertainties in the extrapolation that come just from the variations in the parameters in the assumed power-law mass distribution, before even accounting for uncertainties in the shape of the (unmeasured) low-mass distribution.
    These extrapolated values lie above the estimate in the main text based on the Solar System asteroid population.
    In particular, the upper value would imply that the low-mass asteroids constitute $\sim 0.3\%$ of the total MW stellar mass, which we believe to be unlikely.} %
We then assume (likely with quite large error) that this same fractional abundance holds, on average, everywhere.
Furthermore, stars form a fraction of $\mathcal{O}(10^{-1})$ of all baryonic matter~\cite{Fukugita:1997bi,2004IAUS..220..227F}, so we will take an estimate for the fraction of all baryonic mass in asteroids of the relevant mass range to be $\sim 10^{-8}$.

It is now easy to see that (i) cannot be a relevant foreground: on average over cosmological scales, the DM mass density dominates that of baryons by a factor of $\sim 5$, so the fraction of mass density in relevant-mass asteroids as compared to the average DM mass density can then be estimated as $\lesssim (\text{few}) \times 10^{-9}$.
But our best projected sensitivity is to compact lenses that constitute at least $f_{\textsc{dm}} \sim 10^{-3}$ of the DM mass density.
Therefore, unless our na\"ive estimate of the baryonic mass fraction in relevant-mass asteroids is off by 5 to 6 orders of magnitude, we do not expect to have any sensitivity to SM asteroids randomly scattered outside the MW.%
\footnote{\label{ftnt:altEst}%
    Note that even under the largest alternative estimate discussed in footnote \ref{ftnt:extrapolatePowerLaw}, this foreground would still be too small to be observable, although at the upper end of that estimate it is interestingly close to being observable.} %

To estimate (ii), we note that, in the Solar neighbourhood, the baryonic mass density is a factor of 5--10 larger than the DM mass density~\cite{2015ApJ...814...13M,2019A&A...623A..30W}.
So our estimate of the $\sim 10^{-8}$ fraction of baryonic mass in relevant-mass asteroids then allows us to conservatively estimate that the mass-density in relevant-mass asteroids as compared to the DM mass-density is $\lesssim 10^{-7}$, at least locally in the MW.
Moreover, we also know that even given the large overdensity of PBH DM lenses within the MW, they contribute sub-dominantly to the total picolensing signal as compared to the PBH DM lenses at cosmological distances.%
\footnote{\label{ftnt:localLenses}%
    The line-of-sight-integrated surface density of the PBH DM lenses in the MW (in any direction other than toward the GC) as compared to the integral over the remainder of the line of sight from just outside the MW to a cosmologically distant GRB is similar: $10^5 f_{\textsc{dm}} \bar{\rho}_{\textsc{dm}} \times (10\,\text{kpc}) \sim f_{\textsc{dm}} \bar{\rho}_{\textsc{dm}} \times (\text{Gpc})$, where we took the line of sight in the MW DM overdensity ($\sim 10^5$ as compared to the average DM density $\bar{\rho}$ in the Universe) to be on the order of $\sim 10\,\text{kpc}$ (correct at the order-of-magnitude level assuming a Navarro--Frenk--White [NFW] profile for the MW), and the distance to the GRB to be $\sim \text{Gpc}$.
    But the picolensing cross section in the relevant peak-sensitivity mass range typically scales as $\sigma \sim \chi_E^2 \propto \chi_L \ll \chi_S$ for a cosmologically distant source (see \appref{app:PointS_LargeRsep}), and so the contribution of the MW overdensity of lenses to the total picolensing volume is very small, fractionally of order $10\,\text{kpc}/\text{Gpc} \sim 10^{-5}$.} %
Taken together, these observations again point to the fact that any picolensing signals from asteroids---at least at their estimated local density in the MW---should be many orders of magnitude smaller than any potential PBH signal to which we project any sensitivity.
Of course, the baryon mass density near the MW Galactic Center (GC) is $\sim 2\times 10^3$ larger than the local DM abundance (the GC is also highly baryonically dominated)~\cite{10.1111/j.1365-2966.2011.18564.x}.
As such, there is a relatively larger potential asteroid foreground contribution for GRBs that lie toward the GC; however, it still appears to be an unobservably small foreground, unless the fractional asteroid abundance as compared to all baryons happens to be much larger in the GC than our Solar-System-based estimate.%
\footnote{\label{ftnt:altEst2}%
    Because of the suppression discussed in footnote~\ref{ftnt:localLenses}, not even the largest alternative estimate discussed in footnote \ref{ftnt:extrapolatePowerLaw} would make (ii) an observable foreground.} %

Regarding (b): the x-/$\gamma$-radiation from a GRB is penetrating, so relevant obscuration of a distant source likely requires a solid-density object, as opposed to a diffuse gas/dust cloud. 
Here, a key point to realize is that even if overall source attenuation is possible with diffuse gas or dust clouds, achieving a \emph{differential} obscuration across the relevant tiny angular scales in order to give rise to a foreground for picolensing would be very hard.
Various-sized solid objects could however fully (or partially) obscure GRB sources, especially extended sources, depending on where they lie on the line of sight. 
Something like a star or large planet in a distant galaxy, a random asteroid in the Milky Way, or a very small rocky body in our own Solar System could have the requisite angular size.%
\footnote{\label{ftnt:hockeyPuck}%
    A GRB that is on order of a picoarcsec in size [cf.~\figref{fig:rShist}] could in principle be completely obscured by an object the size of a hockey puck at a distance of $\sim 10^5\,\text{AU}$, which is somewhere in the outskirts of the Oort Cloud.
} %
The probability for a line of sight to accidentally intersect something like a star in a distant galaxy is however very small.
The exact probability for a line of sight to intersect a smaller body like an asteroid somewhere else in the MW that happens to lie on the line of sight, or a small rocky body in the Solar System, is a detailed question of the mass/size distribution and number of such objects that exist in various structures, so we cannot give definitive estimates.
Nevertheless, basic estimates based on conservative assumptions about how much mass is likely to be found in such objects as compared to the mass in compact DM lenses that constitute at least $10^{-3}$ of the average DM mass density, again indicate that such SM obscuration should be rarer than a DM-induced picolensing event.

We do however note that our estimates here are somewhat na\"ive, and that these foregrounds implicate planetary astrophysics beyond the scope of this paper; better estimates for their rates necessarily depend on the unknown and unmeasured size and mass distributions of very low mass objects in the MW.
So, despite our estimates here, it is conceivable that some such foreground event(s) might be detectable, which would be of interest in its own right.
That said, GRBs are essentially isotropically distributed on the sky, whereas any foregrounds associated with asteroids or other objects in the MW likely have a strong directional dependence (i.e., larger near the GC and smaller in other directions); likewise for objects in our Solar System (i.e., larger in the plane of the ecliptic, and smaller in other directions).
Given a large number of picolensing events, spatial information could thus possibly be used as a handle to differentiate any such foreground events from possible PBH-related events (which are dominated by the PBHs outside the MW).

Since our estimates indicate that the foregrounds are negligibly small, a confirmed differential magnification signal detection (or, even more so, a collection of such positive detections) would strongly point toward a DM origin.
We would then be in a position to ask whether we could also say anything more about the identity of the DM.
In this regard, note again that lensing is not a signature unique to a black hole: any object that is contained well within its own Einstein radius (say, for instance, its physical radius is $\mathcal{O}(10\%)$ of its Einstein radius) would give a lensing signal very similar to that of a pointlike PBH.
For instance, massive but compact composite DM states like blobs or nuggets (see, e.g., \citeR[s]{Wise:2014jva,Wise:2014ola,Hardy:2014mqa,Hardy:2015boa,Gresham:2017zqi,Gresham:2017cvl,Gresham:2018anj,Bai:2018dxf,Coskuner:2018are,Grabowska:2018lnd,Hong:2020est,Diamond:2021dth,Hall:2016usm,Baum:2022duc}) may be able to meet this criterion.
As such, it may be challenging to distinguish a picolensing signal arising from a PBH from one arising from such an alternative compact DM state, unless a very large collection of detections were made and subtle differences in the population of lensing events could be teased out.
It may however be possible to distinguish the picolensing signals of more diffuse structures (e.g., those with a size comparable to, or even larger than, their Einstein radius) and we intend to return to this point in future work; cf.~\citeR[s]{Croon:2020wpr,Croon:2020ouk,DeRocco:2023hij,CrispimRomao:2024nbr} in the microlensing context.

One might further wonder about the prospects for parameter estimation beyond simply identifying the nature of the DM state:%
\footnote{\label{ftnt:thanksWill}%
    The authors thank William DeRocco for a highly stimulating discussion on this and related points.
} %
e.g., what is the mass of the lens $M$?
Here, one would need to face up to various parameter degeneracies that may exist (in various parameter regimes) between, e.g., the lens mass, the lens distance, the lens location in the lens plane, and the source size.
Particularly where finite GRB sizes are relevant (i.e., at low $M$), the large uncertainties on individual GRB sizes would also likely be a confounding factor in such an analysis.
On the other hand, when GRB sizes are relatively unimportant, but observer-separation baselines are relevantly large (i.e., at large $M$), exploiting multiple observation baselines to study the lensing parallax dependence on the separation distance between the observers, may be a handle (albeit a monetarily costly one) that could be exploited.

Using temporal evolution of the picolensing event to extract information or break parameter degeneracies na\"ively seems harder than it is in the context of microlensing (see, e.g., \citeR{Johnson:2021jib} and references therein).
However, especially for longer GRBs there is in principle some hope that the lensing signal could evolve nontrivially over the duration of the burst~\cite{Jung:2019fcs}: e.g., over a $T_{90}\sim 100\,\text{s}$ burst, a dark object moving at $v\sim 10^{-3}c$ (roughly the average DM speed in our galaxy, which we take here merely as an indicative speed) would traverse a distance of $\sim 5\times 10^{-2} R_{\odot}$. 
Comparing to the lens-plane-projected Einstein radii in \tabref{tab:REproj}, we see that a distant lens with a mass $M \sim 10^{-12}M_{\odot}$ might then move $\mathcal{O}(10\%)$ of its own Einstein radius during the GRB event duration, which may give rise to some time-evolution of the lensing signal.%
\footnote{\label{ftnt:LensInMW}%
    The corresponding timescale for a close lens (e.g., one within the MW) to move across its own Einstein radius would be shorter by a factor of $\sim \mathcal{O}(\sqrt{10\,\text{kpc}/\text{Gpc}})$.
    Timescales for significant evolution of the picolensing signal for lenses of this mass are then $\sim$ few seconds, which is shorter than many GRB durations.
    However, such lenses contribute negligibly to our projected sensitivity; cf.~footnote~\ref{ftnt:localLenses}.} %
The movement on the scale of the Einstein radius would be even larger for smaller lenses: for instance, a $M\sim 10^{-14}M_{\odot}$ lens might move by an amount on the order of its own Einstein radius during a particularly long GRB event.
Note however that this low-mass regime is also where most sources (and the overwhelming majority of the long-GRB sources) behave as extended objects on the scale of the lens Einstein radii (at least for $k_S=1$).
It is therefore not immediately obvious that motion of the lens that is sizeable on the scale of its own Einstein radius, but still small on the scale of the source size projected into the lens plane, gives rise to a significant change to the magnification of the source.%
\footnote{\label{ftnt:staticLens}%
    Indeed, our entire analysis is undertaken on the assumption that the lens can always be treated as having a fixed, static location throughout the duration of the GRB event.
} %
Nevertheless, there might be some hope that studying the temporal dependence of many detected picolensing events of long GRBs might give additional handles here.
A dedicated study of parameter estimation in the context of a positive picolensing detection is ultimately beyond the scope of this paper, but would be an interesting topic to investigate. 

Finally, let us make a comment regarding larger optical-depth scenarios.
Suppose that one made a positive detection at an inferred value of $f_{\pbh} \sim 1$ in the asteroid-mass window such that $\bar{\tau} \sim 1$ (or, in the alternative, that a non-negligible tail of the source distribution behaves in such a way as to give $\tau_j>1$, even if the population average optical depth is still below unity). 
While such a scenario is perhaps unlikely [cf.~the discussion in \secref{sec:probAndExclusion} around \eqref{eq:opticalDepth2}], in this hypothetical regime we would have the expectation that statistical fluctuations in the number of lenses located within the picolensing volume would be an additional handle that, e.g., might help break parameter degeneracies.
For instance, with $\bar{\tau} \sim 1$, $\sim37$\% of average picolensing volumes would see no lensing, while $\sim37$\% would see a single lens in the picolensing volume and would therefore look like the standard single-lensing picolensing case (i.e., one line of sight is lensed).
However, the remaining $\sim 26$\% would see 2 or more lenses in the picolensing volume.
For the 2-lenses case (which would be $\sim18$\% of the overall cases), we would expect a roughly 50-50 split between having one lens close to each line of sight (which suppresses the differential-intensity signal) and having two lenses close to a single line of sight (which likely enhances the signal, or gives it measurably different properties compared to single lensing).
By studying how a large number of differential-magnification events differ in their morphology, additional information could thus be gleaned in this scenario.

\section{Conclusion}
\label{sec:Conclusion}
In this paper, we have revisited picolensing of cosmologically distant gamma-ray bursts as an observable to probe the asteroid-mass window for primordial black hole dark matter.
We have confirmed the main finding of \citeR[s]{Jung:2019fcs,Gawade:2023gmt}: this observable is a viable way to probe this window.
However, we also showed that uncertainties in GRB sources sizes can significantly degrade the potential reach for this observable, with the result that somewhat more demanding mission parameters are required to be able to robustly explore the window.

More specifically, under conservative assumptions about the distribution of the observed angular sizes of GRBs on the sky (i.e., our $k_S = 10$ results), we showed that a future mission that deploys two space-based\linebreak x-/$\gamma$-ray detectors in the class of at least the \emph{Swift}/BAT instrument must utilize inter-spacecraft separations \emph{somewhat larger} than the Earth--Moon distance in order to be able to robustly exclude PBHs as all of the DM at 95\% confidence in a significant fraction of the PBH asteroid-mass window, if no picolensing events are detected at an SNR of 5 or above having observed 3000 GRBs; see \figref{fig:sourceSize}.
For more aggressive assumptions about the distribution of the observed GRB angular sizes (e.g., our $k_S = 0.1$ or 1 results), an Earth--Moon inter-spacecraft separation baseline would suffice, but that conclusion is however not robust to GRB size uncertainties; see \figref[s]{fig:baseline}--\ref{fig:sizeExclusions}.
We also showed that scatter in the GRB size distribution is less of a concern than a systematic shift in the whole GRB source-size distribution; see \figref{fig:sourceSize}.

We showed that a baseline on the order of the Earth--Lagrange-Point-2 distance (roughly $4$ times the Earth--Moon distance) can probe the PBH asteroid-mass window at $f_{\pbh} = 1$ for $M \gtrsim 10^{-14}M_{\odot}$ even under the most conservative GRB size assumptions we have made; see \figref{fig:sourceSize}. 
An AU-length baseline would also probe the window above $M \gtrsim 10^{-14}M_{\odot}$, and does so to much smaller subcomponent (fractional) DM abundances; moreover, its reach at higher masses would be impressive, surpassing existing microlensing constraints to reach new parameter space up to $M\sim 2\times 10^{-8}M_{\odot}$.
Overall, it would probe below $f_{\pbh}=1$ over some 11 orders of magnitude in mass, $10^{-14}M_{\odot} \lesssim M \lesssim 10^{-3} M_{\odot}$; see \figref{fig:sourceSize}.
These conclusions for either the L2 or AU baselines are also separately robust to variations in the shape of the fiducial GRB size distribution that arise from, e.g, restricting the latter to physical sizes $D_S^{\text{phys}} \gtrsim R_{\odot}$; see \figref{fig:sizeExclusions}.

We also showed that our results are robust to whether one takes the GRB intensity profile on the sky to be a flat disk, or a Gaussian profile that contains 90\% of its intensity within the same radius as the disk; see \figref{fig:baseline}.
For an AU baseline, a factor-of-10 larger background level than that achieved with, e.g., the \emph{Swift}/BAT instrument would begin to inhibit the ability to probe below the microlensing constraints for $M\gtrsim 3\times 10^{-10}M_{\odot}$, but would not qualitatively inhibit its ability to access the majority of the asteroid mass-window at $f_{\pbh}=1$; moreover, it would only mildly degrade the overall mass range over which such a mission could probe PBH DM (see \figref{fig:bgnd}).
Similarly, raising the SNR threshold for individual picolensing detection from 5 to 10 would degrade the ability of this probe to reach new parameter space only for $M\gtrsim 3\times 10^{-10}M_{\odot}$ for AU baselines; see \figref{fig:bgnd}.

The reach projections we have given in this paper are not restricted only to PBH DM: they apply equally well to any asteroid-mass dark objects that are sufficiently compact.
That is, for dark objects whose spatial extent fits within some $\mathcal{O}(10\%)$ fraction of their physical Einstein radius, the picolensing signal should not be significantly altered.
As such, our constraints may also apply to some classes of large composite DM states (e.g., blobs, nuggets, etc.) in the appropriate mass range.

Overall, we believe that our study, taken together with \citeR[s]{Jung:2019fcs,Gawade:2023gmt} and older work~\cite{Nemiroff:1995ak,Kolb:1995bu,Marani:1998sh,Nemiroff:1998zi}, strongly motivates further exploration of the technical feasibility of a picolensing mission to probe for PBH dark matter, either as a stand-alone mission, or via the incorporation of additional instruments on other missions.
Specifically, such a mission would provide access to the asteroid-mass window, which has to-date stubbornly evaded any robust probes.
In parallel, it may also be useful to investigate in more detail the feasibility of reprocessing the large volume of existing GRB data from the existing long-baseline IPN detector network to perform a search for a picolensing signal.

We intend to return in future work to the question of whether this picolensing observable would have interesting implications for probing more diffuse lensing structures, such as axion stars or axion miniclusters (cf., e.g., \citeR[s]{Croon:2020wpr,Croon:2020ouk,DeRocco:2023hij,CrispimRomao:2024nbr} in the context of microlensing).

Finally, we hope our work also further motivates in-depth study of GRBs themselves, both observationally and from a modeling/theoretical perspective, with a view to reducing the currently large uncertainties in their angular sizes; this would allow the refinement of projections for a future picolensing mission, and could even show the situation to be more favourable than our conservative assumptions suggest.

\subsection*{Note Added}
\label{noteAdded}
After the appearance of the \texttt{arXiv v1} preprint of this paper, we became aware of recent OGLE microlensing constraints from high-cadence observations of the Magellanic Clouds~\cite{Mroz:2024wia}, which appeared while we were in the concluding stages of our work.
These results improve constraints on planetary-mass PBH DM above the asteroid-mass window, exceeding previous microlensing limits for $10^{-8}M_{\odot} \lesssim M_{\pbh}\lesssim 10^{-3}M_{\odot}$, thereby closing off some of the parameter space \emph{above} the asteroid-mass window that a future picolensing mission using AU-scale baselines could reach.
We have updated the relevant figures and discussions to reflect these new constraints.
We still believe that the projected reach for an AU-scale picolensing mission is interesting above the asteroid-mass window as it would probe extragalactic lenses, in contrast to the galactic ones probed by microlensing observations.

\acknowledgments
We thank William DeRocco, Junwu Huang, Andrey Katz, Jamie Robinson, Aaron Tohuvavohu, and Wei Xue for useful discussions.
We thank Przemek Mr{\'o}z for bringing the recent OGLE high-cadence results~\cite{Mroz:2024wia} to our attention.

High-performance computing for this research was undertaken on the \textsc{Symmetry} cluster at Perimeter Institute~(PI).
We thank the research computing staff at PI for their support.

Research at PI is supported in part by the Government of Canada through the Department of Innovation, Science and Economic Development and by the Province of Ontario through the Ministry of Colleges and Universities.
The work of M.A.F.~was performed in part at the Aspen Center for Physics, supported by National Science Foundation (NSF) Grant No.~PHY-2210452.
The work of S.S.~is supported by the Natural Sciences and Engineering Research Council (NSERC) of Canada.

\appendix
\section{Lensing Potential}
\label{app:lensingPot}
Further to the discussion in \secref{sec:pointSource}, note that the lensing potential defined at \eqref{eq:lensingPotential} satisfies~\cite{Bartelmann:2010fz} 
\begingroup
\allowdisplaybreaks
\begin{align}
  &\vec{\nabla}^2 \psi \nonumber \\* &= 2 \int_0^{\chi_S} \frac{\chi\cdot(\chi_S-\chi)}{\chi_S} \bm{\nabla}_{\perp}^2 \phi( \bm{\chi}_\perp = \chi \vec{\theta} , \chi^3 = \chi )d\chi \\
  &= 2 \int_0^{\chi_S} \frac{\chi\cdot(\chi_S-\chi)}{\chi_S} \bm{\nabla}^2 \phi( \bm{\chi}_\perp = \chi \vec{\theta} , \chi^3 = \chi )d\chi,\label{eq:line2} \\
  &= 2 \int_0^{\chi_S} \frac{[a(\chi)]^2 \chi\cdot(\chi_S-\chi)}{\chi_S} 
    \nl \qquad\qquad 
  \times \bm{\nabla}_{\text{phys}}^2 \phi( \bm{\chi}_\perp = \chi \vec{\theta} , \chi^3 = \chi )d\chi, \label{eq:line3}\\[2ex]
  &= 8\pi G_{\text{N}} \int_0^{\chi_S} \frac{[a(\chi)]^2 \chi\cdot(\chi_S-\chi)}{\chi_S}\nl\qquad\qquad\times \rho( \bm{\chi}_\perp = \chi \vec{\theta} , \chi^3 = \chi )d\chi, \label{eq:line4} \\[2ex] 
  &= 8\pi G_{\text{N}} \int_0^{\chi_S} \frac{\mathcal{D}_\chi \cdot \mathcal{D}_{\chi S}}{\mathcal{D}_S} \rho( \bm{\chi}_\perp = \chi \vec{\theta} , \chi^3 = \chi ) a(\chi) d\chi\:,\label{eq:line5}
\end{align}
\endgroup
where
\begin{itemize}
    \item at \eqref{eq:line2}, we extended $\bm{\nabla}_\perp^2 \rightarrow \bm{\nabla}^2$ using that the $\chi$-integral turns the difference into a term that is negligible if the Newtonian potential is well localized between the source and observer (thin-lens approximation)~\cite{Bartelmann:2010fz};
    \item at \eqref{eq:line3}, we used $\bm{\nabla} = a(\chi) \bm{\nabla}_{\text{phys}}$; 
    \item at \eqref{eq:line4}, we used Einstein's equations [to linear order in $\phi$], $\bm{\nabla}_{\text{phys}}^2 \phi \approx 4 \pi G_{\text{N}} \rho(\bm{\chi})$; and 
    \item at \eqref{eq:line5}, we defined the angular diameter distances $\mathcal{D}_\chi \equiv a(\chi) \chi = \chi/[1+z(\chi)]$, $\mathcal{D}_{\chi S} \equiv a_S ( \chi_S - \chi) = ( \chi_S - \chi)/(1+z_S) $ and $\mathcal{D}_S \equiv a_S \chi_S = \chi_S/(1+z_S)$.
\end{itemize} 

A point mass $M$ located at $\bm{\chi}_{\perp}=0$ and $\chi = \chi_L$ has a physical mass density $\rho = M \delta^{(2)}( \bm{\chi}_{\perp,\, \text{phys}} )\delta( \chi_{\text{phys}}^3 - a(\chi_L) \chi_L ) = [M/a(\chi_L)] \delta^{(2)}( \bm{\chi}_{\perp,\, \text{phys}} )\delta( \chi - \chi_L )$, so 
\begingroup
\allowdisplaybreaks
\begin{align}
    \vec{\nabla}^2 \psi &= 8\pi G_{\text{N}}M  \frac{\mathcal{D}_L  \mathcal{D}_{LS}}{\mathcal{D}_S} \delta^{(2)}( a(\chi_L)\chi_L \vec{\theta}\, )\\
    &= 8\pi G_{\text{N}} M \frac{\mathcal{D}_L  \mathcal{D}_{LS}}{\mathcal{D}_S} \delta^{(2)}( \mathcal{D}_L \vec{\theta}\, )\\
    &= 8\pi G_{\text{N}} M \frac{\mathcal{D}_{LS}}{\mathcal{D}_L \mathcal{D}_S} \delta^{(2)}( \vec{\theta}\, )\:,
\end{align}
\endgroup
where we used that $\bm{\chi}_{\perp,\, \text{phys}} = a(\chi_L) \chi_L \vec{\theta}$.
But recall that $\vec{\nabla}^2 f = \delta^{(2)}( \vec{\theta} ) \Rightarrow f(\theta) = \ln(\theta) / (2\pi)$, so 
\begin{align}
    \psi &= 4 G_{\text{N}} M \frac{\mathcal{D}_{LS}}{\mathcal{D}_L \mathcal{D}_S} \ln(\theta)\: .
\end{align}
This implies that
\begin{align}
    \vec{\nabla} \psi &= 4 G_{\text{N}} M \frac{\mathcal{D}_{LS}}{\mathcal{D}_L \mathcal{D}_S} \frac{\hat{\theta}}{\theta} = \frac{\theta_E^2}{\theta} \hat{\theta}\: ,
\end{align}
as in the main text; cf.~\eqref{eq:compToAppendixExpr}.

\begin{widetext}
\section{Magnification for extended sources}
\label{app:Gaussian}
In this appendix, we first provide some more details about the computation of the magnification of a disk source intensity profile, the cognates of which were discussed in \secref{eq:GaussianCase} for the Gaussian profile.
We then provide analytical approximations for the magnification for the Gaussian profile; for similar expressions for the disk profile, see \citeR[s]{1994ApJ...430..505W,Jung:2019fcs,Gawade:2023gmt}.

\subsection{Disk Source Integral}
\label{app:muIntegralDisk}
In this subsection, we perform the radial integral in \eqref{eq:muBar} for the disk source intensity profile \eqref{eq:IDisk}.
Since $I_0 = 1$, we must evaluate
\begin{align}
    \bar{\mu}(\bm{\chi}_L) &= \frac{1}{\pi \delta^2 } \int_0^{2\pi} d\Delta\varphi \int_0^\infty d\Xi_S \cdot \Theta\Big[ \delta^2 - \lb( \Xi_L^2 + \Xi_S^2 + 2\Xi_S\Xi_L \cos \Delta\varphi \rb)  \Big] \frac{2+\Xi_S^2}{\sqrt{4+\Xi_S^2}}\:,
\end{align}
\end{widetext}
where we shifted the angular variable $\Delta \varphi \equiv \varphi_S - \varphi_L$ as compared to \eqref[s]{eq:muBar} and (\ref{eq:IDisk}), and used the periodicity of the cosine to shift the integration domain back to $\Delta \varphi\in[0,2\pi)$.
The domain of support of the $\Xi_S$ integral here is of course not actually semi-infinite because the Heaviside function acts to cut it off at finite endpoints, where $\delta^2 = \Xi_L^2 + \Xi_S^2 + 2\Xi_S\Xi_L \cos\Delta\varphi$. 

In order to evaluate the $\Xi_S$ integral, we will therefore need the following definite integral result: 
\begin{align}
    \int_a^b dy \frac{2+y^2}{\sqrt{4+y^2}} &= \frac{b\sqrt{4+b^2} - a\sqrt{4+a^2}}{2}\:.
\end{align}
There are two cases to consider (1) $\delta \geq \Xi_L$, and (2) $\delta < \Xi_L$.

\emph{Case (1)---}%
The argument of the Heaviside function is positive at $\Xi_S = 0$, so the lower limit of the $\Xi_S$ integral remains $\Xi_S=0$. 
The upper limit of $\Xi_S$ depends on $\Delta\varphi$: $\Xi_S^{\text{max}}(\Delta\varphi) = - \Xi_L \cos\Delta\varphi + \lb[ \delta^2 - \Xi_L^2\sin^2\Delta\varphi \rb]^{1/2}$, but the angle $\Delta\varphi$ can run unrestricted from 0 to $2\pi$.
We therefore have
\begin{align}
    \bar{\mu}(\bm{\chi}_L) &= \frac{1}{2 \pi \delta^2 } \int_0^{2\pi} d\Delta\varphi\cdot \Xi_S^{\text{max}}(\Delta\varphi) \sqrt{4+[\Xi_S^{\text{max}}(\Delta\varphi)]^2} \:. 
\end{align}
This angular integral must be evaluated numerically.\\

\emph{Case (2)---}%
The argument of the Heaviside function runs negative at both small and large $\Xi_S$. 
This necessarily implies an angular restriction on $\Delta\varphi$: $\pi-\kappa \leq \Delta\varphi \leq \pi+\kappa$ where $\kappa>0$ and $\sin \kappa = \delta/\Xi_L < 1 \Rightarrow \kappa<\pi/2$.
To see why this is the case, note that if we solve for the zeros of the argument of the Heaviside function subject to $\Xi_S>0$, we find $\Xi_S^{\text{min}}(\Delta\varphi) = - \Xi_L \cos\Delta\varphi - \lb[ \delta^2 - \Xi_L^2\sin^2\Delta\varphi \rb]^{1/2}$ and $\Xi_S^{\text{max}}(\Delta\varphi) = - \Xi_L \cos\Delta\varphi + \lb[ \delta^2 - \Xi_L^2\sin^2\Delta\varphi \rb]^{1/2}$.
The restriction on $\Delta\varphi$ keeps the argument of the square-root positive, and $\cos\Delta\varphi<0$ which leads to $\Xi_S^{\text{min}}(\Delta\varphi)>0$; specifically, $\Xi_S^{\text{min}}(\Delta\varphi)$ reaches its minimum value at $\Delta\varphi = \pi$: $\Xi_S^{\text{min}}(\pi) = +\Xi_L - \delta >0$ (the latter inequality by assumption in this case).
We thus have 
\begin{widetext}
\begin{align}
    \bar{\mu}(\bm{\chi}_L) &= \frac{1}{2 \pi \delta^2 } \int_{\pi-\arcsin(\delta/\Xi_L)}^{\pi+\arcsin(\delta/\Xi_L)} d\Delta\varphi\cdot \lb[ \Xi_S^{\text{max}}(\Delta\varphi) \sqrt{4+[\Xi_S^{\text{max}}(\Delta\varphi)]^2} - \Xi_S^{\text{min}}(\Delta\varphi) \sqrt{4+[\Xi_S^{\text{min}}(\Delta\varphi)]^2}\rb] \:.
\end{align}
This angular integral must again be evaluated numerically.

\subsection{Gaussian Source}
\label{app:muGApprox}
In this subsection, we give reasonably good approximations to the source-averaged magnification for the Gaussian case.
The result for $\bar{\mu}$ at \eqref{eq:muG} can be approximated as
\begin{align}
    \bar{\mu}(\tilde{z},\delta) &\approx \begin{cases}
        1 + \sqrt{\dfrac{\pi}{2}} \dfrac{1}{\delta} \lb[ 1 + \lb( \sqrt{\dfrac{\pi}{2}} \tilde{z} \lb\{  1 + \lb[ \dfrac{1}{2} \lb( \tilde{z} \delta \rb)^3 \rb]^{1/2} \rb\}^{\!2\, } \rb)^{\!\!4} \ \rb]^{-1/4} &;\: \delta \ll 1\:; \\[5ex]
        1 + \dfrac{\exp[-\tilde{z}^2/2]}{\delta^2} + \dfrac{2}{\delta^4} \lb( \dfrac{\tilde{z}^4}{1+\tilde{z}^8}\rb) &;\: \delta \gg 1\:.
    \end{cases}\label{eq:muGApprox}
\end{align}
\end{widetext}
These approximations were obtained via the analytical consideration of the integral at \eqref{eq:muG} in various extreme limits (i.e., $\gg 1$ or $\ll 1$) of the ratios of angular scales $\delta$, $\tilde{z}$, and $\tilde{z}\delta$; we then found numerically reasonably accurate smooth interpolations between those limiting cases.

We recall here that $\tilde{z} \equiv k_G \beta_L/\theta_S \sim 2\beta_L/\theta_S $ measures the location of the lens (relative to the source centroid) on the scale of the angle subtended by the source, while $\delta \equiv \theta_S/(k_G \theta_E)\sim \theta_S/(2 \theta_E)$ measures the angular size subtended by the source on the scale of the Einstein angle (here, we used $k_G \sim 2$ as appropriate for $f_{\text{source}}= 0.9$).
Note then that $\tilde{z}\delta = \beta_L/ \theta_E = y_L$ measures the lens location relative to the source centroid on the scale of the Einstein angle.

\section{Analytic Scalings of the Picolensing Cross Section}
\label{app:sigmaGaussianAnalyt}
In this appendix we consider three limiting cases for the picolensing cross section that are most relevant for the L2 and AU observer baselines [cf.~\figref{fig:sourceSize} and the discussion in \secref{sec:results}]: (1) pointlike sources with widely separated observers, relevant for intermediate lens masses; (2) pointlike sources with more closely spaced observers, relevant for large lens masses; and (3) extended sources with widely separated observers, relevant for small lens masses.
For each case, we analytically derive the parametric scalings of the PBH limit projection $f_{\pbh}^{\text{limit}}$ with various parameters such as the lens mass $M$, observer separation $R_O$, and source size $\theta_S$. 

Our derivations in this section are phrased in terms of the Gaussian magnification results from \appref{app:muGApprox}; note however that because the first two cases involve pointlike sources, those results apply equally to disk sources.
Coincidentally, the results of case (3) also work very well for the disk case; see comments in the final paragraph of \secref{para:DiskCase}.

While there are also other cases we could consider (e.g., small observer separations with extended sources), they are in disadvantageous regimes for picolensing detection (e.g., the LEO baseline at small lens masses), so we do not consider them here.

\subsection{Pointlike Source, Large Observer Separation}
\label{app:PointS_LargeRsep}
The first case we consider is where the lens-plane-projected source size is much smaller than the Einstein radius, which is in turn much smaller than the lens-plane-projected observer separation.
That is 
\begin{align}
    \Delta \chi_{S,\perp} \xi \ll 2 \chi_E \ll R_{O}' (1-\xi)\:, 
    \label{eq:hierarchyCase1}
\end{align}
where $R_O' \equiv R_O |\sin\theta_O|$ and $\xi \equiv \chi_L/\chi_S$.
This implies $\theta_S \ll \theta_E \ll (R_{O}'/(2\chi_S))\cdot(1-\xi)/\xi$.
In this case, we expect that the picolensing volume consists of two narrow tubes around each line of sight, with only one line of sight significantly magnified given a single lens; that is, the picolensing region in a fixed lens plane will be two disjoint, approximately circular regions with one circle centered on each line of sight [except very near the source where $\xi \approx 1$, which breaks the assumed hierarchy at \eqref{eq:hierarchyCase1}].

Suppose without loss of generality that observer 1 has their view of the source significantly lensed, and that observer 2 does not: $\mu_1 = \mu_*$ and $\mu_2 \approx 1$.
At the detectability threshold, we need [cf.~\eqref{eq:SNR}]
\begin{align}
    \varrho_* \approx \frac{ ( \mu_* - 1 ) \bar{S} }{ \sqrt{ 2\bar{B} + (1+\mu_*) \bar{S}} }\:, \label{eq:rhoStar}
\end{align}
which can be inverted to find the minimum required single-observer magnification for detection:
\begin{align}
    \mu_* &= 1 + \frac{\varrho_*^2}{2\bar{S}} \lb[ 1 + \sqrt{ 1 + \frac{8(\bar{S}+\bar{B})}{\varrho_*^2} } \rb]\:. \label{eq:muStar}
\end{align}

Because lensing is strong when the lens is within approximately an Einstein radius of the line of sight, we expect that the boundary of the approximately circular picolensing region around the observer-1 line of sight for $\varrho_* \sim 5$ will actually lie on a circle with angular size $\beta^*_L \sim (\text{few}) \times \theta_E$.
Moreover, because $\theta_S \ll \theta_E$, we also have $\beta^*_L \gg \theta_S$.
In terms of the variables $\tilde{z}$ and $\delta$ used in \appref{app:muGApprox}, we thus expect to have $\delta \ll 1$, $\tilde{z} \gg 1$, and $\tilde{z}\delta \gg 1$, in which limit we have $\bar{\mu} \approx 1 + 2/(\tilde{z}\delta)^4 = 1 + 2\theta^4_E/ \beta_L^4$ [cf.~\eqref{eq:pointLPointSmag}].
Therefore,
\begin{align}
    \beta^*_L &\approx \theta_E\lb\{ \frac{\varrho_*^2}{4\bar{S}} \lb[ 1 + \sqrt{ 1 + \frac{8(\bar{S}+\bar{B})}{\varrho_*^2} } \rb] \rb\}^{1/4} \equiv \kappa_* \theta_E\:,
\end{align}
where $\kappa_* = \kappa_*(\bar{S},\bar{B},\varrho_*)$ is a constant.
The total picolensing cross section is then (the leading factor of 2 is to account for the fact that the lens can be closer to either line of sight)
\begin{align}
    \sigma(\xi) &\approx 2 \times \pi ( \chi_L \beta^*_L)^2 \label{eq:factorof2}\\
    &= 2\pi \kappa_*^2 \chi_E^2 \\
    &= 8\pi \kappa_*^2 G_{\text{N}}M\chi_S [1+z_L(\xi\chi_S)] \xi ( 1- \xi )\:,
\end{align}
where we have written $z_L = z_L(\xi\chi_S)$ to remind the reader that the lens redshift is a (nonlinear) function of $\chi_L = \xi \chi_S$.
If we define 
\begin{align}
    1 + \bar{z}_L &\equiv \frac{ \int_0^1  [1+z_L(\xi\chi_S)]  \xi(1-\xi)d\xi}{ \int_0^1 \xi(1-\xi)d\xi} \\
    &= 6 \int_0^1  [1+z_L(\xi\chi_S)]  \xi(1-\xi)d\xi\:, 
\end{align}
we have for the picolensing volume
\begin{align}
    \mathcal{V} &= \int_0^{\chi_S} d\chi_L\, \sigma(\chi_L)\\
        &= \chi_S \int_0^1 d\xi\,  \sigma(\xi) \\
        &= \frac{4\pi}{3} \kappa_*^2 G_{\text{N}}M\chi_S^2 (  1 + \bar{z}_L )\:,
\end{align}
where we have assumed that the hierarchy at \eqref{eq:hierarchyCase1} holds over the entire integration domain (which is a good approximation everywhere except near the source, ${\xi\sim 1}$).

Because $\kappa_*$ depends on $T_{90}$ and $F_S$ through $\bar{S}$, we see that $\mathcal{V}$ here depends on $T_{90}$, $F_S$, and $z_S$ (via $\chi_S$ and $\bar{z}_L$).
But it does not depend on $R_O'$, at least so long as \eqref{eq:hierarchyCase1} is well satisfied.
Moreover, $\mathcal{V} \propto M$, which implies that $f_{\pbh}^{\text{limit}} \propto M / \overline{\mathcal{V}}$ would be both $R_O$- and $M$-independent in this regime if most sources satisfy \eqref{eq:hierarchyCase1}.

To clarify when \eqref{eq:hierarchyCase1} is satisfied, let us evaluate it at $\xi = 1/2$ and $z_S \sim 1$, which implies $\chi_S \sim 1.13 (h/H_0) \sim 0.766 H_0^{-1}$ and $z_L \sim 0.43$:
\begin{align}
    \frac{H_0/h}{12.9} (\Delta \chi_{S,\perp} )^2 \lesssim 2 G_{\text{N}} M \lesssim \frac{H_0/h}{12.9} (R_{O}')^2\:.
\end{align}
Supposing that $\Delta \chi_{S,\perp} \sim 10 R_{\odot} (1 + z_S ) \sim 20 R_{\odot}$ and $R'_O \sim 1\,\text{AU}$, this gives $5 \times 10^{-11}M_{\odot} \lesssim M \lesssim 6\times 10^{-9}M_{\odot}$; meanwhile, for $\Delta \chi_{S,\perp} \sim 2 R_{\odot}$ the lower limit would move to $5\times 10^{-13}M_{\odot}$.

\subsection{Pointlike Source, Small Observer Separation}
\label{app:PointS_SmallRsep}
Now consider the case where the source remains pointlike, but the Einstein radius grows to exceed the lens-plane-projected observer separation.
That is, we have the hierarchy
\begin{align}
    \Delta \chi_{S,\perp} \xi  \ll R_{O}' (1-\xi)\ll 2 \chi_E\:, 
    \label{eq:hierarchyCase2}
\end{align}
We thus still have $\delta \sim \theta_S/(2 \theta_E) \ll 1$ and $\tilde{z} \sim (2\beta_L/\theta_S) \gg 1$.
However, we make no \emph{a priori} assumption about $\tilde{z}\delta = \beta_L / \theta_E = y_L$, because its value on the contour where $\varrho = \varrho_*$ is not \emph{a priori} as clear as it was in the case of well-separated observers.
In this set of limits, we have from \eqref{eq:muGApprox} that
\begin{align}
    \bar{\mu} \sim 1 + \lb[ y_L \lb( 1 + \sqrt{y_L^3/2} \rb)^2 \rb]^{-1}\;. 
    \label{eq:muApproxClose}
\end{align}
However, because $ R_{O}' (1-\xi)\ll 2 \chi_E$, both observers will see the lens near enough to their respective lines of sight to the source to cause a non-trivial magnification, so now the difference of magnifications that enters \eqref{eq:SNR} will be
\begin{align}
    |\mu_1 - \mu_2| &\sim \lb|\frac{\partial\bar{\mu}}{\partial y_L}\rb| |\Delta y_L| \:,
\end{align}
where 
\begin{align}
    |\Delta y_L| &= |y_1 - y_2 | \sim \frac{ R'_{O} |\cos\phi_L|}{\chi_S \theta_E} \frac{ 1-\xi }{ \xi } \ll 1\:,
\end{align}
where we have taken the values of $y_i$ from \eqref{eq:yi} in the limits $R_O'/\chi_E^O \ll 1 , \Delta \chi_{L,\perp}/\chi_E^L$.

Because of the suppression by $|\Delta y_L|$, we expect that the magnifications for each observer must be individually quite large in order to see a large picolensing SNR, so we will assume $y_L \lesssim 1$ (but not necessarily $y_L \ll 1$) in \eqref{eq:muApproxClose}, leading to $\bar{\mu} \sim 1 + 1/y_L$ [cf.~\eqref{eq:pointLPointSmag}] and $\lb|\partial\bar{\mu}/\partial y_L\rb| \sim 1/y_L^{2}$.
If we let $y_L^*$ be the value of $y_L$ such that a threshold detection is achieved, $\varrho = \varrho_*$, we have [cf.~\eqref{eq:SNR}]
\begin{align}
    \varrho_* \sim \frac{|\Delta y_L|}{(y_L^*)^{2}} \frac{\bar{S}}{\sqrt{2(\bar{B}+\bar{S}) + 2\bar{S}/y_L^*}}\:.
\end{align}
Provided that $y_L^* \gtrsim \bar{S}/(\bar{B}+\bar{S})$, 
\begin{align}
    (y_L^*)^{2} &\sim \frac{|\Delta y_L|}{\varrho_*} \frac{\bar{S}}{\sqrt{2(\bar{B}+\bar{S})}}\\
    &\sim \frac{R_{O}'(1-\xi) |\cos\phi_L|}{\varrho_*\xi \chi_S \theta_E} \frac{\bar{S}}{\sqrt{2(\bar{B}+\bar{S})}}\:.
\end{align}
The validity criterion $y_L^* \gtrsim \bar{S}/(\bar{B}+\bar{S})$ will therefore fail near the source ($\xi \sim 1$), or when the lens is close to the vertical reflection symmetry axis between the two observers ($|\phi_L| \sim \pi/2$).

Ignoring any violations of that validity criterion, it follows that $\sigma$ is given approximately by
\begin{align}
    \sigma &\sim \frac{1}{2}\chi_E^2  \int_0^{2\pi} (y_L^*)^2 d\phi_L \label{eq:nofactorof2} \\
    &= 2 \chi_E^2 (y_L^*)^2|_{\phi_L=0} \\
    &= R_{O}'\chi_S \theta_E \frac{\sqrt{2} \bar{S}}{\varrho_*\sqrt{\bar{B}+\bar{S}}} \xi (1-\xi)  \\
    &= \frac{\bar{S} R_O' \sqrt{8G_{\text{N}} M \chi_S } }{\varrho_*\sqrt{\bar{B}+\bar{S}}} \sqrt{[1+z_L(\xi\chi_S)] \xi(1-\xi)^3 }\:.
\end{align}
Note that there is no extra factor of 2 in \eqref{eq:nofactorof2} since this estimate already correctly accounts for a lens that is nearer to either of the two lines of sight; cf.~\eqref{eq:factorof2}.

We would then have the picolensing volume
\begin{align}
    \mathcal{V} &\sim \frac{\bar{S} R_O'\chi_S \sqrt{8G_{\text{N}} M \chi_S } }{\varrho_*\sqrt{\bar{B}+\bar{S}}} \nl\quad \times
    \int_0^{1} d\xi \sqrt{[1+z_L(\xi\chi_S)] \xi(1-\xi)^3 }\: .
\end{align}
Defining 
\begin{align}
    \sqrt{1+\bar{z}'_L} &\equiv \frac{ \int_0^{1} d\xi \sqrt{[1+z_L(\xi\chi_S)] \xi(1-\xi)^3 }}{ \int_0^{1} d\xi \sqrt{\xi(1-\xi)^3 } }\\
    &= \frac{16}{\pi} \int_0^{1} d\xi \sqrt{[1+z_L(\xi\chi_S)] \xi(1-\xi)^3 }\:,
\end{align}
we can write
\begin{align}
    \mathcal{V} &\sim \frac{\pi \bar{S} R_O'\chi_S^{3/2} \sqrt{2G_{\text{N}} M } }{8 \varrho_*\sqrt{\bar{B}+\bar{S}}} \sqrt{1+\bar{z}'_L} \: .
\end{align}
If we average this over all possible sky locations of the GRB source with respect to the observer-separation baseline (i.e., we average over $\theta_O$ in $R_O' = R_O |\sin\theta_O|$), we obtain
\begin{align}
    \mathcal{V} &\sim \frac{\pi^2 \bar{S} R_O\chi_S^{3/2} \sqrt{2G_{\text{N}} M } }{32 \varrho_*\sqrt{\bar{B}+\bar{S}}} \sqrt{1+\bar{z}'_L} \: .
\end{align}

While this estimate is rough at the level of the numerical factors, its parametric scaling is the important result: $\mathcal{V} \propto R_O\sqrt{2G_{\text{N}} M}\chi_S^{3/2}$.
This tells us that, if the majority of sources are in this limit, $f_{\pbh}^{\text{limit}} \propto M / \overline{\mathcal{V}} \propto \sqrt{M} / R_O$.

\subsection{Extended Source, Large Observer Separation}
\label{app:ExtendedS_LargeRsep}
The last case we consider explicitly is when the observers are widely separated, but the source size becomes large compared to the source-plane-projected Einstein radius:
\begin{align}
    2 \chi_E  &\ll \Delta \chi_{S,\perp} \xi \ll R_{O}' (1-\xi)\\
    \Rightarrow \theta_E &\ll  \theta_S \ll (R_{O}'/(2\chi_S)) \cdot (1-\xi)/\xi \:.
    \label{eq:hierarchyCase3}
\end{align}
Note that $\theta_S / \theta_E \gg 1 \Rightarrow  \delta \gg 1$.

Because we assume that the observers are widely separated on the scale of the source size (both as projected to the lens plane), which is in turn larger than the Einstein radius, it will be the case that a lens location that gives a large magnification for one observer will give almost no magnification ($\bar{\mu} \sim 1$) for the other observer, since it will necessarily be true that $\beta_L/\theta_E \gg \beta_L/\theta_S \gg 1$ with respect to the second observer.
On the other hand, for the observer that does see significant lensing, $\bar{\mu} \sim 1+1/\delta^2 $ for $\tilde{z} \lesssim 1$ where $\delta \sim \theta_S/(2\theta_E)$ and $\tilde{z} \sim 2\beta_L/\theta_S$, while $\bar{\mu}-1 \ll 1$ for larger $\tilde{z}$.
We therefore have that 
\begin{align}
    |\mu_1 - \mu_2 | \approx
    \begin{cases} 
        \delta^{-2} & ;\:  \tilde{z}_1 \lesssim 1 \text{ or } \tilde{z}_2 \lesssim 1 \:; \\
        0           &;\:   \tilde{z}_1 \gtrsim 1 \text{ and } \tilde{z}_2 \gtrsim 1\:.
    \end{cases}
\end{align}

Because we are again in the limit of widely separated observers previously discussed in \appref{app:PointS_LargeRsep}, the results at \eqref[s]{eq:rhoStar} and (\ref{eq:muStar}) can be applied, provided we set $\mu_* \sim 1+1/\delta_*^2$.
This leads to the additional constraint for picolensing detectability
\begin{align}
    \delta^2 \lesssim \delta_*^2 &= \frac{2\bar{S}}{\varrho_*^2} \lb[ 1 + \sqrt{ 1 + \frac{8(\bar{S}+\bar{B})}{\varrho_*^2} } \rb]^{-1}\:,\label{eq:thresholdSNRcase3}
\end{align}
which applies when $\tilde{z}_1 \lesssim 1$ or $\tilde{z}_2 \lesssim 1$.
Note that $\delta_*$ depends only on $\bar{S}$, $\bar{B}$, and $\varrho_*$.

Together, these results imply that the picolensing cross section in this case will be
\begin{align}
    \sigma &\approx \begin{cases}
        2 \times \pi [ ( \theta_S / 2 ) \times \chi_L ]^2             &;\:   1\lesssim \delta\lesssim\delta_{*}\:; \\[1ex]
        0                                       &  ;\: \delta\gtrsim\delta_{*}
    \end{cases} \\[1ex]
    &    \approx \begin{cases}
        \dfrac{\pi}{8} (\Delta \chi_{S,\perp})^2 \xi^2  &;\:   1\lesssim \delta\lesssim\delta_{*}\:;  \\[1ex]
        0                                       &   ;\: \delta\gtrsim\delta_{*}\:.
    \end{cases} \label{eq:sigmaCase3}
\end{align}

The restriction $\delta\lesssim \delta_*$ implies a restriction on the range of the $\chi_L$ integral over this picolensing cross section to obtain the picolensing volume via \eqref{eq:VlensConst}.
This is because $\delta \lesssim \delta_*$ implies
\begin{align}
    \delta_* \gtrsim \frac{ \Delta \chi_{S,\perp} }{ 4 \chi_S \theta_E } &\approx \frac{ \Delta \chi_{S,\perp} }{ 4 \chi_S \theta'_E  \sqrt{ (1-\xi) / \xi} } \\
    \Rightarrow  \sqrt{ \frac{1-\xi}{\xi} } &\gtrsim \frac{ \Delta \chi_{S,\perp} }{ 4 \chi_S  \theta'_E   \delta_* }\:\\
    \Rightarrow 0 \lesssim \xi &\lesssim \lb[ 1 + \lb(\frac{ \Delta \chi_{S,\perp} }{ 4 \chi_S  \theta'_E   \delta_* }\rb)^2 \rb]^{-1} \equiv \xi_{\text{max}}\:,
\end{align}
where we defined $\theta'_E \equiv \theta_E(\xi = 1/2)$, and ignored the $\xi$-dependence in the factor of $[1+z_L(\xi\chi_S)]$ that appears in the definition of $\theta_E(\xi)$ [cf.~\eqref{eq:thetaE1}] by approximating $z_L(\xi) \approx z_L(\xi=1/2)$, which leads to an error in this estimate by a factor of $\mathcal{O}(1)$.

Therefore, we have the picolensing volume
\begin{align}
    \mathcal{V} 
    &\approx \dfrac{\pi}{24} \chi_S (\Delta \chi_{S,\perp})^2 \xi_{\text{max}}^3 \label{eq:VleadingFactor} \\
    &= \dfrac{\pi}{24} \chi_S (\Delta \chi_{S,\perp})^2\lb[ 1 + \lb(\frac{ \Delta \chi_{S,\perp} }{ 4 \chi_S \theta_E'  \delta_* }\rb)^2 \rb]^{-3} \:.
\end{align}
Supposing that we are deep in the regime $\theta_E \ll \theta_S$, the $[\ \cdots]$-bracket is dominated by the second term appearing inside the bracket, and 
\begin{align}
    \mathcal{V}
    &\sim \dfrac{\pi}{24} \chi_S (\Delta \chi_{S,\perp})^2  \lb(\frac{ 4 \chi_S \theta_E'  \delta_* }{ \Delta \chi_{S,\perp} }\rb)^6 \\
    &\sim \dfrac{\pi}{24} \chi_S (\Delta \chi_{S,\perp})^2  \lb(\frac{ 64 \chi_S \delta_*^2G_{\text{N}}M(1+z_L'')}{ (\Delta \chi_{S,\perp})^2 } \rb)^3\\
    &\propto \frac{ \chi_S^4 (G_{\text{N}} M)^3 }{ (\Delta \chi_{S,\perp})^4 } \delta_*^6 \\ 
    &\propto \frac{ (G_{\text{N}} M)^3 }{ \theta_S^4 } \delta_*^6
    \:, \label{eq:inverseFourthPower}
\end{align}
where $z_L'' \approx z_L(\xi \sim 0.5)$ is a representative lens redshift, but we have ultimately discarded the numerical prefactor to focus only on the parametric dependence of the result.

The picolensing volume $\mathcal{V}$ as given at \eqref{eq:inverseFourthPower} is an extremely steep function of both $M$ and $\theta_S$.
Therefore, as $M$ is decreased, and $\theta_E$ drops below $\theta_S$ for each source in turn, individual sources will rapidly cease to contribute to the source-averaged picolensing volume, suppressing $\overline{\mathcal{V}}$.
As such, the GRB-population distribution of $\theta_S$ becomes critically important in setting the $M$-dependence of $f_{\pbh}^{\text{limit}}$ at small $M$, and we cannot give a simple scaling law for how $f_{\pbh}^{\text{limit}}$ will behave in this regime; i.e., the $M$-dependence of the limit in this regime is a convolution of the $M$-scaling in \eqref{eq:inverseFourthPower} and the source-population distribution of $\theta_S$.

However, at very small $M$, there is a rescaling rule that can be derived analytically.
In the low-mass regime, where $\theta_S^j \gg \theta_E(M)$ for nearly all sources $j$ (and the observers are widely separated), one can write the average picolensing volume as $\overline{\mathcal{V}} = \aleph^{-1} \sum_j \mathcal{V}_j = \aleph^{-1} \mathcal{V}_* \sum_j (\theta_S^*/\theta_j)^4$, where $\mathcal{V}_* \propto (G_{\text{N}} M)^3/(\theta_S^*)^4$ is the largest picolensing volume to any individual source, corresponding roughly to the source whose angular size $\theta_S^*$ is the smallest. 
However, because $\theta_S$ ranges over some orders of magnitude (cf.~\figref{fig:rShist}), the sum is dominated by only a small number of sources with sizes $\theta_S^j \sim \theta_S^*$; let us call that number of sources $N_*$.
It then follows that $\overline{\mathcal{V}}\sim (N_*/\aleph)\times \mathcal{V}_*  \propto (N_*/\aleph) ( G_{\text{N}} M )^3 /(\theta_S^*)^4$. 

Now, rescaling $\theta_S \rightarrow \kappa \theta_S$ simultaneously for all sources clearly leaves $N_*$ unchanged provided that \emph{the majority of the sources} are in the regime where $\theta_S^j \gg \theta_E(M)$ (i.e., $M$ is small enough) because, in that regime, $N_*$ just counts the number of sources that satisfy $(\theta_S^*/\theta_j)^4\sim 1/N_*$.
This rescaling alone thus sends $\mathcal{V}_* \rightarrow \kappa^{-4} \mathcal{V}_*$.
If we however simultaneously rescale $M\rightarrow \kappa^2M$, the net rescaling of $\mathcal{V}_* \propto  (G_{\text{N}}M)^3/(\theta_S^*)^4$ is $\mathcal{V}_* \rightarrow \kappa^2 \mathcal{V}_*$.
In turn, we then have $\overline{\mathcal{V}} \rightarrow \kappa^2 \overline{\mathcal{V}}$.
However, this rescaling is exactly compensated by the rescaling of the factor of $M$ in the numerator of \eqref{eq:fDMlimit}: $M/\overline{\mathcal{V}} \rightarrow M/\overline{\mathcal{V}}$.
It follows that $f^{\text{limit}}_{\pbh}$ is unchanged under this combined rescaling.

With reference to the alternative explanation and discussion given in \secref{sec:results}, note that under $\theta_S \rightarrow \kappa \theta_S$ and $M \rightarrow \kappa^2 M \Rightarrow \theta_E \rightarrow \kappa \theta_E $, we have $\xi_{\text{max}} \rightarrow \xi_{\text{max}}$; therefore, the only rescaling of $\mathcal{V}$ is by the overall factor of $\theta_S^2$ implicit in the factor of $(\Delta\chi_{S,\perp})^2$ that appears in \eqref{eq:VleadingFactor}, leading in turn to $\mathcal{V} \rightarrow \kappa^2 \mathcal{V}$, as claimed in \secref{sec:results}.

\section{Alternative Criterion for Exclusion-Limit Projections}
\label{app:differentApproach}

\begin{figure*}[t]
    \centering
    \includegraphics[width=\textwidth]{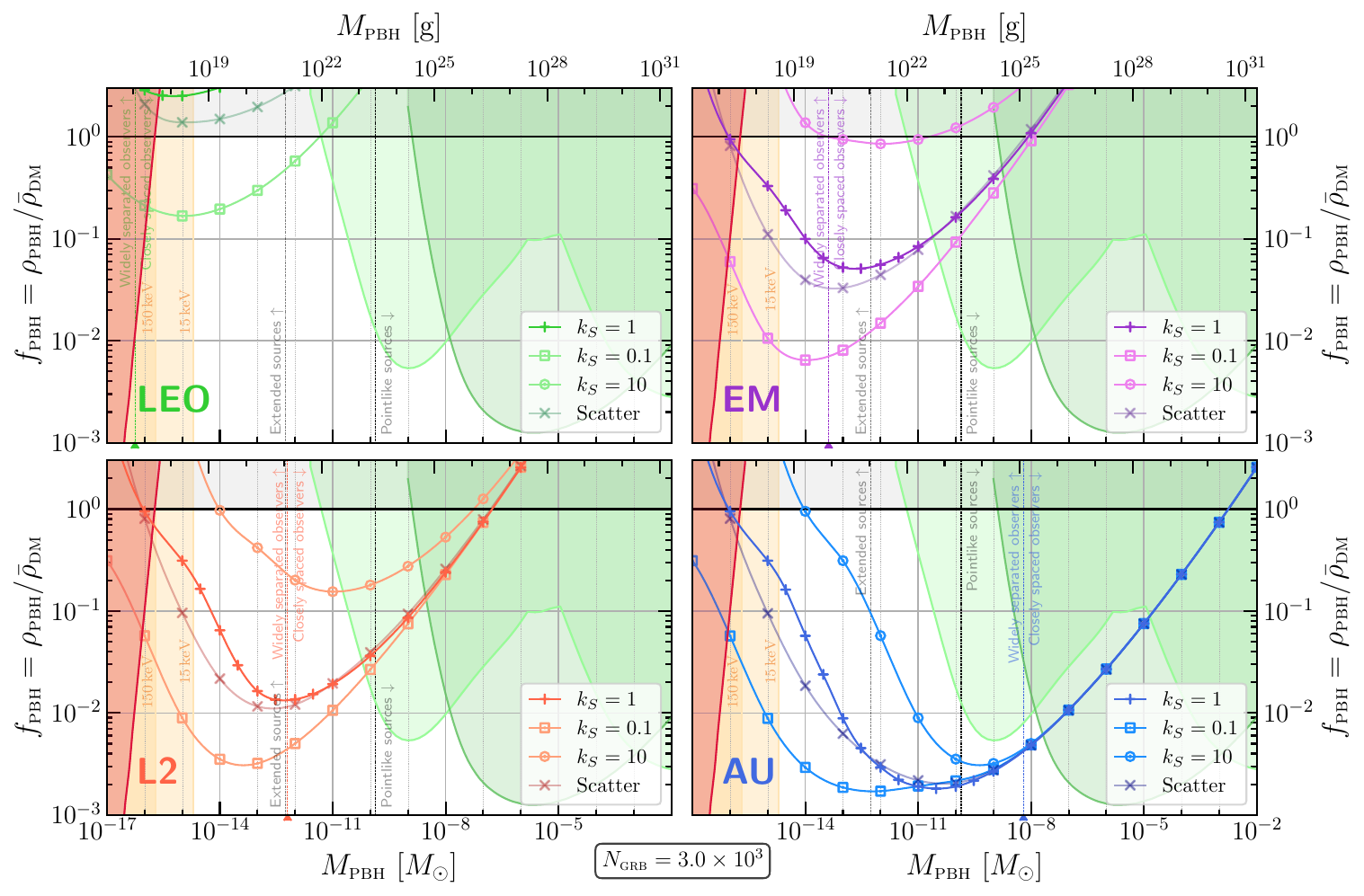}
    \caption{\label{fig:sourceSize-altfDM}%
    As for \figref{fig:sourceSize}, except that we have used the alternative criterion $P^{(N_{\grb})}_{\text{no single lensing}} = 1-\alpha_{95}$ as defined at \eqref[s]{eq:PnosingleLensA}--(\ref{eq:altCriterion}) to make the exclusion-limit projections.
    }
\end{figure*}

In the analysis in the main text (\secref{sec:probAndExclusion}) we used as the exclusion-limit criterion the absence of \emph{any} lenses in the combined picolensing volume of $N_\grb$ sources: $P^{(N_{\grb})}_{\text{no lens}} \equiv [ P^{(\aleph)}_{\text{no lens}} ]^{N_{\grb}/\aleph} = 1-\alpha_{95}$, from which we arrived at \eqref{eq:fDMlimit}. 
The picolensing volume we used for this computation was derived from a picolensing cross section that is only accurate if there is a single lens in the picolensing volume. 
As a result, this computation does not accurately handle the case of multiple lenses being present in the picolensing volume.
Some multiple-lens configurations would be expected to still give rise to detectable differential magnification, while others would not (e.g., exactly one lens lying exactly on each line of sight).

An alternative approach avoiding the complications with multiple lenses is to demand instead the absence of any picolensing volumes containing \emph{exactly one} lens~\cite{Jung:2019fcs,Gawade:2023gmt}.
Let us apply this procedure.
We define
\begin{align}
    P^{(\aleph)}_{\text{no single lensing}} &= \Pi_{j=1}^{\aleph} \Big[ 1- \text{Pr}(\mathfrak{n}=1,j) \Big] \label{eq:PnosingleLensA} \\
    &= \Pi_{j=1}^{\aleph} \lb[ 1 - \tau_j(M,\varrho_*) e^{-\tau_j(M,\varrho_*)} \rb] \:,
\end{align}
which is the probability for the absence of any single-lens picolensing volumes in a representative sample of $\aleph$ GRB events.
We then set a 95\%-confidence exclusion-limit projection for $N_{\grb}$ events as
\begin{align}
    P^{(N_{\grb})}_{\text{no single lensing}} &\equiv \lb( P^{(\aleph)}_{\text{no single lensing}} \rb)^{N_{\grb}/\aleph} 
    = 1-\alpha_{95}\:. \label{eq:altCriterion}
\end{align}
We show the results of this alternative analysis in \figref{fig:sourceSize-altfDM}.

Comparing to the cognate results in \figref{fig:sourceSize}, the largest differences between the two approaches in all physically relevant parts of parameter space are seen to be $\mathcal{O}(20\%)$ and then only for $M \lesssim 10^{-14}M_{\odot}$, where fewer GRBs are relevant for setting limits.
The larger-$M$ results are essentially unchanged. 

We can understand this as follows.
Provided that $N_{\grb} \gg 1$ sources are relevant in setting the exclusion-limit projection, the discussion below \eqref{eq:opticalDepth2} makes clear that that exclusion-limit projection automatically lies in the $\bar{\tau} \ll 1$ regime.
Because the two approaches to setting limits differ only in their handling of the case with $\mathfrak{n}\geq 2$ lens in the picolensing volume, which is rare whenever $\tau \ll 1$, we therefore expect these approaches to give similar results.
More specifically, in the limit $\tau_j \ll 1$, we have $1 - \text{Pr}(\mathfrak{n}=1,j) \approx 1 - \tau_j(M,\varrho_*) \approx P^{(1),j}_{\text{no lens}}$ [cf.~\eqref{eq:statisticalLimitCriterion}].
If, for the sake of argument, were we to take all $N_{\grb}$ sources to have the same optical depth $\bar{\tau}$, then $P^{(N_{\grb})}_{\text{no single lensing}} / P^{(N_{\grb})}_{\text{no lens}} \approx 1 + N_{\grb} \bar{\tau}^2/2$; however, $\bar{\tau}\sim 3/N_{\grb}$ at the limit projection [see below \eqref{eq:opticalDepth2}], so the fractional difference in the limit projections would be of order $\sim 5/N_{\grb}\ll 1$.

However, particularly at very small $M$, it is possible that the relevant number of GRBs that meaningfully contribute to the exclusion-limit projection is not very large, in which case the results from these two approaches might differ somewhat.
From the discussion in \appref{app:ExtendedS_LargeRsep}, we know this to be the relevant case at smaller masses, explaining why the largest differences occur there.

Overall, we judge the differences in our results arising from this difference in statistical approaches to be essentially negligible.
A more careful analysis would have to consider how multiple-lens scenarios give rise to differential-magnification events, but we judge this to be unnecessary for our application.
This would be more relevant when considering positive detection at larger optical depth; see \secref{sec:beyondConstraints}.

\bibliographystyle{JHEP.bst}
\bibliography{bibliography}

\end{document}